\documentclass[preprint2]{aastex}

\usepackage{graphicx}
\usepackage{txfonts}  
\usepackage{epsfig}

\usepackage{natbib}
\usepackage{times}
\usepackage{units}
\newcommand{\msun}{{\,\rm M}_{\odot}}
\newcommand{\lsun}{{\,\rm L}_{\odot}}

\slugcomment{To appear in The Astrophysical Journal Supplement Series}

\shorttitle{Looking into the hearts of Bok globules}
\shortauthors{Launhardt et al.}

\begin{document}

\title{Looking into the hearts of Bok globules:\\ 
MM and submm continuum images of isolated star-forming cores}

%\titlerunning{Looking into the hearts of Bok globules}

\author{R. Launhardt}
\affil{Max Planck Institute for Astronomy, K\"{o}nigstuhl 17,
       D-69117 Heidelberg, Germany}
\email{rl@mpia.de}

\and 

\author{D. Nutter, D. Ward-Thompson}
\affil{School of Physics and Astronomy, Cardiff University, 
	Queens Buildings, The Parade, Cardiff, CF24\,3AA, UK}

\and 

\author{T. L. Bourke}
\affil{Harvard-Smithsonian Center for Astrophysics, 
       60 Garden Street, Cambridge, MA 02138, USA}

\and 

\author{Th. Henning, T. Khanzadyan\altaffilmark{1}, M. Schmalzl, S. Wolf\altaffilmark{2}}
\affil{Max Planck Institute for Astronomy, K\"{o}nigstuhl 17,
       D-69117 Heidelberg, Germany}

\and 

\author{R. Zylka}
\affil{Institut de Radio Astronomie Millim\'etrique, 300 Rue de la Piscine, 
    Domaine Universitaire, 38406 Saint Martin d’H\`eres, France}

\altaffiltext{1}
	{Dept. Experimental Physics, National University of 
	Ireland Galway, Galway, Ireland}
\altaffiltext{2}
	{University of Kiel, Institute of Theoretical Physics and Astrophysics, 
   Leibnizstrasse 15, 24098 Kiel, Germany}

%\authorrunning{Launhardt et al.}

%\offprints{Ralf Launhardt, \email{rlau@mpia.de}}

\newpage
%%%%%%%%%%%%%%%%%%%%%%%%%%%%%%%%%%%%%%%%%%%%%%%%%%%%%%%%%%%%%

\begin{abstract}
We present the results of a comprehensive infrared, submillimetre, and 
millimetre continuum emission study of isolated low-mass star-forming cores in 32 Bok globules, 
with the aim to investigate the process of star formation in these regions.
The submillimetre and millimetre dust continuum emission maps together with the
spectral energy distributions are used to model and derive the physical 
properties of the star-forming cores, such as luminosities, sizes, masses, densities, etc. 
Comparisons with ground-based near-infrared and space-based mid and far-infrared images 
from Spitzer are used to reveal the stellar content of the 
Bok globules, association of embedded young stellar objects with the submm dust cores, and 
the evolutionary stages of the individual sources.
Submm dust continuum emission was detected in 26 out of the 32 globule cores observed.
For 18 globules with detected (sub)mm cores we derive evolutionary stages and physical parameters 
of the embedded sources. 
We identify nine starless cores, most of which are presumably prestellar,
nine Class\,0 protostars, and twelve Class\,I YSOs. 
Specific source properties like bolometric temperature, core size, and central densities are discussed 
as function of evolutionary stage.
We find that at least two thirds (16 out of 24) of the star-forming globules 
studied here show evidence of forming multiple stars on scales between 1,000 and 50,000\,AU.
However, we also find that most of these small prototstar and star groups 
are comprised of sources with different evolutionary stages, 
suggesting a picture of slow and sequential star formation in isolated globules.
\end{abstract}

%%%%%%%%%%%%%%%%%%%%%%%%%%%%%%%%%%%%%%%%%%%%%%%%%%%%%%%%%%%%%


%%%%%%%%%%%%%%%%%%%%%%%%%%%%%%%%%%%%%%%%%%%%%%%%%%%%%%%%%%%%%

\maketitle

\section{Introduction}                   \label{sec-intro}

%{\it 
%Star formation general, low-mass star formation, standard picture - 
%binarity, turbulence,\\
%structure of pre-protostellar and protostellar cores\\
%What is known about star formation in globules\\
%Short guide through chapters of the paper}

Different aspects of star formation can be studied on different size scales
and in different environments.
The large-scale distribution of star-forming regions and the relation between
molecular cloud life cycles, galactic spiral density waves, and star 
formation can be studied by observing nearby galaxies 
\citep[e.g.][]{1998A&A...333...92B,2009A&A...494...81S}. 
The stellar initial mass function (IMF), which is needed to interpret these 
data, is usually derived from rich young stellar clusters in our own Galaxy 
\citep[e.g.][]{2002Sci...295...82K,2003ApJ...586L.133C}. 
Dense star-forming dark cloud complexes such as the Orion and Ophiuchus molecular clouds 
are the places to study the relation between the molecular core mass spectrum (CMF) 
and the interstellar IMF 
\citep[e.g.][]{1998A&A...336..150M,2007MNRAS.374.1413N,2008MNRAS.391..205S,2008A&A...477..823G}. 

Nearby and more isolated star-forming cores, such as Bok globules, 
are the best places to study in detail the initial properties of 
individual star-forming cores,
their chemical evolution, kinematic structure, and the physics of
their collapse and fragmentation 
\citep[e.g.,][]{1988ApJS...68..257C,1995MNRAS.276.1052B,1997A&A...326..329L,1997MNRAS.288L..45L,
1998A&A...338..223H,2007prpl.conf...33W,2008ApJ...687..389S}.
Bok globules are small, simply-structured, relatively isolated, opaque
molecular clouds that often contain only one or two star-forming core. 
They are often not completely isolated, but reside in the filamentary 
outskirts of larger dark cloud complexes \citep{1997A&A...326..329L}, 
a fact that may tell something about their origin.
With their size, mass, densities, etc., Bok globules resemble small clumps in larger molecular clouds 
\citep[cf. ][]{2007ARA&A..45..339B}, only that they lack the surrounding cloud.
Table \ref{tbl-globprop} summarizes the average general properties of typical
Bok globules and their star-forming cores. 

\placetable{tbl-globprop}

Although they are the most simple star-forming molecular clouds,
many globules deviate considerably from spherical geometry.
They are often cometary or irregularly shaped. 
The dense star-forming cores are not always located at the center of the 
globule, but in cometary-shaped globules are often located closer to the sharper rim 
at the ``head'' side \citep[e.g.,][]{Launhardt:1996}.
Similarly, pre-stellar cores, which are the earliest stage of star formation 
\citep{1994MNRAS.268..276W,2007prpl.conf...33W} often appear to be fragmentary and filamentary. 
However, the protostellar cores and envelopes of the more evolved Class 0 \& I YSOs 
\citep{1987IAUS..115....1L,1993ApJ...406..122A}
are more spherically symmetric, which can be understood 
as a result of the gravitational collapse of the inner dense $R\sim5000$\,AU 
region. Many of these isolated cores were found to be the sources of bipolar 
molecular outflows, indicating the presence of embedded protostars 
\citep[e.g.,][]{1994ApJS...92..145Y,1995MNRAS.274.1219W,1996A&A...314..477B}.

In order to investigate the star-forming potential and 
evolutionary stages of Bok globules in the solar neighbourhood, we had 
surveyed a large number of globules for signs of star-forming cores, using 
as tracers, e.g., the mm dust continuum emission 
\citep{Launhardt:1996,1997A&A...326..329L,1998A&A...338..223H,1998ApJS..119...59L}, 
NH$_3$\ \citep{1995MNRAS.276.1067B}
or CS line emission \citep{1998ApJS..119...59L}.

In this paper we present a submillimetre and 
millimetre -- hereafter (sub)mm -- continuum study of 32 Bok globules, 
which were identified from these previous surveys 
as promising candidates for globules with currently ongoing star formation. 
The (sub)mm maps are complemented by deep near-infrared (NIR) images and NIR to mm 
spectral energy distributions (SEDs).
In Sect.\,\ref{sec-obs}, we describe the observations and data reduction. 
In Sect.\,\ref{sec-res}, we present the NIR and (sub)mm images, SEDs, and 
results on multiplicity, physical parameters, and evolutionary stages.
In Sect.\,\ref{sec-dis1}, we discuss in particular the source properties as 
function of evolutionary stage and the results on multiplicity.
In Sect.\,\ref{sec-dis}, we describe and discuss the individual globules, 
and Sect.\,\ref{sec-sum} summarizes the main results of this study.

%%%%%%%%%%%%%%%%%%%%%%%%%%%%%%%%%%%%%%%%%%%%%%%%%%%%%%%%%%%%%

\section{Observations and data reduction}           \label{sec-obs}

%%%%%%%%%%%%%%%%%%%%%%%%%%%%%

A total of 32 Bok globules were studied using a combination of NIR 
and (sub)mm continuum observations. 
Coordinates, distances, and observation summary of these sources are
given in Table~\ref{tbl-sourcelist}. 
Although many of the globules appear in different catalogues and
thus often have several names assigned, we use throughout this paper 
the CB name and number introduced by
\citet{1988ApJS...68..257C} for northern sources, and the BHR name and number
introduced by \citet{1995MNRAS.276.1052B} for southern sources.

\placetable{tbl-sourcelist}

%%%%%%%%%%%%%%%%%%%%%%%%%%%%%%%%%

\subsection{(Sub)millimetre continuum observations}   \label{sec-obs-mm}

Submillimetre and millimetre dust continuum observations at 450\,$\mu$m,
850\,$\mu$m, and 1.3\,mm were performed during several observing runs
between 1994 and 2002 at the 15-m James Clerk Maxwell Telescope{\footnote{JCMT is 
operated by the JAC, Hawaii, on behalf of the UK STFC,
The Netherlands OSR, and the Canadian NRC.}}
(JCMT) on Mauna Kea, Hawaii,
at the 30-m Institut de Radio Astronomie Millimetrique (IRAM) telescope on Pico Veleta, Spain, 
and at the 15-m Swedish-ESO Submillimetre Telescope (SEST) on La Silla, Chile.

\placetable{tbl-obsmm}

Observations at 1.3\,mm of the northern sources were carried out during four 
observing
runs between April 1994 and March 1997 using the MPIfR 7-channel (1994--95) and
19-channel (1996--97) bolometer arrays \citep{1998SPIE.3357..319K} at the IRAM 30-m 
telescope.
The maps were obtained with the standard double-beam technique described
by \citet{1979A&A....76...92E}.
The telescope was scanned continuously at 4\arcsec/sec in azimuth while
the wobbling secondary, operating at 2~Hz with a chopper throw of 30\arcsec\
along the scan direction, generated the dual beams. Scan lengths and map sizes were
adapted to source sizes to provide sufficient basline for background determination. 
Individual scan maps of a source, which were later combined,
were obtained at different hour angles in order to reduce scanning effects.

Observations at 1.3\,mm of southern sources were carried out in September and
November 2002 at the Swedish/ESO Submillimetre Telescope (SEST) at La Silla in Chile, 
using the 37-channel bolometer array SIMBA (SEST Imaging Bolometer Array).
Since SIMBA was used without a wobbling secondary mirror, several fast-scanning
maps (at 80\arcsec/sec) were obtained for each field at different hour angles
to reduce scanning effects.
Source BHR\,71 was already observed in 1995 with the MPIFR single-channel 
bolometer at SEST \citep{1997ApJ...476..781B}.
Individual maps at different hour angles were obtained with the same 
double-beam mapping technique used at the 30-m telescope, but by employing 
the SEST focal plane chopper.

The submillimetre observations were carried out
during three observing runs between August 1997 and September 2001
using the Submillimetre Common User Bolometer Array (SCUBA) on the JCMT.
The observations were conducted simultaneously at 850 and 450\,$\mu$m
using a 64-point jiggle pattern to create fully sampled maps of 2.3\arcmin\ 
diameter fields around the coordinate centers
\citep[for more details, see ][]{1999MNRAS.303..659H}.
Time-dependent variations in the sky emission were removed by
chopping the secondary mirror at 7.8\,Hz with a chop throw of 120\arcsec\
in azimuth.
During the 2000 and 2001 observations, we also obtained 850\,$\mu$m
polarization maps of six globule cores
\citep[see][]{2001ApJ...561..871H,2003ApJ...592..233W}.
In addition, the JCMT archive, operated by the Canadian Astronomy Data Centre (CADC),
was searched for additional data for the sources observed at 1.3\,mm. 
One of the sources (CB224) was observed using SCUBA in the scan-mapping mode. 
In this observing mode, the array is rastered across the sky to produce a rectangular map 
several arcminutes in extent. Variations in sky emission are again removed by chopping the secondary 
mirror, though six chop configurations are used, chopping in 
30, 44 and 68\arcsec\ in both RA and Dec 
\citep{1998SPIE.3357..548J}.

During all observing runs, the atmospheric transmission was determined
using the `skydip' method, which were performed every 2--3 hours.
At the JCMT, we also used polynomial fits to the 225~GHz atmospheric opacity data
which was measured by the Caltech Submillimeter Observatory (CSO) radiometer.
Average zenith optical depths for each run are listed in Table~\ref{tbl-obsmm}.

Telescope pointing and focus were checked regularly by observing strong
standard continuum point sources. Pointing was found to be repeatable within
$\approx$3\arcsec\ at the IRAM 30-m telescope and the JCMT, and 
$\approx$5\arcsec\ at SEST.
For flux calibration, maps of Uranus, Mars, Neptune, or different 
secondary calibrators \citep{1994MNRAS.271...75S}
were obtained during each shift and at different elevations.
Observing parameters for all runs are summarized in Table\,\ref{tbl-obsmm}.

%%%%%%%%%%%%%%%%%%%%%%%%%%%%%

\subsection{Reduction and calibration of the (sub)millimetre data}   \label{sec-datan-mm}

1.3\,mm bolometer maps observed using the IRAM 30-m and SEST telescopes were 
reduced and analyzed with the
MOPSI{\footnote{MOPSI (Mapping On-Off Pointing Skydip Infrared)
is a software package for infrared, millimetre,
and radio data reduction that has been developed and is updated by
R. Zylka.}} software package.
The chopped dual-beam scan maps were flat-fielded, corrected for atmospheric 
extinction, de-spiked,
base-line subtracted, and processed with an advanced sky noise algorithm.
The double-beam maps were then restored into single-beam maps using the 
Emerson-Klein-Haslam algorithm \citep{1979A&A....76...92E}. The individual maps were
averaged with corresponding weights and converted into the 
equatorial coordinate
system.
The SIMBA fastscan maps were reduced according to the instructions in the
SIMBA Observer's Handbook (2002; \citealp[see also][ Appendix A]{2003A&A...409..235C}).

The SCUBA 450\,$\mu$m and 850\,$\mu$m jiggle maps were reduced with the
SURF package \citep[SCUBA User reduction Facility,][]{1998adass...7..216J}.
The individual maps were flat-fielded, corrected for extinction, processed
with a sky-noise routine \citep{1998SPIE.3357..548J}, and combined.
The SCUBA scan maps were processed in a similar manner, before removing the
dual-beam function. Finally, the maps from the six different chop configurations were 
combined in Fourier space, weighting each to minimize 
noise on spatial scales corresponding to the chop throw 
\citep{1998SPIE.3357..548J,1999MNRAS.303..659H}.

The data were calibrated using nightly observations of either primary planetary calibrators,
or secondary calibrators when a planet was unavailable.
Main beam flux calibration factors for all maps were computed by integrating
over the main antenna beam in the planet maps, assuming planet brightness
temperatures listed by \citet{1986Icar...67..289O} and \citet{1993Icar..105..537G} 
and correcting for the error beams (see below).
The flux densities of the secondary calibrators were taken from \citet{1994MNRAS.271...75S}.
The statistical scatter of the 1.3\,mm flux calibration factors derived from all
IRAM 30-m and SEST data was found to be $\approx$6\%.
Due to the uncertainty of the adopted planet temperatures and uncertainties
in the flux integration procedure, we estimate that the 
total calibration uncertainty
is $\pm$15\% at 1.3\,mm, $\pm$20\% at 850\,$\mu$m, and 
$\pm$30\% at 450\,$\mu$m.
Typical 1$\sigma$\ noise levels (off-source) in the final maps are 
$5.5\pm1.3$\,mJy/beam (IRAM 30-m) and $20\pm5$\,mJy/beam (SEST) at 1.3\,mm,
$\sim$20\,mJy/beam at 850\,$\mu$m, and $\sim$200\,mJy/beam at 450\,$\mu$m, respectively.

For each instrument and observing run, the average beam size full-width at half
maximum (FWHM) was determined from the flux calibration maps, mostly on Uranus,
taking into account the intrinsic size of the planet disk 
(see Table\,\ref{tbl-obsmm}). The secondary calibrators are less well characterized, and so 
on nights when planetary beam-maps were unavailable, 
the average beam FWHM for the entire data-set is quoted in Table \ref{tbl-obsmm}.
Error beams at 1.3\,mm and 850\,$\mu$m were found to be negligible,
but at 450\,$\mu$m a significant error beam was found to contribute
up to $\sim$50\% of the flux density. Note that the main beam flux calibration accounts 
for this (to first order). Flux density measurements in the final maps 
were done using the GILDAS{\footnote{http://www.iram.fr/IRAMFR/GILDAS} software 
and are are described in Sect.\,\ref{sec-res-mm}.

For weak and extended sources, the integrated flux density and the derived source size
also depend on the map size, the curvature of the extended
flux density distribution, and the method of baseline subtraction, since
the sources are always observed against a more extended emission
background originating from the parental molecular cloud.

%%%%%%%%%%%%%%%%%%%%%%%%%%%%%%%%%

\subsection{Near-infrared observations}                       \label{sec-obs-nir}

We use the NIR images only for overlays with the (sub)mm maps, to 
identify which core is associated with a star or NIR nebula, and for 
discussion of evolutionary stages. 
NIR photometry is not be described in this paper, but we use the 
fluxes for the SEDs where relevant.
To this end, NIR observations of the globules listed in 
Table~\ref{tbl-sourcelist}
were performed during five observing runs between December 1993 and 
October 2003 at
the 3.5m telescope on Calar Alto, Spain and at the MPG/ESO 2.2-m telescope 
on La Silla, Chile.
Observing parameters for all runs are compiled in Table~\ref{tbl-obsnir}. 
Images were obtained in standard dithering mode with 4 to 5 positions per 
field.
Flat field calibrations were obtained each night via dome flats using 
difference frames of 
integrations with a halogen lamp on and off.
Sets of photometric standard stars from lists by
\citet{1982AJ.....87.1029E} and \citet{2001MNRAS.325..563H} were observed during
each night, but the calibration for most sources was done 
by using 2MASS catalogue fluxes for stars in the fields.

The data were reduced using a combination of the
IRAF{\footnote{IRAF is distributed by the NOAO, which are operated by
the Association of Universities for Research in Astronomy, Inc., under
cooperative agreement with the National Science Foundation.}}
``Experimental Deep Infrared Mosaicking Software''
(XDIMSUM{\footnote{XDIMSUM is a variant of the DIMSUM package
(V3: August 19, 2002) developed firstly by \citet{dimsum}}})
and Starlink Project packages.
For photometric calibration, zero-points were derived from observed
standard stars of each night. Then, magnitudes for individual objects were
derived using
GAIA{\footnote{GAIA is provided by Starlink, and is a derivative of the Skycat catalogue and image
display tool, developed as part of the VLT project at ESO. Skycat is free
software under the terms of the GNU copyright.}}
with the embedded PHOTOM{\footnote{The PHOTOM photometry package is provided by Starlink}} package. 
The average 3\,$\sigma$\ point source detection limit in all three NIR bands
is $\sim $\,19\,mag.
%The flux densities of individual sources, together with detection limits
%are listed in Table~\ref{tbl-resnir}. 
Only sources that are associated with
a dense globule core (i.e. a (sub)mm source) are considered in this paper. 
%The NIR data will be published in full elsewhere.

For astrometric calibration of mosaicked frames we used the
Guide Star Catalogue II{\footnote{The Guide Star Catalogue II is a joint
project of the STSI and the Osservatorio Astronomico di Torino. STSI is
operated by the Association of Universities for Research in Astronomy, for
the NASA under contract NAS5-26555.}}
and the
SIMBAD{\footnote{SIMBAD is an online database, operated at CDS, 
Strasbourg, France}}.
The resulting typical positional accuracy is better 0{\mbox{$.\!\!^{\prime\prime}$}}8 (1\,$\sigma$\ rms).

\placetable{tbl-obsnir}

%%%%%%%%%%%%%%%%%%%%%%%%%%%%%
\subsection{{\it Spitzer} data}                       \label{sec-obs-spitzer}

All sources listed in Table\,\ref{tbl-sourcelist} were observed by {\it Spitzer} 
within different programs, most with {\it IRAC} and {\it MIPS}, except 
CB\,17\,SW, BHR\,58, BHR\,137, and CB\,224. 
The relevant data were obtained from the Spitzer Science
Center (SCC) Archive. In particular, we retrieved, where available, IRAC
maps at 3.6, 4.5, 5.8, and 8\,$\mu$m and MIPS scan maps at 24, 70, and 160\,$\mu$m.
The data were processed by the SCC using their standard pipeline to
produce Post-Basic Calibration Data (P-BCD) images. Photometry was done
in IDL by applying aperture photometry, where the aperture corrections
available at the instruments webpage were used. The conversion from
P-BCD units (MJy\,sr$^{-1}$) to flux density was achieved by applying Quick
Point Source Photometry as described online. The overall flux
calibration and flux measurement accuracy is estimated by
the SSC to be accurate to within 20\%.

%%%%%%%%%%%%%%%%%%%%%%%%%%%%%%%%%%%%%%%%%%%%%%%%%%%%%%%%%%%%%

\section{Results}                          \label{sec-res}

%%%%%%%%%%%%%%%%%%%%

\subsection{(Sub)millimetre continuum maps}             \label{sec-res-mm}

In Figs. \ref{fig-cb6} through \ref{fig-cb246}, we present 
the (sub)mm maps, together with NIR K-band images 
for each of the sources listed in Table~\ref{tbl-sourcelist} (sources with
only upper limits are not shown).
Digital Sky-Survey (DSS2)\footnote{The Digitized Sky Survey was produced at the Space Telescope
Science Institute under US Government grant NAG W-2166. The
images of these surveys are based on the photographic data obtained
using the Oschin Schmidt Telescope on Palomar Mountain and the
UK Schmidt Telescope.} optical images of the globules are also shown,
together with the submm contours (850\,$\mu$m or 1.3\,mm) and the boundaries 
of the submm maps,
to illustrate their overall morphology and the relative location
of the dense cores within the globules.
Note that in the case of SCUBA jiggle maps the entire mapped region is indicated, 
while in the case of IRAM (dual-beam) or SEST scan maps only the area from which the mm 
emission was restored is marked.
In addition, SEDs, including fluxes from optical wavelengths up to 3\,mm, are shown for 
those sources were we could compile sufficient data and associate them unambiguously 
to one source. 

\placetable{tbl-resmm}

(Sub)mm flux densities are summarized in Table \ref{tbl-resmm}. 
The center coordinates for each dust core were derived from gaussian fits to the 
1.3\,mm intensity maps after decomposition into extended envelope and individual 
compact components.
For some sources (marked in Table \ref{tbl-resmm}), the peak position was derived from 
the 850\,$\mu$m maps. The peak intensity $I_{\nu}^{\rm peak}$ of each compact component 
is given in Columns 3, 5 and 7 of Table \ref{tbl-resmm} in units of Jy/beam. The beam size 
is also quoted here in parentheses. 
Note that the maps of some extended sources (e.g., CB\,246) were convolved with a larger 
gaussian beam to increase the SNR.
%The beam size is quoted as an average 
%over the entire dataset, because a number of sources were observed on multiple nights. 
We also quote the total integrated flux density, $S_{\nu}^{\rm tot}$, for each source. 
This was derived by integrating within a polygon enclosing the (closed) 2\,$\sigma$\ contour.
If there is one or more compact sources within a larger envelope, we also give the integrated 
flux densities of the compact sources, derived after fitting and subtracting the extended 
envelope from the maps. 
We do not list individual error bars for the flux densities, because uncertainties are 
mostly dominated by the calibration uncertainties (see Sect.\,\ref{sec-datan-mm}). 
However, for extended sources, the uncertainty is dominated by the size of the area 
from which emission could be recovered above the 2-$\sigma$-level, which in turn depends on 
the curvature of the emission background,
the ratios of the angular sizes of the beam, the source, and the chop throw, 
as well as on the SNR in the map.
In general, the relative fraction of recovered to total actual flux increases from 
faint extended to strong compact, and from nearby to distant sources.

%%%%%%%%%%%%%%%%%%%%%%%%%%%%%

\subsection{Association of sub(mm) and infrared sources}     \label{sec-res-stars}

When compiling SEDs and attempting to model and interpret the observational results, 
it is important to reveal which of the sources and flux densities detected at different 
wavelengths, with different angular resolution, apertures, and astrometric precision, 
are actually physically associated and which ones have different origins. 
For this reason, we overlay the (sub)mm maps on the optical (DSS) and NIR images,  
show the IRAS point source position error ellipses where applicable, and also inspect 
the {\it Spitzer} maps where available (see \ref{sec-obs-spitzer}).

We found the following associations between NIR/MIR sources and mm cores:
No NIR sources brighter than 0.02\,mJy ($\approx$19\,mag) at 2.2\,$\mu$m 
or $\approx$1\,mJy at 4.5\,$\mu$m are detected in 
CB\,17-SMM1 and 2, CB\,130-SMM2, CB\,232-SMM, CB\,243-SMM1 and 2, and CB\,246.
However, since local dust extinction, source morphology, and projection effects 
 (e.g., the alignment between the line of sight and outflow cavities) 
play a major role at these early evolutionary stages, this detection limit cannot be 
directly converted into an upper limit for the mass or luminosity of possible embedded sources.
No K-band emission, but faint NIR sources with steeply rising NIR SEDs peaking 
at $\approx 4\ldots 6\,\mu$m have been found in 
BHR\,55, BHR\,86, CB\,68, and CB\,199.
Extended NIR nebulosities, in several cases NIR jets, but no star-like objects were detected in
CB\,26, BHR\,12-SMM1 and SMM2, BHR\,36, BHR\,71, CB\,130-SMM1, CB\,224-SMM1, CB\,230, and CB\,244-SMM1.
A single red star or compact star-like NIR source, in some cases with extended nebulosity, was detected at 
the mm peak position in 
CB\,6, CB\,188, and CB\,224-SMM2.
Several NIR sources or clusters, respectively, are associated with the (sub)mm cores in 
CB\,34, CB\,54, CB\,205, and CB\,240. 
The situation is somewhat inconclusive in 
CB\,58 and BHR\,137, where it is not clear if the NIR sources are actually 
associated with the mm core or not.

The following (sub)mm cores have very nearby, but most likely not directly associated 
NIR/MIR sources and the flux measurements could not always be disentangled:
CB\,17 has an IRAS point source and a {\it Spitzer}-detected low-luminosity IR point 
source that is clearly not associated with the mm core (Fig.\,\ref{fig-cb17}).
CB\,130-SMM1 is a strong and slightly extended submm source with a small NIR nebula 
at the mm peak position. A very nearby (14\arcsec) strong and pointlike IR source (IRS) could not be 
completely disentagled from SMM1, but does not seem to exhibit detectable mm emission on its own.
CB\,232 and CB\,243 both have bright reddened stars within the area of the mm emission, 
but clearly offset and not associated with the main mm peak.

%%%%%%%%%%%%%%%%%%%%%%%%%%%%%

\subsection{Multiplicity}     \label{sec-res-mult}

Although the (sub)mm maps are not sensitive to spatial scales that cover the peak of the 
binary star separation distribution \citep[20\,--\,100\,AU; ][]{1991A&A...248..485D}, 
we find that at least 16 out of 24 globules with detected (sub)mm emission show signatures 
of multiplicity at scales larger than $\approx$1,000\,AU, out to 80,000\,AU 
(0.4\,pc; see Table\,\ref{tbl-resmult}).
This includes prestellar and protostellar double cores resolved in the mm dust continuum 
(e.g., CB\,17, CB\,130, CB\,246, BHR\,12),
unresolved protostellar cores where  NIR or MIR observations have 
revealed a secondary embedded source  (e.g., CB\,230, BHR\,71),
prestellar or protostellar cores where NIR/MIR observations have 
revealed sources that are significantly offset from the (sub)mm peak and 
that are most likley not embedded in these cores (e.g., CB\,17, CB\,130, CB\,232, 
CB\,243; see Fig.\,\ref{fig-spitzermaps}),
as well as large dust cores that are associated with clusters 
of NIR sources, presumably young stars that formed from the same globule 
(e.g., CB34, CB\,54, CB\,205). 
Nine of these 24 globules were resolved to have two or more cores  
in the (sub)mm dust emission maps (see Table\,\ref{tbl-resmult}). 
However, only three of these double cores have projected 
separations $<$10,000\,AU (CB\,17, BHR\,12, and CB\,130). 
Separations $>$10,000\,AU are probably not relevant for the formation of bound 
stellar systems. 
%Even for the smaller separations, we cannot make definite statements 
%about how many which globules form true binary stars. 

Altogether, we find that at least two thirds (16 out of 24) of the star-forming globules 
studied here show evidence of forming multiple stars, either in multiple star-forming cores, 
wide embedded binaries, or small star clusters. 
The fraction of closer binaries formed from unresolved (sub)mm cores 
might be higher, but remains unknown from this study \citep[cf.][]{2010arXiv1001.3691M}.

\placetable{tbl-resmult}

%Compare to Spitzer results for Perseus and Ophiuchus (Joergensen et al)

%%%%%%%%%%%%%%%%%%%%%%%%%%%%%

\subsection{Spectral Energy Distributions}    \label{sec-res-seds}

Spectral Energy Distributions (SEDs) were compiled from aperture photometry
in the (sub)mm maps (Table\,\ref{tbl-resmm}) and NIR images 
(Figures \ref{fig-cb6}--\ref{fig-cb246})
presented in this work (Sects. \ref{sec-res-mm} and \ref{sec-res-stars}),
the 2MASS catalogue, the IRAS point source catalogue, pipeline-calibrated {\it Spitzer} 
IRAC and MIPS maps, and
other publicly available databases
and literature papers. We do not explicitly list all flux values 
which we have compiled, but show in Figs.\,\ref{fig-cb6} through \ref{fig-cb244} 
the SEDs for those sources 
where we are confident that the measurements represent one source or, in some cases, 
physically related, mostly unresolved double sources (e.g., CB\,230). 
Note that we actually show the energy density ($\nu\,S_{\nu}$) and not the flux density ($S_{\nu}$).
We do not show SEDs for sources where we either do not have enough data points 
(e.g., BHR\,137, CB\,246) or where the (sub)mm emission is associated with multiple 
NIR sources (e.g., CB\,34, CB\,54, CB\,58, CB\,205, CB\,240).
For the other sources, we indicate in the figure captions and in 
Table\,\ref{tbl-physprop} whether we use the total (sub)mm fluxes or only 
one compact component.
Non-detections are indicated by down-arrows as 
3\,$\sigma$\ upper limits.
Flux values that derive from apertures smaller than the expected 
emission region (e.g.,
single-point measurements with the JCMT-UKT14 bolometer) or interferometric 
measurements,
which are not sensitive for extended emission, are indicated by up-arrows as
lower limits.
A comprehensive overview of all SEDs, ordered by evolutionary stage 
(see Sect.\,\ref{sec-res-evol}) is shown in Fig.\,\ref{fig-sed-all}.

In order to derive luminosities and bolometric temperatures,
we first interpolated and then integrated the SEDs,
always assuming spherical symmetry. In the case of, e.g., bipolar NIR 
reflection nebulae, this may lead to either an under-estimate
of the NIR luminosity (e.g., CB\,26, Fig. \ref{fig-cb26}) or an
overestimate (e.g., CB\,230, Fig. \ref{fig-cb230}), depending on the 
inclination with respect to the line-of-sight. 
However, all sources studied in this paper emit the 
majority of their total
energy at much longer wavelengths, where the radiation is reprocessed by dust, 
so that the general results derived here 
are not
significantly affected by the simplifying assumption of spherical symmetry.
Interpolation between the measured flux densities was done by a $\chi^2$\ 
grey-body
fit to all points at $\lambda\ge 100\,\mu$m and by simple logarithmic 
interpolation between
all points at $\lambda\le 100\,\mu$m.
Free parameters in the grey-body fits were the dust temperature, source size,
and optical depth. The (sub)mm spectral index of the dust emissivity was fixed to 
$\beta = 1.8$\ (see Sect. \ref{sec-res-physprop}). 
Fitted dust temperatures were in the range 15\,K (CB\,17-SMM) 
to 53\,K (CB\,224-SMM2), with mean values of 33\,K for Class\,0 sources 
and 42\,K for Class\,I sources (see Sect.\,\ref{sec-res-evol}).

Note, however, that we use these fits only to derive an approximate 
interpolation for integrating the SEDs. 
In particular, we do not use the dust temperatures or source size 
from these grey-body fits because the derived effective temperatures 
are usually dominated by small contributions from either embedded heating 
sources or externally heated very small grains in the envelopes.
Thus, they usually do not represent the mass-averaged dust temperature.
Instead, we use fixed mass-averaged dust temperatures as function of 
evolutionary stage (see Sect.\,\ref{sec-res-physprop})
and derive source sizes directly from the (sub)mm maps.

%%%%%%%%%%%%%%%%%%%%%%%%%%%%%

\subsection{Evolutionary stages} \label{sec-res-evol}

We use a combination of different diagnostics to derive the evolutionary 
stages of the globule cores. 
We do not attempt to classify those cores of distant globules that are associated 
with clusters of NIR sources and for which the NIR, FIR (IRAS), and (sub)mm emission 
could not be uniquely associated (e.g., CB\,34, CB\,54, CB\,58, CB\,205, CB\,240). 
Sources for which we have barely more information than a mm flux  (e.g., BHR\,58, BHR\,137) 
are also not considered here. 
For the other sources, we analyze the following tracers of evolutionary stage:
\begin{enumerate}
\item
The ratio of submm to bolometric luminosity, $L_{\rm smm}\,/\,L_{\rm bol}$,
where $L_{\rm smm}$\ is measured longward of 350\,$\mu$m, 
which roughly reflects the ratio of envelope to stellar mass, 
$M_{\rm env}\,/\,M_{\ast}$\ \citep{1993ApJ...406..122A}.
\cite{2000prpl.conf...59A} propose to use $L_{\rm smm}\,/\,L_{\rm bol} > 0.5$\%
as a threshold for Class\,0 protostars, reflecting $M_{\rm env}\,/\,M_{\ast} > 1$.
\item
The bolometric temperature, $T_{\rm bol}$\ \citep{ML1993}, which  
is defined by the temperature 
of a blackbody having the same mean frequency (ratio of the first and the zeroth 
moment of the SED) as the observed continuum spectrum.
\cite{1995ApJ...445..377C} and \cite{1998ApJ...492..703M} derive the boundary 
between Class\,0 and Class\,I YSOs to be at $T_{\rm bol} \sim 70$\,K.
Note that we use the interpolated SEDs (see Sect.\,\ref{sec-res-seds})
with flux values at all wavelengths to compute $T_{\rm bol}$\ and luminosities, 
rather than just the actual data points. The latter approach could lead to 
significant biases if the SED is not well-sampled \citep[cf.][]{2009ApJ...692..973E}.
\item
The IR spectral index, 
$\alpha_{\rm IR} = - d\,{\rm log}(\nu\,F_{\nu})\,/\,d\,{\rm log}\,\nu$\ 
\citep{1987IAUS..115....1L}, which characterizes the IR excess with respect 
to stellar blackbodies. Where sufficient data are available, we calculate 
$\alpha_{\rm IR}$\ between $\lambda = 2.2$\ and 24\,$\mu$m, thus avoiding 
to caracterize only the NIR/MIR nebulosity ``bump'' in the Class\,0 SEDs (see below), 
the possible influence of the silicon absorption, and the low-resolution 
$IRAS$\ measurements at 12 and 25\,$\mu$m, which are prone to confusion problems. 
Quantitatively, we follow \citet{1994ApJ...420..837A} and 
use $\alpha_{\rm IR} = 0$\ as boundary between Class\,I and II.
No well-defined boundary exists for Class\,0 sources since only 
{\it Spitzer} lead to the detection of the NIR/MIR emission in many of them.
\end{enumerate}

Along with these three quantitative classification criteria, we evaluate 
the morphology of the NIR emission and the presence of molecular outflows.
The first one can be either direct star light, indicating that the circumstellar emission 
originates either from an optically thin envelope or a phase-on disk, 
or diffuse nebulosity, indicating, e.g., scattered light from outflow cavities 
in an otherwise optically thick envelope. 
The presence of a molecular outflow is an indicator of an embedded source 
that distinguishes protostars from prestellar cores. Two notes of caution are:
{\it (i)} the non-detection of an outflow does not prove the absence 
  of an embedded protostar, in particular as we have no homogeneous 
  high-sensitivity, high-angular resolution outflow survey at hand, and 
{\it (ii)} the driving source of an outflow is not necessarily identical 
 with the (sub)mm source - this must be verified in each individual case.

%% Discussion of evolutionary tracers:

The applicabilty and drawbacks of the classification criteria mentioned above 
has been discussed 
in detail by \cite{2009ApJ...692..973E}. An analysis of our data confirms the finding 
of these authors in that $\alpha_{\rm IR}$\ does not correlate well with $T_{\rm bol}$\ 
and embeddedness for Class\,0 and embedded Class\,I sources. 
The deep NIR images, in particular those from Spitzer, show that most of 
even the youngest Class\,0 sources already have a small NIR nebulosity 
associated. This can also be clearly seen in the compilation of SEDs 
shown in Fig.\,\ref{fig-sed-all}.
We attribute this to the fact 
that the amount of NIR/MIR flux seen from such an embedded object largely depends 
on the alignment between the line of sight and the outflow cavity, 
allowing or not a direct view onto the hot dust in the inner protostellar core 
and/or forward-scattered light in the outflow cavity.
Therefore, we only list the $\alpha_{\rm IR}$\ values in Table\,\ref{tbl-physprop}, 
but do not use them for the classification of evolutionary stages.

We find that all sources, for which we could compile meaningful SEDs, 
have $L_{\rm smm}\,/\,L_{\rm bol} \ge 0.8$\% 
(Table\,\ref{tbl-physprop}), i.e., according to the criterion proposed by 
\citet{2000prpl.conf...59A} they would all be Class\,0 sources. 
The bolometric temperature, although reflecting similar properties of the SEDs, 
appears to be a more reliable indicator of the evolutionary stage than the 
$L_{\rm smm}\,/\,L_{\rm bol}$\ ratio, 
as it is less sensitive to the amount of extended 
envelope emission recovered in the submm maps (see Sect.\,\ref{sec-dis1-evol}). 
We therefore use the bolometric temperature as the main discriminator 
between Class\,0 protostars and Class\,I YSOs, along with the other 
supporting indicators mentioned above \citep[cf.][]{2009ApJ...692..973E}. 

%% Results on evolutionary stages:

The derived evolutionary stages, along with other source properties, 
for 18 globules are summarized in Table\,\ref{tbl-physprop}. 
Figure\,\ref{fig-tbol-lsubmm} shows the $L_{\rm smm}\,/\,L_{\rm bol}$\ ratio 
vs. $T_{\rm bol}$\ for these sources. 
In total, we identified 
nine starless cores, presumably prestellar, in six globules,
nine Class\,0 protostars in nine globules, and  
twelve Class\,I YSOs in eleven globules. 
Interestingly, seven globules contain co-existing sources at different 
evolutionary stages (see discussion in Sect.\,\ref{sec-dis1-mult}).
Six globules were not detected at (sub)mm wavelengths, and 
for the submm cores in eight globules we could not or did not attempt to derive 
an evolutionary stage because we either had insufficient data, or the available 
data indicated that several sources with possibly different evolutionary stages contribute 
to the partially unresolved flux measurements.

%%%%%%%%%%%%%%%%%%%%%%%%%%%%%

\subsection{Physical parameters of the star-forming cores} 
\label{sec-res-physprop}

The following physical source quantities are derived from the 
(sub)mm continuum maps and the SEDs:
bolometric and submm luminosities (Sect.\,\ref{sec-res-evol}),
source sizes, masses, beam-averaged peak column densities, 
and source-averaged volume densities.

FWHM mean source sizes ($\sqrt{a\,b}$) are simply derived by measuring the 
angular diameters of the 50\% contours in the 850\,$\mu$m or 1.3\,mm 
maps (after decomposition into envelope and compact core, where applicable), 
deconvolving them with beam size, and applying the distances listed in 
Table\,\ref{tbl-sourcelist}. The results are listed in Table\,\ref{tbl-physprop}.

\placetable{tbl-physprop}

To calculate masses, densities, and column densities, we adopt mass-averaged mean 
dust temperatures that depend on the evolutionary stage of the source 
(Sect.\,\ref{sec-res-evol}). 
Quantitatively, we follow \citet{2001A&A...365..440M}, 
\citet{2002ApJ...575..337S}, and \cite{2003ApJS..145..111Y}, 
who have derived mean mass-averaged isothermal dust temperatures 
from radiative transfer models for samples of starless cores, 
Class\,0 protostars, and envelopes of Class\,I YSOs. 
We use here 
$\langle T_{\rm d}\rangle = 10$\,K for starless (prestellar) cores, 
$\langle T_{\rm d}\rangle = 15$\,K for Class\,0 protostars, 
and $\langle T_{\rm d}\rangle = 20$\,K for the envelopes of Class\,I YSOs. 
The first dust temperature map directly derived from new {\it Herschel} data 
of the Bok globule CB\,244 with its two embedded cores confirms the validity of 
these temperature assumptions \citep{2010Stutzetal}.
The adopted temperatures and resulting masses for each source are listed in Table\,\ref{tbl-physprop}.

Assuming optically thin isothermal dust emission, the hydrogen gas mass
$M_{\rm H} = M({\rm H}) + 2\,M({\rm H}_2)$\
is calculated from the (sub)mm flux density by 
\begin{equation}                      \label{eqmass1}
M_{\rm H} = \frac{S_{\nu}\, D^{2}}{\kappa_{\rm d}(\nu )\,
B_{\nu}(\nu,T_{\rm d})}\,
\left(\frac{M_{\rm H}}{M_{\rm d}}\right)\qquad .
\end{equation}
Here, $\nu$\ is the observing frequency,
$S_{\nu}$\ the observed flux density from the region of interest,
$D$\ is the distance to the source,
$\kappa_{\rm d}(\nu)$\ is the dust opacity,
$B_{\nu}(\nu,T_{\rm d})$\ the Planck function,
$T_{\rm d}$\ the dust temperature,
and $M_{\rm H}/M_{\rm d}$\ the hydrogen-to-dust mass ratio.
Throughout this paper, we use a
hydrogen-to-dust mass ratio of $M_{\rm H}/M_{\rm d} = 110$\ 
and a dust opacity of
$\kappa_{\rm d}(\lambda) =
\kappa_{\rm d}(1.3{\rm mm})\times \left(\lambda/1.3{\rm mm}\right)^{-1.8}$\
with $\kappa_{\rm d}(1.3{\rm mm}) = 0.5$\,cm$^2$\,g$^{-1}$\ 
\citep[][uncoagulated MRN dust grains with thin ice mantles]{1994A&A...291..943O}, 
a that is value typical for dense and cold molecular cloud cores and 
intermediate between the opacities usually adopted for unprocessed interstellar 
grains in the diffuse gas  \citep[e.g.,][]{1984ApJ...285...89D}
and for processed and coagulated grains in protoplanetary disks
\citep[e.g.,][]{2000prpl.conf..533B}.

We derive total Hydrogen masses 
of 1\,--\,20\,M$_{\odot}$\ for prestellar cores,
0.5\,--\,10\,M$_{\odot}$\ for Class\,0 protostars,
and  0.11\,--\,1\,M$_{\odot}$\ for Class\,I YSOs (not counting the extended envelopes; 
see Table\,\ref{tbl-physprop}).
The total gas mass, accounting for helium and heavy elements, amounts then to 
$M_{\rm g} \approx 1.36\,M_{\rm H}$.

The beam-averaged peak Hydrogen column density
$N_{\rm H} = N({\rm H}) + 2\,N({\rm H}_2)$\ was derived from the observed
peak flux densities by
\begin{equation}                      \label{eqcold}
N_{\rm H} = \frac{I_{\nu}}{\kappa_{\rm d}(\nu )\,\Omega_{\rm B}\,B_{\nu}
(\nu,T_{\rm d})}\,
\frac{1}{\rm m_H}\,\left(\frac{M_{\rm H}}{M_{\rm d}}\right)\qquad ,
\end{equation}
where $I_{\nu}$\ is the peak flux density (Table\,\ref{tbl-resmm}),
$\Omega_{\rm B} = 1.133\,\theta_{\rm B}^2$\ the beam solid angle,
$\theta_{\rm B}$\ the beam FWHM size, and 
${\rm m_H}$\ is the the proton mass. 
The resulting beam-averaged peak column densities are also listed in Table\,\ref{tbl-physprop}.

Assuming that the sources have the same extent along the line of sight
as the mean size perpendicular to the line of sight, the 
source-averaged volume density of hydrogen atoms, 
{$n_{\rm H} = n({\rm H}) + 2\,n({\rm H}_2)$}, 
was calculated by
\begin{equation}                      \label{eqdens}
n_{\rm H} = \frac{M_{\rm H}}{m_{\rm H}\,V}\qquad ,
\end{equation}
with $V \sim \pi/6\,d_{\rm S}^3$\ being the volume of the source, 
and $d_{\rm S}$\ the FWHM size listed in Table \,\ref{tbl-physprop}.
The resulting densities are listed in Table\,\ref{tbl-physprop}.
Note that these are source-averaged densities and the local peak densities 
in sources with density gradients can be orders of magnitude higher.

In an earlier paper, we had also derived radial density profiles for most of the globule 
cores by fitting ray-traced and beam-convolved model cores to the radially averaged 
(sub)mm data at all three wavelegths \citep{2005YLU_rl}.
The extended dataset presented in this paper would certainly permit to derive 
better-constrained density profiles for many globules cores.
However, as in this previous paper, such an analysis would be hampered by the 
fact that we have to assume temperature profiles, since the (sub)mm data provide 
only weak constraints. 
In particular for the prestellar cores, which do not have a central heating source and 
may have a positive radial temperature gradient, 
this can introduce significant biases and may lead to an underestimation of the 
density gradient.
Since we expect to obtain soon Herschel data for many of the sources presented here, 
which will provide direct constraints on the temperature profiles, we postpone 
the derivation and analysis of the radial density profiles to a succeeding paper.

%%%%% Table 7: Source properties
%%%%% End Table 7

%%%%%%%%%%%%%%%%%%%%%%%%%%%%%%%%%%%%%%%%%%%%%%%%%%%%%%%%%%%%%%%%%%%%%%%%%%%%%%%

\section{Discussion}        \label{sec-dis1}

%%%%%%%%%%%%%%%%%%%%%%%%%%%%%

\subsection{Source properties as function of evolutionary stage} \label{sec-dis1-evol}

Being relatively nearby and isolated, the cores of Bok globules 
are the best-suited places to observe the properties and evolution of individual 
low-mass star-forming cores. However, this isolation, 
which makes them such good targets, may also alter the core properties 
and evolutionary process as compared to the more representative (for general star formation), 
but harder to observe, star-forming cores in clustered regions. 
Some of the most obvious differences in environmental conditions are 
the interstellar radiation field, which is not shielded by the surrounding cloud,
the missing external pressure due to the lack of inter-core gas, 
and the lack of turbulence input from the surrounding cloud.

The most obvious changes in core properties during the evolution from prestellar cores, 
through collapsing protostars, towards Class\,I YSOs are expected to be seen 
in the core size, central core density, ratio of core mass to envelope mass, and contribution of the 
central (proto)star to the source luminosity at shorter wavelengths.
The latter two effects have been used in the form of parameters 
$L_{\rm smm}\,/\,L_{\rm bol}$\ and $T_{\rm bol}$\ 
as main criteria to identify the evolutionary stage of the sources 
(Sect.\,\ref{sec-res-evol}).

\placefigure{fig-tbol-lsubmm}

In Figure\,\ref{fig-tbol-lsubmm}, we show a plot of $L_{\rm smm}/L_{\rm bol}$\ 
vs. $T_{\rm bol}$\ for the sources compiled in Table\,\ref{tbl-physprop}. 
We find that all sources with $T_{\rm bol} <70$\,K, the empirical 
boundary between Class\,0 protostars and Class\,I YSOs proposed by \cite{1998ApJ...492..703M},
have $L_{\rm smm}\,/\,L_{\rm bol} >2$\%.
Within the Class\,0 phase, there seems to be no significant correlation between the two parameters 
and no obvious evolution from lower to higher values of $T_{\rm bol}$. 
This may reflect the fact that during this early phase, the optically thick envelope dominates 
over the still very low-mass and low-luminosity central protostar. 
However, there is a relatively large spread in $L_{\rm smm}\,/\,L_{\rm bol}$\ ratios, 
probably reflecting the fact that during this phase the mass of the infalling envelope 
is relatively quickly reduced. 

The more evolved (Class\,I) YSOs in our sample have bolometric temperatures between 
100\,K and 350\,K and $L_{\rm smm}/L_{\rm bol} \sim 0.8\ldots3.5$\% (see Fig.\,\ref{fig-tbol-lsubmm}).  
All sources with NIR nebulosities but no visible stars have $L_{\rm smm}/L_{\rm bol} \ge 1.6$\% 
and extended (resolved) (sub)mm emission. These sources are probably less-evolved, envelope-dominated 
Class\,I YSOs. 
The three sources with $L_{\rm smm}/L_{\rm bol} \le 1.3$\% (CB\,188, CB\,224-SMM2, CB232-IRS1) 
all have visible stars and compact (sub)mm emission arising presumably from circumstellar disks. 
These sources probably represent the more evolved, disk-dominated Class\,I YSOs. 
We take these findings as tentative indication that the $L_{\rm smm}\,/\,L_{\rm bol}$\ ratio, 
as indicator of envelope dispersal, may better trace the evolution within the Class\,0 
and Class\,I phases than $T_{\rm bol}$.

Two of the Class\,I sources in our sample have double-peaked SEDs with NIR nebulosities and 
$\alpha_{\rm IR}<0$\ (CB\,26, CB\,230), which would indicate Class\,II. 
However, the IR emission seems completely originating from scattered light 
and the IR fluxes and morphology seem dictated by projection effects.
For CB\,26, a massive edge-on protoplanetary disk with a remnant envelope, 
\cite{2001ApJ...562L.173L} derive an age of $\approx 1$\,Myr. 
CB\,230, with a more massive envelope and less-developed disk(s) \citep{2001IAUS..200..117L}, 
is most likely younger. We therefore classify these two sources as
``Class\,I'' in Table\,\ref{tbl-physprop}, despite the fact that they have $\alpha_{\rm IR}<0$.

To verify these relations between $T_{\rm bol}$, $L_{\rm smm}/L_{\rm bol}$, and evolutionary stages, 
we have compared our sample of globule cores to 
the sample of Class\,0 protostars from \cite{2000prpl.conf...59A} .
Plotting these sources into the same diagramme (not shown here) 
shows a different picture.
While none of the globules sources have both $T_{\rm bol} < 70$\,K and 
$L_{\rm smm}/L_{\rm bol} < 2$\%, numerous protostars from the Andr\'e sample 
do populate this region.
The two different source samples have similar distance distributions,
and in each case, both nearby and more distant sources are distributed
evenly around the $L_{\rm smm}/L_{\rm bol}$ vs. $T_{\rm bol}$ plot. Therefore
there is no evidence to suggest that limited resolution or sensitivity
for either of the samples is causing the discrepancy.
Similarly, there is no evidence that any particular aspect of the
morphology of a Bok globule is causing this discrepancy between 
the two samples.
As none of the above characteristics can possibly cause a systematic shift 
of the Bok globules in the $L_{\rm smm}/L_{\rm bol}$\ vs. $T_{\rm bol}$\ diagramme, 
we conclude that the absence of globule cores with both  
low $T_{\rm bol}$\ and low $L_{\rm smm}/L_{\rm bol}$-ratios 
is a real effect, caused by the systematic environmental differences between 
the cores of Bok globules and protostars located in larger molecular clouds 
mentioned at the beginning of this section.

The outer boundary of a protostellar core that is embedded in a larger molecular cloud is
generally judged to be where it's envelope merges with the surrounding
cloud material, or the envelope of an adjacent object.  
In contrast, the column density profile of a globule core 
continuously decreases towards the "edge" of the small cloud, 
making it more difficult to distinguish between the core and the 
surrounding cloud, both physically and observationally, as the 
envelope (cloud) emission is usually not chopped away in the small globules. 
This has the effect of making the detected envelopes larger and 
elevating the submm luminosity as compared to dense cores embedded 
in larger clouds.
It is therefore possible that the cutoff in $L_{\rm smm}/L_{\rm bol}$, used
to differentiate between Class\,0 protostars and Class\,I YSOs should be
raised from $0.5\%$\ in the case of protostars in larger molecular clouds, to
$\approx2\%$\ in the case of isolated Bok globules, to account for their
larger envelope of cold dust. The value of $2\%$\ is chosen as this is the lowest
value of $L_{\rm smm}/L_{\rm bol}$\ in the observed sample of 
globule cores with $T_{\rm bol} <70$\,K. 
Although we cannot present a modified quantitative physical model 
\citep[cf.][]{1993ApJ...406..122A}, 
we think that this different cut-off value for $L_{\rm smm}/L_{\rm bol}$\ 
still represents the point where the mass of the infalling protostellar envelope is 
equal to that of the accreting protostar, but it accounts for the contribution from the
extended envelope which can no longer be clearly distinguished from the protostellar core.

We conclude that $T_{\rm bol}$\ appears to be the most reliable parameter to 
discriminate between Class\,0 protostars and Class\,I YSOs, 
with no hint for a systematic difference between sources in isolated globules and 
those embedded in larger molecular clouds. 
On the other hand, the evolution during the Class\,0 phase and during the Class\,I phase 
may probably be better traced by the $L_{\rm smm}\,/\,L_{\rm bol}$\ ratio. 
Despite the fact that the two parameters are not completely independent, 
this can be understood as a consequence of $T_{\rm bol}$\ being more sensitive 
to a flux contribution at shorter (NIR) wavelength from a central heating source,  
while $L_{\rm smm}\,/\,L_{\rm bol}$\ is more sensitive to the size and mass of the 
cold envelope.

\placefigure{fig-glob_prop}

We have also determined the mean FWHM core sizes and the mean source-averaged 
volume densities (see Table\,\ref{tbl-physprop}) 
of the three evolutionary subgroups.
We find that the mean core sizes decrease from 
$\approx$8,000\,AU for prestellar cores, to $\approx$3,000\,AU for Class\,0 protostars, 
to $<$2,000\,AU for Class\,I YSOs (Fig.\,\ref{fig-glob_prop}). 
We also find that the Class\,I YSOs usually have extended envelopes 
with a mean diameter of 20,000\,AU.
The source sizes of the Class\,0 protostars might be somewhat underestimated 
since we did not decompose the sources into central point source (the warm protostar) 
and envelope before measuring the FWHM size.
The source-averaged volume densities, $n_{\rm H}$, increase from 
$1\times 10^7$\,cm$^{-3}$\ for prestellar cores, to $7\times10^7$\,cm$^{-3}$\ for Class\,0 protostars.
At these high densities, one can in general assume that gas and dust temperatures are coupled.
For the cores of the Class\,I YSOs, which are mostly unresolved and presumably contain 
significant flux contrinutions from circumstellar disks, we derive a lower limit to 
the mean density of $>8\times10^7$\,cm$^{-3}$. 
The extended remnant envelopes of the Class\,I YSOs have a mean density of $7\times 10^4$\,cm$^{-3}$.

%%%%%%%%%%%%%%%%%%%%%%%%%%%%%

\subsection{Multiple and sequential star formation in globules} \label{sec-dis1-mult}

One interesting aspect of the combined results for multiplicity and evolutionary stages 
is that the large majority of 
the globules with signs of multiple star formation shows evidence 
of non-coeval, possibly sequential star formation.
This includes neighbouring mm sources with obviously 
different evolutionary stages (e.g., BHR\,12, CB\,130, CB\,224), 
prestellar or protostellar cores with nearby IR sources, 
presumably more evolved protostars or Class\,I YSOs 
(e.g., CB\,17, CB\,130, CB\,232, CB\,243), 
as well as NIR star clusters next to 
large (sub)mm cores with the potential to form more stars (e.g., CB\,34, CB\,54, BHR\,137) 
(see Table\,\ref{tbl-physprop}).
In only three globules we find (approximately) coeval pairs, 
ranging from multiple pre-stellar cores in CB\,246, 
embedded Class\,0 protostars in BHR\,71, to a pair of embedded Class\,I YSOs in CB\,230.

\placefigure{fig-spitzermaps}

The most interesting sources in this respect are shown in Fig.\,\ref{fig-spitzermaps}.
CB\,17 contains two prestellar cores and one low-luminosity Class\,I YSO within 5\,000\,AU (projected separation).
CB\,130 contains three sources within 9\,000\,AU, all with different evolutionary stages. 
A Class\,0 protostellar core is flanked by a prestellar core at 6,000\,AU to the west, 
and by a Class\,I or II YSO at 3,000\,AU to the east.
CB\,232 and CB\,243 both contain a prestellar core with a Class\,I YSO within 7\,000\,AU.

This widespread non-coevality suggests that the multiple sources observed here 
have in general not formed by gravitational fragmentation of collapsing protostellar cores 
\citep[e.g.,][]{2006MNRAS.370..488B}, 
but rather by initial turbulent fragmentation prior to the collapse phase \citep[e.g.,][]{2002ApJ...576..870P}, 
followed by a relatively independent evolution of the individual fragments.
Whether or not the first forming star can trigger collapse of the other fragment(s) and thus lead 
to small-scale sequential star formation cannot be directly answered by this study. 
The lack of close coeval pairs can be understood as an observational bias as our study 
is in general not sensitve to size scales $<$1000\,AU, which would be more relevant to the first mechanism.
Consequently, we cannot draw direct conclusions on the frequency of bound binary stars 
formed in isolated Bok globues \citep[cf.][]{2010arXiv1001.3691M}.

These findings also call for special attention when compiling 
SEDs and attempting to derive source properties from flux measurements with insufficient 
angular resolution. One may easily end up classifying the combined SED of a prestellar core 
and a nearby, more evolved YSO as a Class\,0 protostar (see, e.g., CB243, Fig.\,\ref{fig-cb243}f).

%%%%%%%%%%%%%%%%%%%%%%%%%%%%%%%%%%%%%%%%%%%%%%%%%%%%%%%%%%%%%%%%%%%%%%%%%%%%%%%

\section{Description of individual globules}        \label{sec-dis}

%%%%%%%%%%%%%%%%%%%%%%%%%%%%%

In this section, we describe in somewhat more detail the individual globules, 
their morphology and physical properties, and discuss the evolutionary 
stages of the embedded sources.
Distance references are not given in the text, but are compiled in Table\,\ref{tbl-sourcelist}.
Globule sizes and morphology are summarized in Table\,\ref{tbl-resmult} 
and properties and evolutionary stages of the embedded sources are summarized in 
Table\,\ref{tbl-physprop}.

%%%%%%%%%%%%%%%

{\bf CB\,4} is a small, roundish, and relatively isolated dark 
globule at a somewhat uncertain distance. Although \citet{1994ApJS...95..457W} 
assumed a distance of 200\,pc to the Cas\,A dark clouds in this region of the sky, 
we follow the analysis of \citet{1983ApJ...271..143D} who derived a distance of 600\,pc.
This latter value is also close to the distance of the ``-12\,km\,s$^{-1}$'' clouds 
\citep[800\,pc;][]{1987ApJ...322..706D} with which \citet{1997A&A...326..329L} 
associated the globules CB\,4 and CB\,6. 
CB\,4 is associated with an IRAS point source, located at its southern rim, 
which is detected at 100\,$\mu$m only.
There is no further evidence of an embedded source, including the {\it Spitzer} 
maps up to 160\,$\mu$m.
Deep IRAS aperture photometry and molecular line studies showed
that the globule is cold ($\sim 7$\,K) and has a relatively low mean density 
\citep{1991ApJS...75..877C,1994ApJ...433L..49K,1997ApJ...483..235T}.
Despite its compact and dark appearance it thus rather resembles a cirrus cloud.
CCD polarimetry of background stars revealed a faint and well-ordered magnetic 
field with a position angle (P.A.) of $71^\circ$\ \citep{2005MNRAS.361..177S}, 
which is apparently coupled to the surrounding Galactic field \citep{1995ApJ...445..269K}.
Our non-detection of compact thermal dust emission confirms the assumption that 
this globule does not seem to form stars.

%%%%%%%%%%%%%%%

{\bf CB\,6} 
(LBN\,613; Fig.\,\ref{fig-cb6}) is a small, cometary-shaped globule 
with a long diffuse tail, visible as scattered light cloudshine. 
It is located only a few degrees from CB\,4 and has the same radial velocity. 
We therefore assign CB\,6 the same distance (600\,pc) as CB\,4.
At its centre, the globule harbours an embedded YSO with circumstellar dust emission 
and a conical reflection nebula \citep[RNO\,3;][]{1980AJ.....85...29C,1994A&AS..108...73E}, 
likely representing scattered light in an outflow cone from the embedded YSO.
Indeed. broad line wings in CO\,(2-1) \citep{1991ApJS...75..877C} and 
CS\,(2-1) \citep{1998ApJS..119...59L} indicate the presence of a molecular outflow 
from the YSO, but no outflow maps exist yet.
\citet{1996MNRAS.282..587S} searched for, but did not detect NH$_3$ and C$_2$S.
The evolutionary stage of the YSO is controversial: 
while \citet{1988ApJS...68..257C} described it as Class\,III YSO, 
\citet{1997MNRAS.288L..45L} derived a L$_{\rm bol}$ / L$_{\rm smm}$\ ratio indicative 
of a Class\,0 protostar \citep{1993ApJ...406..122A}.
Our (sub)mm continuum maps (Fig\,\ref{fig-cb6}) show that the thermal dust emission 
is dominated by a compact source centered at the origin of the NIR nebula, 
indicating that the bulk of the mm dust emission does not originate from an optically 
thick envelope, but rather from a disk. The disk and the more extended envelope 
probably obscure the southern outflow lobe which is presumably pointed away from us.
This morphology, combined with the bolometric temperature of 
$T_{\rm bol} \sim 180$\,K let us conclude that this is a Class\,I YSO. 
%The relatively high $L_{\rm smm}\,/\,L_{\rm bol}$\ ratio (2.2\%) may reflect the fact that, 
%in particular at 850\,$\mu$m, we recover substantial amounts of emission from 
%the extended envelope that is not part of the infall region.

CB\,6\,N is a small globule located $\approx$10\arcmin\ north of CB\,6. 
It has no IR source associated and we did not detect it at mm wavelengths, 
indicating that it does not contain a star-forming core.

%%%%%%%%%%%%%%%

{\bf CB\,17}
(L\,1389; Fig.\,\ref{fig-cb17}) is a small and slightly cometary-shaped globule,  
located near Perseus and associated with the Lindblad ring \citep{1973A&A....24..309L}.
It has a long diffuse tail, pointing north-east and visible as cloudshine reflection.
The distance of CB\,17 was derived by \citet{1997A&A...326..329L}
via association in projected space and radial 
velocity with other Lindblad Ring clouds. In the direction of CB\,17, 
the Lindblad Ring structures have a mean distance from the Sun of 
$\approx 300$\,pc \citep{1987ApJ...322..706D}. 
HD\,25347, a bright G5\,III star at a distance of $210\pm40$\,pc 
\citep{2007ASSL..350.....V} is located about $\approx$11\arcmin\ (0.65\,pc at 200\,pc) 
south of CB\,17 and could be responsible for the cometary shape and 
diffuse cloudshine from the rim and tail of CB\,17 if it is located at the same 
distance as the globule. 
%Since the Lindblad Ring is not a well-defined structure, it is possible that 
%CB\,17 is somewhat closer than 300\,pc even if it is associated with the 
%Lindblad Ring structures. 
Combining the possible associations of CB\,17 with both the Lindblad Ring 
and HD\,25347, and the distances and uncertainties involved, 
we adopt a distance of 250$\pm$50\,pc for CB\,17.

The roundish cloud core of CB\,17 is associated with a dense submm core and a 
faint and cold IRAS point source 
(IRAS\,04005+5647) that is detected only at 60 and 100\,$\mu$m. 
CB\,17 has been studied extensively by various authors using different 
mm molecular line transitions.
The core was found to have a mean kinetic gas temperature of
$T_{\rm kin}\sim 10$\,K and a virial mass
of $\approx$3\,M$_\odot$
\citep[e.g.,][ scaled to $D=250$\,pc]{1996A&A...312..585L,2002ApJ...572..238C}.
\citet{2006ApJ...645.1212P} compared a number of different molecular line maps 
of CB\,17 with chemodynamical models. They derive a chemical age of $\approx$2\,Myr, 
as well as the relative rotational, thermal and turbulent energies, concluding that 
the core will most likely fragment.

Indeed, our deep 1.3\,mm continuum map (Fig.\,\ref{fig-cb17}) shows a double core with 
14\arcsec\ separation and a common envelope, located at the south-western rim of the globule head. 
The low-SNR 850\,$\mu$m SCUBA map can, however, not confirm the double-core nature, and the 
source remains undetected at 450\,$\mu$m.
The {\it Spitzer} {\it IRAC} maps show a 8\,$\mu$m shadow that agrees in position and shape with the 
double core (Fig.\,\ref{fig-spitzermaps}). 
No emission is detected at wavelengths shorter than 160\,$\mu$m, 
indicating these cores have no, or only very low-luminosity internal heating sources.
Since the double core is resolved only in the 1.3\,mm map, we compile the SED 
only for the combined core, which we call SMM.
However, a faint point-like NIR/MIR source (IRS), that is clearly not directly detected in 
the (sub)mm continuum, is located $\approx$10\arcsec\  northwest of SMM and just outside 
the 8\,$\mu$m shadow, indicates the presence of a third, more evolved source in CB\,17 
(Fig.\,\ref{fig-spitzermaps}).
This source dominates the emission at wavelengths shorter 160\,$\mu$m,
while SMM dominates at longer wavelengths.
Although the MIPS3 map at 160\,$\mu$m does not resolve the two positions, 
a slight shift of the peak position from IRS towards SMM indicates detectable emission 
from SMM at this wavelength. For compiling the SEDs, we assigned 15\% of the total 
160\,$\mu$m flux to SMM, but our results depend only weakly on the adopted flux splitting.
The morphology and SED of SMM (T$_{\rm bol}\approx 15$\,K, $L_{\rm smm}\,/\,L_{\rm bol}>30$\%) 
suggest this is a prestellar (double) core \citep[cf.][]{2002ApJ...572..238C,2006ApJ...645.1212P}.
Source IRS, which is located at a projected separation of $\sim 2000$\,AU from SMM2  
within the same globule core, 
is tentatively classified as a low-luminosity Class\,I YSO protostar 
(T$_{\rm bol}> 55$\,K, $L_{\rm smm}\,/\,L_{\rm bol}< 2.6$\%), 
although an outflow remains to be detected (Chen et al., in prep.).
A faint and diffuse NIR nebula at the position of SMM2 is speculated to be 
related to IRS rather than to a source embedded in SMM2, 
although the nature of a faint compact NIR knot at the peak position of SMM2 
remains unknown.
We are currently analyzing interferometric molecular line maps, attempting to 
better constrain evolutionary stages and the kinematic relation of the three sources 
(Schmalzl et al., in prep.).
CB\,17 is thus one of those globules that contain two or more sources of different evolutionary stage 
with a few thousand AU (see Sect.\,\ref{sec-dis1-mult} and Table\,\ref{tbl-resmult}).

CB\,17\,SW (L\,1388) is a small elliptical globule of similar size as CB\,17, 
located $\approx$12\arcmin\ south-west of CB\,17. 
It has no IR source associated and we did not detect it at mm wavelengths, 
indicating that it does not contain a star-forming core.

%%%%%%%%%%%%%%%

{\bf CB\,26} 
(L\,1439: Fig.\,\ref{fig-cb26}) 
is a small, cometary-shaped, double-core Bok globule located
$\sim$\,10$^\circ$ north of the Taurus-Auriga dark cloud, at a
distance of $\sim$\,140\,pc.  
An IRAS point source and dense core with signatures of star
formation is located at the south-west tip of the western globule core
\citep{1997A&A...326..329L}. While the eastern globule core 
(at 4:56:20, 52:01:30, B1950, in Fig.\,\ref{fig-cb26}) is 
visible as 8\,$\mu$m shadow in the {\it Spitzer IRAC} maps, 
the IR source is located $\approx$4\arcmin\ further west.
The eastern globule core was not searched for mm emission, but the 
{\it Spitzer} maps also do not indicate the presence of an embedded source.
OVRO observations at the IRAS position of the mm dust continuum
emission and of the $^{13}$CO\,(1-0) line have revealed a nearly
edge-on circumstellar disk of radius 200\,AU with Keplerian rotation
\citep{2001ApJ...562L.173L}, surrounding a very young (obscured) low-mass 
T\,Tauri star.  It is associated with a small bipolar near-infrared (NIR)
nebula bisected by a dark extinction lane at the position and
orientation of the edge-on disk \citep{2004ApJ...617..418S}. 
The star and disk are 
surrounded by an optically thin asymmetric envelope with a well-ordered
magnetic field directed along P.A.\,$\sim 25^\circ$
\citep{2001ApJ...561..871H}.  Furthermore, an Herbig-Haro object
(HH\,494) was identified by H$\alpha$ and S[II] narrow-band imaging,
6.15 arcmin northwest of CB\,26 at P.A.\,=\,$145 ^\circ$
\citep{2004ApJ...617..418S}.  The HH object is thus perfectly aligned
with the symmetry axis of the disk and the bipolar nebula.
% (P.A.\,=\,$145$\degr). 
Recently, \citet{lau2009} reported the discovery of a small collimated molecular 
outflow along the same direction, that appears to rotate about its polar axis. 
It is speculated that this rotation is related to the possible binarity of the 
central star.
We are currently analyzing higher resolution interferometric CO maps, attempting to 
reveal the mechanism that drives this outflow rotation 
(Launhardt et al., in prep.).
Due to the edge-on configuration, the 
bolometric luminosity listed in Table\,\ref{tbl-physprop} is only a lower limit, 
while the $L_{\rm smm}\,/\,L_{\rm bol}$\ ratio is an upper limit.
Although the IR spectral index indicates a Class\,II source, 
the other evolutionary tracers  and the estimated age of 1\,Myr \citep{lau2009} 
suggest it is a Class\,I YSO.

%Fig.\,\ref{fig-cb26} shows strong centrally concentrated emission at (sub)mm 
%wavelengths that is coincident with the T-Tauri star. The SED (Fig.\,\ref{fig-cb26}f) 
%shows a bimodal distribution, due to the contribution from the hot photosphere of the 
%T-Tauri star and the cool surrounding material. The contribution from the central 
%source gives the core a high T$_{\rm bol}$, and a relatively low $L_{\rm smm}\,/\,L_{\rm bol}$.

%%%%%%%%%

{\bf CB\,34} 
(Fig.\,\ref{fig-cb34})
has the appearance of a Bok globule, but, 
with a distance of $\sim 1.5$\,kpc, it is rather atypical and more 
massive than other Bok globules. 
It has three dense cores and is associated with 
numerous young stars that seem to have formed from this cloud 
\citep{1997A&A...326..329L,1997AJ....113.1395A,2000A&A...362..635H}.
A water maser \citep{2006AJ....132.1322G} as well as several collimated 
outflows testify ongoing star formation activity. 
A bipolar CO outflow from cool gas \citep{1994ApJS...92..145Y}, 
shocked optical knots of atomic emission lines from radiative shocks (Herbig-Haro objects), 
as well as H$_2$\ infrared jets 
\citep{1994AJ....108..612Y,2002A&A...383..502K} have been reported. 
The chemical age of the globule
was estimated to be $>10^5$\,yr \citep{1998MNRAS.298.1092C}.
The presence of a pre-main-sequence star, CB34FU, with an age of
$\sim$\,10$^6$\,yr supports this young age \citep{1997AJ....113.1395A}.
\citet{2003MNRAS.344.1257C} studied the molecular gas of this globule 
and showed that the current star-formation activity is concentrated in the 
three main clumps which have sizes of $\sim 0.25$\,pc.
Because of this complexity, we do not further investigate the individual sources 
in this globule, but only present our data.

%not a great deal to say that isn't clear from the figure.

%%%%%%%%%%

{\bf CB\,39} 
is a small and not very opaque globule, located 
in direction of the Gemini\,OB1 cloud. It is associated with the well-known 
Herbig Ae/Be star HD\,250550, the distance of which was derived with 700\,pc.
No mm dust continuum emission has been detected from this star, indicating 
that its circumstellar disk, if still present, cannot be very massive.
\citet{1994ApJS...92..145Y} found a small, bipolar, low-velocity  ($\sim 3$\,km\,s$^{-1}$) 
CO outflow with not well-separated lobes.  
CB\,39 has been observed in different molecular lines.
It was detected in SO by \citet{1997MNRAS.291..337C} but not detected in
CS by \citet{1998ApJS..119...59L}. 
\citet{2005MNRAS.361..177S} measured a magnetic field P.A. of $150^\circ$, based on the polarization of
a number of background stars along the line of sight of the globule. 

%no figure to talk about

%%%%%%%%%%

%{\bf CB\,52}  

%%%%%%%%%%

{\bf CB\,54} 
(LBN\,1042; Fig.\,\ref{fig-cb54}) 
is a relatively compact molecular cloud with the appearance of 
a Bok globule, but seems to be associated with the Vela\,OB1 cloud complex at 
a distance of 1.1\,kpc. 
It contains a cluster of embedded young stars, visible on NIR images
\citep{1994AJ....108..612Y,1996AJ....111..930Y},
as well as a cold IRAS point source (07020-1618)
at the position of the mm peak and the embedded star cluster. 
CO observations revealed the presence of a collimated
bipolar molecular outflow with well-separated lobes and centered 
on the IRAS point source \citep{1994ApJS...92..145Y}. 
This source is also associated with cm continuum emission 
\citep[VLA\,AY0073;][]{1996AJ....111..841Y}.
C$^{18}$O observations suggests the presence of multiple cores with infall signatures 
\citep{1995ApJ...454..217W}. \citet{2006AJ....132.1322G} detected a water
maser. 
Polarimetry of background starlight revealed a magnetic field with an average 
P.A. of $116^\circ$ \citep{2005MNRAS.361..177S}.
Because of this complexity, we do not further investigate the individual sources 
in this globule, but only present our data.

%% water maser from YSO - 2006AJ....132.1322G

%%%%%%%%

{\bf CB\,58}  
(Fig.\,\ref{fig-cb58})
is a relatively large, slightly elongated and cometary-shaped globule 
at a distance of 1.5\,kpc. 
It is associated with IRAS point source 07161-2336 and the NIR images indicate 
the presence of several embedded YSOs. 
\citet{1998AJ....115.1111A} concluded from HCN observations that the core 
represents a Class\,I evolutionary stage. 
Our (low-quality) mm dust continuum map suggests the presence of two 
dense cores in the head of the globule, where also the IRAS source and 
a number of reddened stars with associated nebulosities are located.
However, none of the IR sources can be directly associated with one of the 
two dust clumps.
Because of this complexity, we do not further investigate the individual sources 
in this globule, but only present our data.

% The average P.A. of the magnetic field has been measured at $101^\circ$ \citep{2005MNRAS.361..177S}.
%Although is classified as a cool and quiescent cloud \citep{1991ApJS...75..877C},
%\citet{2000A&AS..141..175S} found out that the polarization vectors in CB\,58
%are scattered and tend not to align among themselves. The polarization strength
%shows low to high polarization of 4-5\%.

%% Class I core - 1998AJ....115.1111A
%% distance 1.5kpc - 1997A&A...326..329L
%% 7.8x3.4' - 1988ApJS...68..257C
%% 4-5% pol - 2000A&AS..141..175S

%%%%%%%%

{\bf BHR\,12} 
(CG\,30, DC\,253.3-1.6; Fig.\,\ref{fig-bhr12}) 
is a small, opaque, and very pronounced cometary globule with a bright rim, 
located in the Gum nebula region. 
The distance towards the Gum nebula clouds is somewhat uncertain, with estimates 
ranging from 200\,pc \citep{1999A&A...350..985K} to 450\,pc \citep{1983A&A...117..183R}.
For consistency with earlier papers \citep[e.g.,][]{1998A&A...338..223H}, we use here 400\,pc. 
The globule has an angular diameter of $\sim \unit[2.3]{\arcmin}$ and its 
tail (which merges with the tail of CG\,31) has a length of $\unit[25]{\arcmin}$ 
\citep{1983A&A...117..183R}.
At its head side, CG\,30 harbors two protostellar cores with a projected separation of $\sim $20\arcsec\ 
($\sim $8000\,AU), embedded in a common envelope \citep{2000IAUS..200P.103L}. 
The northern source (SMM\,1, CG\,30N) is associated with the IRAS point source 08076-3556 
and a NIR source, which drives the Herbig-Haro flow HH\,120 
\citep[see][ and references therein]{1995ApJ...453..715H}.
The southern core \citep[SMM\,2, CG\,30S;][]{2001ASPC..235..134L} is the 
origin of a protostellar jet with position angle 
$44^\circ$ \citep{1995ApJ...453..715H}, but no NIR counterpart is seen at this position.
Both cores are detected in all IRAC bands
\citep{2008ApJ...683..862C}. The IRAC 4.5\,$\mu$m image also shows bipolar jets, 
which have been traced in CO(2--1) using the SMA \citep{2008ApJ...686L.107C}.
Fig.\,\ref{fig-bhr12} shows that the two cores are well resolved at 450 and 
850\,$\mu$m, though they are surrounded by a common envelope. 
The cores are unresolved in our 1.3\,mm map. 
The SED is well sampled for both cores, and shows that SMM\,2 is colder or 
more heavily extincted than SMM\,1.  
We classify SMM\,1 as a Class\,I YSO, based on its bolometric temperature 
(see Table\,\ref{tbl-physprop}).
The low bolometric temperature of 50\,K 
suggests that SMM\,2 is a Class\,0 protostar and thus younger than SMM\,2.
BHR\,12 is thus one of those globules that contain two or more sources of different evolutionary stage 
within a few thousand AU (see Sect.\,\ref{sec-dis1-mult} and Table\,\ref{tbl-resmult}).

%%%%%%%% 

{\bf BHR\,36}  
(DC\,267.4-7.5; Fig.\,\ref{fig-bhr36})
is a cometary-shaped globule in the Vela sheet at a
distance of $\unit[400]{pc}$. 
A dense core, mapped at 1.3\,mm continuum, is located at the head side 
of the globule. The mm emission peaks at the position of the IR source 
detected both by IRAS and {\it Spitzer} 
and at the center of a bipolar jet seen on NIR images.
Two Herbig-Haro objects, HH46/47, are associated with a collimated bipolar 
outflow centered at this position \citep{1982ApJ...263L..73D}.
From their observations of $^{13}$CO(2--1), \citet{2007ApJS..171..478L} derive a total mass
of $255\pm60\msun$ and a size of $\sim$\,1.3\,pc for the entire globule.
In their ammonia survey, \citet{1995MNRAS.276.1067B} obtain a mass of $\sim \,11\msun$\ 
within the central 0.3\,pc for the dense core. 
They also derive a kinetic temperature of $T_k=\unit[18]{K}$\ in the centre.
The  chemical age is of the globule was estimated to be 
$\sim \unit[0.1]{Myr}$ \citep{2001MNRAS.326.1255C}. 
BHR\,36 is too far south for the JCMT to observe.
Like for the other southern globules, we are therefore limited 
to the 1.3\,mm maps. The fit to the SED of this source 
indicates T$_{\rm bol}$ of $\approx$150\,K (see Table\,\ref{tbl-physprop}). 
We therefore classify the embedded YSO as a Class\,I source.

%% d=400pc (Vela sheet) - 1998A&A...338..223H 
%% DC\,267.4-07.5
%% IRAS 08242-5050
%Its size was determined to be 5.6\arcmin$\times$2.4\arcmin \citep[ DSS blue]{1999ApJS..123..233L}. 
% 12'x15' 13CO(2--1), 11.5'x12' 12CO(4--3) - 2007ApJS..171..478L
%% 255\pm60 M\sol, (1.4x1.2pc size, 13CO(2-1)) Tkin<22K - 2007ApJS..171..478L
%% 11 M\sol (0.3pc NH3) 1995MNRAS.276.1067B
%% 0.1 Myr chemical age - 2001MNRAS.326.1255C
%% Tk=18K - 1995MNRAS.276.1067B
%% shock-excited HH46/47 objects in a collimated bipolar flow - 1982ApJ...263L..73D
%% DSS 5.6'x2.4' - 1999ApJS..123..233L

%%%%%%%% not yet finished - need spitzer points for SED

{\bf BHR\,55} 
(DC\,275+1.9; Fig.\,\ref{fig-bhr55}) is an average-sized, low-extinction globule,
associated with the annulus of cometary globules in the Vela-Gum complex \citep{1995MNRAS.276.1052B}, 
though it has not a very pronounced cometary appearance.
\citet{1995MNRAS.276.1067B} estimate the distance to the globule to be 300\,pc 
by looking at the reddening of background stars. 
From ammonia observations, \citet{1995MNRAS.276.1067B} 
derive kinetic and rotational temperatures of 13\,K and 12\,K respectively, 
a central volume density of ${\rm 8.1\times10^3\,cm^{-3}}$, and a core mass of 3.7\,M$_{\odot}$.
The globule is too far south to observe with JCMT, therefore only the 1.3\,mm data 
constrain the long-wavelength end of the SED.
The very compact (unresolved) mm source is associated with an IRAS point 
source (09449-5052), but has no near-infrared counterpart. 
Due to the lack of data at submm and MIR wavelengths, the SED is not well-constrained.
However, the low T$_{\rm bol}$ of $\approx$40\,K suggests this is a Class\,0 protostar. 

% listed as isolated & complex in BHR.
% coords 275, 1.8 place it in the vela annulus.
% IRAS 09449-5052, DC\,275+1.9
% 9x9' 13CO(2--1), 9.5x6.5 12CO(4--3) - 2007ApJS..171..478L

%%%%%%%%

%{\bf BHR\,58}  

%%%%%%%%

{\bf BHR\,71}
(DC 297.7-2.8; Fig.\,\ref{fig-bhr71})  
is a very opaque, elongated globule located near the southern Coalsack 
at an estimated distance of $\sim $\,200\,pc.
%
% dense core and embedded sources 
At its center, the globule contains a dense protostellar core, 
which emits strong mm dust continuum emission and is detected 
by various instruments down to MIR wavelengths.
%IRAS 11590-6452 
Observations with higher angular resolution revealed that this is actually a 
double core \citep{1998ASPC..132..173M} with two embedded IR sources, IRS\,1 and IRS\,2, 
separated by 17\arcsec\ \citep[3400\, AU --][]{2001ApJ...554L..91B}. 
Outflows are seen in the ${\rm H_2}~v=1-2~{\rm S}(1)$ line 
emanating from both sources. 
A highly collimated molecular outflow has been detected \citep{1997ApJ...476..781B} 
which is associated with two Herbig-Haro objects \citep[HH\,320 and HH\,321;][]{1997IAUS..182P..85C}.
\citet{2001ApJ...554L..91B} and \citet{2006A&A...454L..79P} resolved it into two outflows, 
mapping CO and methanol, 
and found that the IRS\,2 outflow is significantly hotter than the IRS\,1 outflow.
%Centimetre emission has been detected towards IRS\,1. 
%
%In this jet, methanol and silicon monoxide are enhanced by factors of 40 to 350 compared to
%typical dark clouds \citep{1998ApJ...509..768G}.
%
From $^{13}$CO(2--1) measurements, \citet{2007ApJS..171..478L} derived a mass
of $\unit[12\pm4]{\msun}$\ for the entire cloud.
From their ammonia measurements, \citet{1995MNRAS.276.1067B}
derived a mass of $\unit[2.2]{\msun}$\ for the dense core 
and \citet{1997ApJ...476..781B} derive $\unit[2.4]{\msun}$\ 
from the 1.3\,mm dust continuum map.
The 1.3\,mm continuum image does not resolve the two sources, 
but the low ${\rm T_{bol}}$\ of the combined SED indicates that at least IRS\,1 
(the brighter source at the mm peak position) is a Class\,0 protostar. 
This is in agreement with earlier work \citep{1997ApJ...476..781B}.

%% Coalsack, 12\pm4 Msol (13CO(2-1)) - 2007ApJS..171..478L
%% 150\pm30pc 1995A&AS..114..105F
%% DC 297.7-2.8, IRAS 11591-6452
%% 6.9'x3.2' - 1999ApJS..123..233L
%% 15'x15' 13CO(2--1), no detection 12CO(4--3) - 2007ApJS..171..478L
%% 2.2Msol NH3 - 1995MNRAS.276.1067B
%% 1.3Msol/beam 1.3mm cont - 1998A&A...338..223H
%% methanol, silicon monoxide enhanced (40 to 350) vs. a typical dark cloud - 1998ApJ...509..768G

%%%%%%%%

{\bf BHR\,86} 
(DC 303.8-14.2; Fig.\,\ref{fig-bhr86})
is a cometary-shaped globule located in the Chameleon dark cloud complex at a 
distance of 150 to 180\,pc. 
\citet{2007ApJS..171..478L} carried out $^{12}$CO(2--1) observations and measured a size of 
$\unit[15]{\arcmin}\times\unit[15]{\arcmin}$ and a mass of $\unit[19\pm4]{\msun}$.
From {\it ISO} observations, \citet{2005A&A...437..159L}
derived central dust temperature of $\unit[14\pm1]{K}$ and a mass of $\unit[2.7]{\msun}$\ 
for the dense core and concluded that the embedded source is a 
Class\,0/I transition object. 
They also showed that the main heating source of the dense core is not the embedded source, 
but the interstellar radiation field.
From CS(2--1), HCO$^+$(1--0) and HCN(1--0) line profiles
\citet{1997A&A...317L...5L} could classify BHR\,86 as a core with simultaneous
infall of dense gas in the centre and a non-collapsing envelope. 
They also detected a bipolar molecular outflow in $^{12}$CO(1--0).
NIR observations in $JHK_s$ showed an exponential extinction profile with
$p=2.29\pm0.08$ \citep{2007A&A...463.1029K}.
From the low bolometric temperature of $\approx$60\,K, the high $L_{\rm smm}\,/\,L_{\rm bol}$\ ratio 
of $\approx$6\%, and the absence of a NIR source we conclude this this is a Class\,0 protostar,
which is in agreement with the ISO observations.

% IRAS 13036-7644
% Chamaeleon Cloud Complex, 19\pm4Msol - 2007ApJS..171..478L
%% 150-180pc - 1997A&A...327.1194W
%% DC 303.8-14.2, IRAS 13037-7644
% 4.7'x2.4' DSS - 1999ApJS..123..233L
% 15'x15' 12CO(2--1), 12'x15' 13CO(4--3) - 2007ApJS..171..478L
%% Tdust 14.6±1K, Class 0/I transition object, externally heated, 2.7Msol from ISO - 2005A&A...437..159L
%% JHKs extinction map profile exp=2.29 p = 2.29 ± 0.08, 2007A&A...463.1029K
%% CS(J=2-1), blue peak stronger
%% HCO+(1-0) red peak stronger - vel grad
%% HCN(J=1-0) both types - velocity grad 
%% 12CO(1-0) jet - 
%% -> infall of the dense gas, non-collapsing envelope - 1997A&A...317L...5L

%%%%%%%%

%{\bf BHR\,87}  

%%%%%%%%

{\bf CB\,68} 
(L\,146; Fig.\,\ref{fig-cb68}) 
is a small, nearby, slightly cometary-shaped opaque Bok globule 
located in the outskirts of the $\rho$\,Oph dark cloud complex
at a distance of $\sim160$\,pc.
The dense core of the globule is associated with a cold IRAS point source 
(16544-1604) and exhibits strong, extended, but centrally peaked
sub(mm) dust continuum emission
\citep{1997A&A...326..329L,1998ApJS..119...59L,2000ApJ...542..352V}.
No NIR source is detected at 2.2\,$\mu$m.
The central source, which was classified as a Class\,0 protostar 
\citep{1997A&A...326..329L},
drives a weak, but strongly collimated bipolar molecular outflow
at P.A. 142$^{\circ}$ \citep{1996A&AS..115..283W,1997ApJ...489..719M,2000ApJ...542..352V}.
\citet{2000ApJ...542..352V} also performed submm continuum polarization measurements at 
850\,$\mu$m and found a large linear polarization of 10\%, indicating that Bok globules can have 
well ordered magnetic fields, and suitable conditions for grain alignment. 
The SED, which  is well-sampled between the 1.3\,mm IRAM data and the NIR IRAC data, 
indicates a 
${\rm T_{bol}}$ of $\sim 50$\,K (Table\,\ref{tbl-physprop}), 
confirming the classification as a Class\,0 protostar (Fig.\,\ref{fig-tbol-lsubmm}). 

% CB68 references:
% 1988ApJS...68..257C - CB88
% 1997A&A...326..329L - Launhardt & Henning 97 - mm continuum survey
% 1998ApJS..119...59L - Launhardt et al 98 - CS survey
% 2000ApJ...542..352V - Vallee etr al. 2000
% 1996A&AS..115..283W - Wu et al. 96 - CO outflow detection
% 1997ApJ...489..719M - Mardones et al. 97 - H2H+, H2CO, CS
% 1995ApJ...454..217W - Wang et al. 95 - C18O, H2CO
% 1996A&A...312..585L - Lemme et al. 96
% 1997MNRAS.291..337C - Codella & Muders 97 - 
% 1982ApJ...258L..75W - Walmslay et al. 82
% 1998AJ....115.1111A - Afonso et al. 98 - HCN

%%%%%%%%

{\bf BHR\,137}  
(DC\,344.6-04.3; Fig.\,\ref{fig-bhr137}) is a small, opaque, and elongated globule 
with a diffuse tail. The distance is somewhat uncertain.
\citet{2000ApJ...532.1038V} associate the core 
with the nearby Scorpius region at a distance of 145\,pc, whereas \citep{1995MNRAS.276.1067B} 
derive a distance of $\unit[700]{pc}$, based on measuring the reddening of background stars. 
We adopt here the latter distance estimate. 
At the northern rim of the globule, outside the dark core, there are several red 
stars with associated nebulosities. An IRAS point source (17181-4405) appears to be 
associated with the southern-most of these stars.
For the dense core, which is located south of the nebulous stars and elongated north-south, 
\citet{1995MNRAS.276.1067B} estimate a lower mass limit $\unit[8.7]{\msun}$\ from 
NH$_3$. We derive a total mass of $\unit[26]{\msun}$\ from the 1.3\,mm dust continuum map.
BHR\,137 was not observed by {\it Spitzer} and our observations do not permit to compile a reliable 
SED for this source. 
Therefore, we cannot evaluate the evolutionary stage of this core, 
although the high mass and lack of NIR sources at the position of the mm peak 
suggest a rather early evolutionary stage. 
Due to the large distance and low angular resolution of our SEST map, we can also 
not evaluate the possible multiplicity of this massive elongated dust core.
%
%% DC 344.6-04.3 
%% 5.0'x3.0' ESO/SERC Southern J survey - 1986A&AS...63...27H
%% 250pc (reddening), 0.4Msol - 1995MNRAS.276.1067B
%% 700pc (reddening), 8.7Msol - 1995MNRAS.276.1067B    **************CORRECTION****PLEASE DOUBLE-CHECK!
%% no IRAS                                             ************IRAS source mentioned in 1999AJ....118..990Y

%%%%%%%%

{\bf CB\,130} 
(L\,507; Fig.\,\ref{fig-cb130}) 
is a small, slightly elongated globule located in the Aquila rift at a distance 
of $\unit[200]{pc}$.
There is no IRAS point source associated with this globule.
\citet{1997A&A...326..329L} detected only weak mm-continuum emission in its centre 
and \citet{1999AJ....118..990Y} concluded from the weakness of the HCN(1--0) emission 
that there is no embedded YSO. Since there is also no evidence of outflows, the source 
was originally classified as a pre-stellar core by \citet{1997A&A...326..329L}.
However, our (sub)mm maps show that CB\,130 is well detected in all bands (see Fig.\,\ref{fig-cb130}). 
The core is resolved into two separate compact sources, 
separated by 30\arcsec\ ($\approx$6000\,AU) and surrounded by a common envelope. 
The brighter source, SMM\,1, is associated with a faint NIR source, which is also detected by {\it Spitzer} up to 24\,$\mu$m 
(there are no useable data at longer wavelengths).
A very red star, which we call IRS, is located 14\arcsec\ ($\approx$3000\,AU) 
east of SMM\,1 (Figs.\,\ref{fig-cb130}b and \ref{fig-spitzermaps}). 
At 24\,$\mu$m, this star is still 1.3 times brighter than the IR source 
associated with SMM\,1. However, as none of the (sub)mm maps shows a local 
maximum at this position, source IRS does not seem to have 
detectable mm dust continuum emission associated. 
We conclude that the spatial coincidence between IRS and the dust core 
is a projection effect and the source is located outside the dust core.
We could compile a SED for SMM\,1, separating the fluxes from IRS at short wavelengths and from SMM\,2 at 
(sub)mm wavelengths. 
The low ${\rm T_{bol}}$\ of $\approx$55\,K and high $L_{\rm smm}\,/\,L_{\rm bol}$\ 
ratio of $\approx$10\% suggest SMM\,1 is a Class\,0 protostar, although 
the NIR/MIR SED is rather indicative of a Class\,I YSO (see Fig.\,\ref{fig-sed-all}).
The IR SED of IRS ($\alpha_{\rm IR}\,=\,-0.7$) suggests this is a Class\,II YSO, but we don't have 
strong constraints on its FIR to (sub)mm SED.
The (sub)mm SED of SMM\,2 has a somewhat steeper slope than SMM\,1 and there is no 
IR counterpart detected, indicating that this core does not have an embedded heating source.
We therefore assume that this is a very low-mass prestellar core.
CB\,130 is thus one of those globules that contain two or more sources with different evolutionary stage 
within a few thousand AU (see Sect.\,\ref{sec-dis1-mult} and Table\,\ref{tbl-resmult}).

%% no IRAS, L507
%% 5.6'x2.2' (POSS) - 1988ApJS...68..257C
%% Mgas 9.4E−2Msol - 1997A&A...326..329L
%% 200pc (Aquila rift) - 1985ApJ...297..751D
%% weak HCN(1--0) -> no YSO - 1999AJ....118..990Y
%% ext = 3.7

%%%%%%%%

{\bf CB\,188}
(L\,673-1; Fig.\,\ref{fig-cb188})
is a small, roundish, and opaque globule with a detached diffuse tail. 
It is located towards Aquila at a distance of $\sim\unit[300]{pc}$\ and is 
associated with the Lindblad ring.
The dense core of the globule is associated with a cold IRAS point source 
(19179+1129) and exhibits strong, extended, but centrally peaked
sub(mm) dust continuum emission.
At the peak of the mm emission and the IRAS position, there is a very red star, 
indicating that the bulk of the compact mm dust emission does not arise from 
an optically thick envelope, but rather from a circumstellar disk.
The source drives a bipolar molecular outflow at P.A.$=75^\circ$\
with a mass loss rate of $\sim\unit[6.6\times10^{-7}]{\msun/yr}$ \citep{1994ApJS...92..145Y}.
The red and blue lobe have a strong overlap, indicating that the source is 
seen almost pole-on. This is consistent with seeing a strong NIR source at the 
peak of the mm continuum emission.
The source shows only weak N$_2$H$^+$(1--0)
emission which can be caused by destruction due to CO, which is released
from grain surfaces in outflows \citep{2007ApJ...669.1058C}.
%
%\citet{2005AJ....130.2166K} classified CB\,188 as an unstable
%core using NIR and C$^{18}$O(1--0) observations and fitting a Bonnor-Ebert-model
%to the observations. From MAMBO observations \citet{2008A&A...487..993K}
%derived a mass of $\unit[2.6]{\msun}$.
%
Our fit to the SED, which is well-sampled between 1.25\,$\mu$m and 3\,mm, 
indicates ${\rm T_{bol}}\approx$300\,K. 
We conclude that CB\,188\,SMM is a Class\,I YSO, exhibiting a relatively massive circumstellar 
disk, surrounded by an extended envelope. We thus confirm earlier classifications 
\citep[e.g.,][]{1998AJ....115.1111A}.

% Faint secondary submm clump 20 arcsec N-NE.
%% NIR Av=27.2mag, unstable C18O(1--0) - 2005AJ....130.2166K
%% Class I - 1998AJ....115.1111A
%% IRAS (19179+1129) source
%% 6.6E-7 Msol/yr mass loss

%%%%%%%

{\bf CB\,199} 
(B\,335, L\,663; Fig.\,\ref{fig-b335}) is a roundish, very opaque, relatively
isolated, and well-studied prototypical Bok globule, located towards Aquila 
\citep[see][ and references therein]{2008ApJ...687..389S}.
It's distance was recently redetermined to be only 90\,--\,120\,pc 
\citep{2009A&A...498..4550}.
The globule contains a single compact core at its center, 
that is detected at far-infrared wavelengths
and is very bright in the submillimetre 
range \citep{1983ApJ...274L..43K,1990MNRAS.243..330C}.
\citet{1993ApJ...417..613Z} have found evidence for mass infall onto the
central source, which was confirmed by observations in a CO rare isotope
\citep{1993ApJ...414L..29C} and by CSS observations 
\citep{1995ApJ...451L..75V}. 
The source drives a well-collimated bipolar molecular outflow \citep{1982ApJ...256..523F}.
Furthermore, a number of HH objects have been detected 
both at optical wavelengths \citep{1986AJ.....92..633V,1992A&A...256..225R} 
and in the 1-0 S(1) line of H$_2$\ \citep{1998ApJ...500L.183H}.
This combination of characteristics that makes the central embedded source in 
B\,335 one of the clearest examples of a Class\,0 protostar.
\citet{2006AJ....132.1322G} found a water maser with a high velocity shift
($\sim\unit[30]{km/s}$) compared to the cloud velocity. This could be evidence
for a protoplanetary or young planetary disk.
The low ${\rm T_{bol}}$ of 37\,K that is derived from our fit to the SED 
corroborates the earlier classification.

%%%%%%%%%

{\bf CB\,205} 
(L\,810; Fig.\,\ref{fig-cb205})
has the appearance of a Bok globule, but, 
with a distance of $\unit[2]{kpc}$, its apparent
size of $\unit[8]{\arcmin}$ corresponds to a diameter of $\unit[4.8]{pc}$.
It is thus much larger and more massive than other Bok globules. 
SCUBA observations revealed the presence of a large submm core with two 
condensations, separated by $\approx \unit[45]{\arcsec}$ \citep{2000A&A...362..635H}.
The first and more massive core (CB205-SMM1) coincides with the IRAS source 19433+2743
and has a mass of $\unit[10]{\msun}$.
The total of mass of the dense core is estimated to be $\unit[120]{\msun}$. 
The cloud is known to harbour many optical and NIR sources.
\citet{1990A&A...231..165N} showed that at least 12 stars are
associated with CB\,205 and are located in the eastern part of the cloud.
A very bright and red
source (L\,810\,IRS) with a bolometric luminosity of $\unit[890]{\lsun}$\ is the
illumination source for a NIR nebula \citep{1993ApJ...408L.101Y,2004A&A...419..241M}.
This nebula is hourglass-shaped and has an extent in NS direction of
$\unit[20]{\arcsec}$. It has the same orientation as the molecular outflow, which
shows a strong overlap of the blue and red lobes. The peak of $^{13}$CO intensity
map coincides with the IRAS source \citep{1994ApJS...92..145Y}.
In the western part, \citet{2006A&A...457..891C} detected three Class\,0
sources with enhanced high velocity outflow activity, suggesting that the star formation
propagates through CB\,205 as the NIR cluster represents a later evolutionary
stage.
\citet{1985A&A...153..253N} also found water maser emission of $\unit[7.5]{Jy}$
towards L\,810, which could not be confirmed in later observations \citep{2006AJ....132.1322G}.
Although the entire source was classified as Class\,I source by \citet{1998AJ....115.1111A}, 
we find that this complexity does not allow us to construct meaningful SEDs or to reliably interpret 
a possible integrated SED. Therefore, we do not further investigate the individual sources 
in this globule or discuss its evolutionary stage, but only present our data.

%% L810
%% 1.5kpc - 1985A&A...153..253N
%% extended NIR nebulosity at the IRAS source - 2004A&A...419..241M
%% IRAS 19433+2743
%% 13CO core coincides with IRAS - 1994ApJS...92..145Y
%% NS-outflow strong overlap red-blue - 1994ApJS...92..145Y
%% Class I - 1998AJ....115.1111A
%% NIR bolometric luminosity of 890 lsol - 
% 20`` nebula NS hourglass shape, NIR jet - 1993ApJ...408L.101Y
%% IR luminosity 1.4E2 lsol - 1999A&AS..139..257L
%% two submm-sources, 45'' separation SCUBA 120 Msol tot, 10 Msol SMM1 - 2000A&A...362..635H
%% 12-15 embedded sources - 1990A&A...231..165N
%
%From IRAS observations, the IR luminosity was determined to be $\unit[1.4\times10^2]{\lsun}$
%\citep{1999A&AS..139..257L}.

%%%%%%%%%

{\bf CB\,224} 
(L\,1100; Fig.\,\ref{fig-cb224}) is a small, roundish globule 
with a large visible cloudshine halo (like most globules).
It is located in the northern Cygnus region at $d\sim 400$\,pc.
The core of the globule contains two mm sources, separated by 
$\approx$35\arcsec, both detected at all three (sub)mm bands.
The northeastern source (SMM\,2) is associated with the cold IRAS
source 20355+6343 and with a red star which is also seen at optical wavelengths.
The SED of this source indicates a ${\rm T_{bol}}$ of 340\,K 
and we classify this source as a Class\,I YSO.
The southwestern source (SMM\,1), which is the brighter mm source, is associated with a very 
faint and diffuse NIR nebula. This source drives a
collimated $^{13}$CO bipolar outflow at P.A. $=-120^\circ$
\citep{2007ApJ...669.1058C}. HCO$^+$(1--0) observations clearly show evidence
of infall towards the core center \citep{2002ApJ...577..798D}.
CB\,224 was not observed with {\it Spitzer}. The SED is therefore not well-constrained at 
MIR/FIR wavelengths and we rely on aperture photometry in IRAS maps \citet{1999yCat.2225....0G}.
The SED of this source indicates a ${\rm T_{bol}}$ of 56\,K 
and we classify it as a Class\,0 protostar.

% 3.9 lsol - 1997A&A...326..329L
% HCO+ infall - 2002ApJ...577..798D

%%%%%%%%%%

{\bf CB\,230} 
(L\,1177; Fig.\,\ref{fig-cb230}) 
is a small, slightly cometary-shaped, bright-rimmed Bok globule located in the 
Cepheus Flare region at a distance of $\sim 400\pm 100$\,pc. 
The dark globule has a low-extinction tail, which is clearly visible on optical 
images by its cloudshine, that connects to the bright nebula Sh\,2-136.
At its southern rim, the globule contains a dense core 
that is associated with two NIR reflection nebulae, 
separated by $\sim\unit[10]{\arcsec}$\ \citep{1996AJ....111..930Y,Launhardt:1996,2001IAUS..200..117L}.
The brighter, western nebula (IRS\,1), has a conical shape with the origin at the mm peak position.
The fainter eastern nebula (IRS\,2) is also located inside the mm core, but seems to 
exhibit much less dust emission than IRS\,1.
An IRAS point source (21169+6804) is located between the two NIR nebula, indicating that 
both sources contribute to the IRAS PSC fluxes.
CB\,230 was classified by \citet{2005ApJS..156..169F} as a Class\,0/I transition object.
CB\,230 was also observed in several molecular lines and was found to exhibit signatures of 
mass infall. Its virial mass was determined to be $\unit[12]{\msun}$ (CS) and $\unit[8]{\msun}$ (C$^{18}$O), 
respectively \citep[][]{1998ApJS..119...59L, 2001IAUS..200..117L}. 
The dense core is associated with a large-scale collimated CO 
outflow with $\unit[6.6\times10^{-7}]{\msun/yr}$ at P.A. $=7^\circ$ 
and a dynamical age of $\sim 2\times 10^4$\,yr \citep{1994ApJS...92..145Y}.
This outflow was resolved by \citet{2001IAUS..200..117L} into two misaligned bipolar flows, 
the stronger one originating at IRS\,1 and the weaker one at IRS\,2.
The NIR nebulae can thus be interpreted as forward-scattered light emanating from those 
outflow cones that are directed towards us.
Observations of [FeII] and weak H$_2$ emission in the jet suggest
the presence of fast, dissociative J-shocks \citep{2004A&A...419..241M}. 
Magnetic field strength and projected direction in the dense core 
were derived by submm polarimetry to be $B = 218\pm 50~\mu$G and P.A. $= -67^\circ$, respectively 
\citep{2003ApJ...592..233W}. Earlier work by \citet{2000ApJ...542..352V} is consistent with this picture.
Since the two embedded sources are only resolved at wavelengths shorter 25\,$\mu$m, 
we cannot compile complete separate SEDs. 
However, the NIR luminosity ratio and the submm maps indicate that IRS\,1 
is the more massive and luminous source that dominates the SED at all wavelengths.
The combined SED and the resulting ${\rm T_{bol}}$ of $\approx$210\,K should thus be representative for IRS\,1.
While the bolometric temperatures and high 
$L_{\rm smm}\,/\,L_{\rm bol}$\ ratio suggests a Class\,I YSO with a significant envelope, 
the $\alpha_{\rm IR}$\ value of the double-peaked SED indicates Class\,II 
(see discussion in Sect.\,\ref{sec-dis1-evol}).
From the similarity of the NIR/MIR SEDs and the spatial proximity, we also conclude that 
IRS\,2 is at the same evolutionary stage, but less massive, and has formed from the same 
initial core as IRS\,1.

%The density peak of the globule is seen clearly at ${\rm 450\,\mu m}$ to 
%1.3\,mm (see Fig.\,\ref{fig-cb230}c-e). The grey-body fit to the SED shows 
%two peaks, at $\sim 3\,\mu$m and $\sim 100\,\mu$m. These peaks are due to the 
%emission from the central protostar and the cold dust envelope respectively.
%
% (IRAS 21169+6804)
% a jet in [FeII] and weak H2 emission suggests the presence of fast,
% dissociative J-shocks.- 2004A&A...419..241M
% 1994AJ....108..612Y
% Class 0/1 2005ApJS..156..169F
% Mv(CS)=12, Mv(C18O)=8 - 1998ApJS..119...59L
% 6.6E-7 Msol/yr, monopolar outflow (red only) - 1994ApJS...92..145Y

%%%%%%%%%

{\bf CB\,232}
(B\,158; Fig.\,\ref{fig-cb232})
is a small, slightly cometary-shaped globule which is associated with the 
Sh\,2-126 cloud at a distance of $\unit[600]{pc}$.
Using SCUBA observations, \citet{1999ApJ...526..833H} detected a submm double core 
with the two components having different evolutionary stages and masses 
of $\unit[0.3]{\msun}$ and $\unit[0.1]{\msun}$ respectively. The total mass was derived to be
$\unit[6]{\msun}$. The more evolved object, which we call here IRS and which presumably can be associated with
a nearby NIR star and IRAS source 21352+4307, was classified to be a Class\,I source \citep{1995AJ....109..742Y}.
The main submm core (with no NIR counterpart; called here SMM) is a candidate Class\,0 protostar. A bipolar outflow
with poor collimation at P.A. $= -30^\circ$ \citep{1994ApJS...92..145Y}
cannot clearly be associated with either of the two sources.
The virial mass of the entire dense core was derived with 
$\unit[42]{\msun}$ (CS) and $\unit[16]{\msun}$ (C$^{18}$O), respectively 
\citep{1998ApJS..119...59L}.
CB\,232 is also currently the smallest
Bok globule that is known to harbour a water maser \citep{2006AJ....132.1322G}.
Although our submm maps, except for the low SNR 450\,$\mu$m map, do not resolve 
the two embedded sources (SMM and IRS1), we tried to decompose the emission into an extended 
envelope and two compact gaussian sources at the positions of the mm peak and the 
red star. The resulting individual source fluxes at 850\,$\mu$m are somewhat higher 
than estimated by \citet{1999ApJ...526..833H} and the relative uncertainty is of the 
order 50\%, but they can be used to constrain the SED of source IRS fairly well.
The {\it Spitzer} maps show that even at 70\,$\mu$m source IRS1 
completely dominates and there is no trace of emission detected from the 
position of source SMM, indicating that the dust core has no or only a very low-luminosity 
internal heating source (see Fig.\,\ref{fig-spitzermaps} for the 24\,$\mu$m image).
Source SMM is thus more likely prestellar than protostellar.
The SED for IRS1, which may contain some unknown but small contribution from SMM around 100\,$\mu$m, 
indicates a ${\rm T_{bol}}$ of $\approx$170\,K. We classify IRS1 as a Class\,I YSO.
A second IR source (IRS2), located $\approx$40\arcsec\ south of SMM, is seen on all {\it Spitzer} 
images and seems to be associated with the globule, but we don't compile an SED or 
classify this source.
CB\,232 is one of those globules that contain two or more sources of different evolutionary stage 
within a few thousand AU (see Sect.\,\ref{sec-dis1-mult} and Table\,\ref{tbl-resmult}).

%% Barnard 158
%% Lbol 11.721 lsun, 7.6' size, previously unreported NIR reflection nebula - 2007AJ....133.1528C
%% 600 pc - 1997A&A...326..329L
%% Class I - 2005ApJS..156..169F,
%% smallest globule to harbour a water maser from YSO - 2006AJ....132.1322G
%% Mv(CS)=42 Mv(C18O)=16 - 1998ApJS..119...59L
%% bipolar outflow, poor collimation, pa=-30, 1994ApJS...92..145Y
%% SCUBA two submm sources with 0.3 and 0.1 Msol, total 6 Msol 1999ApJ...526..833H
%% Class 0 and a more evolved star - multi epoch SF
%
%\citet{2007AJ....133.1528C} detected a previously unreported NIR reflection nebula.
%They determined the optical size to be $\unit[7.6]{\arcmin}$ and the bolometric
%luminosity to be $\unit[11.7]{\lsun}$. 

%%%%%%%%%

{\bf CB\,240} 
(L\,1192; Fig.\,\ref{fig-cb240}) is a small, roundish globule with a diffuse tail, 
located in the Cepheus Flare region. Its distance is estimated to be $\sim 500$\,pc, 
but this is rather uncertain.
IRAS source 22317+5816, which has been identified as a Class II-D YSO \citep{1995AJ....109..742Y}, 
is located outside the visible boundary of the globule, $\approx$2\arcmin\ to the south-east. 
In a NIR survey, \citet{1994AJ....108..612Y} detected a red object at the IRAS position, 
but no nebulosity or evidence of multiplicity. 
\citet{1997A&A...326..329L}, however, identify a cluster of young stars at the IRAS position 
(see Fig.\,\ref{fig-cb240}b), 
and calculate the bolometric luminosity of the associated core to be $7.3\,L_\odot$.
CB\,240 has been detected in a number of molecular transitions, including C$^{18}$O, NH$_3$, HCO$^+$ and SO 
\citep{1995ApJ...454..217W,1995ApJ...444..708T,1995ApJ...455..556T,1995ApJ...449..635T}.
No outflow has been detected from this source \citep{1992ApJ...385L..21Y}.
CB\,240 is detected in our dust continuum maps with a low signal-to-noise ratio at $850\,\mu$m and 1.3\,mm. 
It is not detected at $450\,\mu$m.
Since the IRAS source and mm core is located outside the optical boundaries of the globule and is obviously 
composed of a small cluster of young stars with a diffuse common dust envelope, 
we cannot discriminate and classify individual sources. Therefore, we do not show a SED  
and do not further discuss this source here. 

%L97 detected 1.3 mm emission from a compact source very close to the IRAS source. 
%huard et al 1999 only see extended emission at 850 microns (these data) 
%(I would say that the peak of the smm data is only 6" from the IRAS %source, 
%so we are probably detecting the 850 micron emission from it. but is it a point source??

%%%%%%%%%

{\bf CB\,243}
(L\,1246; Fig.\,\ref{fig-cb240})
is an opaque, very elongated, filamentary-like globule with a long dark tail pointing eastward, 
located at a distance of $\unit[700]{pc}$\ in the Cep\,OB3 region. 
A head-like, very opaque extension at its western end contains a dense core that is 
currently forming one or more stars.
Extended submm dust emission is traced from the entire western part of the globule, 
with two embedded compact sources: 
SMM\,1 in the western head of the globule 
and SMM\,2 towards the center of the globule \citep{2001MNRAS.323..257V}. 
SMM\,1 is associated with the IRAS source 23228+6320 
and was classified a Class\,0/I transition source \citep{2005ApJS..156..169F} 
%with an apparent optical size of $\unit[3.3]{\arcmin}\times\unit[1.1]{\arcmin}$ \citep{1999ApJS..123..233L} and 
with a bolometric luminosity of $\unit[10]{\lsun}$\ \citep{1997A&A...326..329L}.
\citet{2001MNRAS.323..257V} also detected a molecular outflow from SMM\,1 in $^{12}$CO(2--1), 
with a dynamical age of $\unit[4.6\times 10^4]{yr}$ and masses of $\unit[0.004]{\msun}$\ 
in the red and $\unit[0.015]{\msun}$ in the blue lobe, respectively.
Our data confirm the general morphology of the dust emission from this globule.
However, we find that the dust core SMM1 is clearly offset by $\sim10$\arcsec\ to the north-west 
with respect to IRS (see Fig.\,\ref{fig-cb240}b).
This offset is larger than the pointing uncertainty of the mm telescopes and is confirmed 
independently in both the IRAM and SCUBA maps.
Unfortunately, the {\it Spitzer\ IRAC} maps cover only the bulk of the globule containing SMM\,2, 
but not the head with SMM\,1. 
They show that SMM\,2 has an 8\,$\mu$m shadow and does not contain an embedded heating source.
We therefore assume that SMM\,2 is a prestellar core.
However, the MIPS 24\,$\mu$m map shows clearly that 
there is a point-like MIR source at the IRAS position close to SMM\,1, but no emission is detected at the 
center position of SMM\,1, indicating that the dust core has no or only a very low-luminosity 
internal heating source (see Fig.\,\ref{fig-spitzermaps} for the 24\,$\mu$m image).
Since we cannot separate the SEDs of the two sources (SMM\,1 and IRS) and it is unclear from which 
of them the molecular outflow originates, we cannot clearly classify the sources in terms of their 
evolutionary stage. We can only speculate that IRS is a low to intermediate-mass 
Class\,I YSO that has most likely formed in CB\,243, and that SMM\,1 is 
most likely a 5\,--\,10\,M$_{\odot}$\ prestellar core. 
This of course also makes invalid the classification attempts by 
\citet{2002AJ....124.2756V} and \citet{2005ApJS..156..169F}, 
who assumed that all fluxes arise from one source. 
CB\,243 is thus one of those globules that contain two or more sources of different evolutionary stage 
within a few thousand AU (see Sect.\,\ref{sec-dis1-mult} and Table\,\ref{tbl-resmult}).

%% IRAS 23228+6320, LDN 1246
%% 700pc, Lbol 10 lsun - 1997A&A...326..329L
%% 3.3'x1.1' DSS - 1999ApJS..123..233L
%% 2.5 msun Mgas - 1997MNRAS.288L..45L
%% 6.6 lsun Lbol - 0.96 env - class 0/1 transition source - 2005ApJS..156..169F
%% Mtot 40.1 msun, peak Av=56, outflow 12CO(2-1)  t=4.6E4yr, 0.015 msun red, 0.004 msun blue, 
%% circumstellar envelope 2.2msun - 2001MNRAS.323..257V
%From SCUBA observations, \citet{2001MNRAS.323..257V} derived a total mass
%of $\unit[40.1]{\msun}$ and a peak extinction in the central core of $A_v=\unit[56]{mag}$. 

%%%%%%%%

{\bf CB\,244} 
(L\,1262; Fig.\,\ref{fig-cb244}) 
is a medium-sized, slightly elongated globule located in the Cepheus region 
and most likely related to the Lindblad ring at a distance of $\approx 200$\,pc.
The globule contains two (sub)mm cores, surrounded by extended envelopes. 
The brighter (at submm wavelengths) and more compact core (SMM\,1) is associated with 
a faint NIR reflection nebula and a cold IRAS point source (23238+7401).
The SED of this source is well-sampled between NIR and mm wavelengths 
and we derive $T_{bol}\approx70$\,K.
The source drives a collimated bipolar molecular outflow with total mass 
$\unit[0.07]{\msun}$\ \citep{1989ApJ...340..472T,1994ApJS...92..145Y}.
We confirm earlier classifications of CB\,244\,SMM\,1 as a Class\,0 protostar, 
although it is probably more evolved than, e.g., B\,335.
The second core, SMM\,2, which is located $\approx$90\arcsec\ (18,000\,AU) 
north-west of SMM\,1, is much less compact and only detected at low signal-to-noise 
in the 850\,$\mu$m and 1.3\,mm maps. There is no indication of an embedded source 
in this core. Rather, it is seen as a dark shadow in the 8\,$\mu$m {\it IRAC}\,4 map
\citep{2010ApJ...712.1010T} and in the 24\,$\mu$m {\it MIPS} map \citep{2010Stutzetal}.
We therefore assume that SMM\,2 is a prestellar core \citep[cf.][]{2009ApJ...707..137S}.
For the region between the two submm-cores, \citet{2003ApJ...592..233W} observed 
an alignment of the polarization vectors parallel to the gas/dust bridge connecting 
the two cores. This could be a trace of mass infall along field lines in the
early stages of star formation. In the outflow region around SMM\,1, the 
polarization vectors tend to align with the direction of the outflow.

%Its bolometric luminosity and total envelope mass are
%$L_{\rm bol} \sim\unit[1.1]{\lsun}$ and $M_{\rm env} \sim$3.3\,M$_{\odot}$,
%respectively \citep{1997A&A...326..329L}.
%The two (sub)mm peaks are detected at 850\,$\mu$m and 1.3\,mm, whereas only 
%the MM1 has been detected at 450\,$\mu$m. The 450\,$\mu$m data
%of MM2 has a very high noise level and has been removed from Fig.\,\ref{fig-cb244}c.
%% 4.7'x2.1' DSS - 1999ApJS..123..233L
%% 0.07Msol in the outflow - 1989ApJ...340..472T

%%%%%%%%%%%

{\bf CB\,246}
(L\,1253; Fig.\,\ref{fig-cb246}) 
is a medium-sized, dark, roundish globule with a narrow tail pointing southeast.  
It is most likely associated with the Lindblad ring, the distance to which 
in this direction is $\sim 140$\,pc. 
%but other distance estimates also exist \citep[300\,pc;][]{1987ApJ...322..706D}.
%
The globule has an extended complex core that is fragmented into at least two elongated 
sub-cores, but no compact embedded sources have been found. 
{\it Spitzer\ IRAC} images show the two sub-cores as strong 8\,$\mu$m shadows.
\citet{1996A&A...312..585L} and \citet{1998MNRAS.298.1092C}
observed CB\,246 in C$_2$S(2--1) and NH$_3$. They also identified the two sub-cores 
and derived their total mass with $\approx \unit[7]{\msun}$.
They conclude that this source as inactive, with neither a water maser
nor molecular outflow.
\citet{2005ApJ...619..379C} concluded from the observed low N$_2$D$^+$/N$_2$H$^+$-fraction 
that CB\,246 is still in a very early evolutionary phase. 
\citet{2001ApJS..136..703L} classified CB\,246 as
having significant red excess, but no blue asymmetry in the
CS spectra, indicating that the core has not begun to collapse.
Recently, \citet{2005MNRAS.361..177S} and \citet{2009arXiv0906.0248W} 
derived the orientation of the magnetic field by measuring the polarization 
of background stars. 
They find that the magnetic field has a mean angular offset from the short axis of 
the globule core of $28\pm20$\,deg. Like for other globule they have studied, this hints towards 
magnetic fields not governing the star formation in these clouds.
Because the source is not detected at wavelengths shorter than 450\,$\mu$m, 
we cannot compile a meaningful SED. 
We conclude that CB\,246 contains at least two starless cores, possibly prestellar in nature, 
thus confirming earlier classifications. Sub-core SMM\,2 is fairly elongated and might be 
fragmented into a chain of sub-clumps.

%% LDN 1253
%% 140 pc - 1997A&A...326..329L, 300pc - 1987ApJ...322..706D
%% veeeery early stage low N2D+/N2H+ fraction - 2005ApJ...619..379C
%% 3.2'x2.1' DSS - 1999ApJS..123..233L
% similar N2H+/CO morphology - 2004ApJ...614..194W
%% 1Msol core, 7Msol from DSS, close to virial equilibrium, double core - 1998MNRAS.298.1092C
%% NH3 survey double core - 1996A&A...312..585L
%
%\citet{1999ApJ...526..788L} classified CB\,246 as a strong outflow candidate, 
%based on ${\rm CS\,(2-1)}$ and ${\rm N_2H^+ (1-0)}$ observations. 
% No, they have not observed CB246!
%
%\citet{2005MNRAS.361..177S} measured a magnetic field P.A. of $67^\circ$, 
%based on the polarization of a number of background stars. 

%%%%%%%%%%%%%%%%%%%%%%%%%%%%%%%%%%%%%%%%%%%%%%%%%%%%%%%%%%%%%

\section{Summary and conclusions}     \label{sec-sum}

We have studied the dense cores of 32 Bok globules and have obtained 
deep NIR images and (sub)mm dust continuum maps at up to three 
wavelengths (0.45, 0.85, and 1.3\,mm). 
With the exception of a small control sample, all sources were 
selected from earlier surveys which identified them as good 
candidates for having ongoing star formation, i.e., this is not an 
unbiased survey.
We also compiled SEDs, taking special care of separating the flux contributions 
from different neighbouring sources at different wavelengths, 
and fitted them to derive various source quantities and evolutionary stages of the sources.
The main results of this study are:
\begin{enumerate}
\item
We detected (sub)mm dust continuum cores in 26 out of the 32 globules observed.
The tentative, low SNR, single-dish 1.3\,mm continuum detections in CB\,52 
and BHR\,41, published in and earlier paper, could not be confirmed. 
%In four globules with no previous single-dish 1.3\,mm continuum detections, 
%we also did not detect mm dust continuum cores. 

\item
Eight of the 26 globules with detected (sub)mm cores are not studied in further 
detail or evaluated in terms of their evolutionary stage because they were 
either at too large distances ($>1$\,kpc) and multiple embedded sources were not resolved, 
or the (sub)mm maps were of too low quality, or we simply did not have enough data 
to draw any reliable conclusions.

\item
In 18 globules with detected (sub)mm cores, we derived evolutionary stages and physical parameters 
of the embedded sources. In total, we identified 
nine starless cores, presumably prestellar,
nine Class\,0 protostars, and eleven Class\,I YSOs in these 18 globules.

\item
We find that the bolometric temperature is the most reliable tracer to discriminate 
between Class\,0 protostars and Class\,I YSOs and confirm the empirical boundary of 70\,K.
The spread of $L_{\rm smm}\,/\,L_{\rm bol}$\ ratios within the Class\,0 and Class\,I groups 
is relatively large (2\,--\,10\% within Class\,0 and 0.8\,--\,3.5\% within Class\,I), with 
no significant correlation between $T_{\rm bol}$\ and $L_{\rm smm}\,/\,L_{\rm bol}$\ within the groups.
However, the three most evolved Class\,I sources, with visible stars and compact (sub)mm emission 
arising presumably from circumstellar disks, also have the lowest $L_{\rm smm}\,/\,L_{\rm bol}$\ ratios 
($\le 1.3$\%). We take this as tentative indication that the $L_{\rm smm}\,/\,L_{\rm bol}$\ ratio, 
as indicator of envelope dispersal, may better trace the evolution within the Class\,0 
and Class\,I phases. 

\item
The mean FWHM core sizes decrease from 
8,000\,AU for prestellar cores, to 3,000\,AU for Class\,0 protostars, 
to $<$2,000\,AU for Class\,I YSOs. 
The latter group also exhibits extended remnant envelopes with diameters of order 20,000\,AU.
Source-averaged volume densities, $n_{\rm H}$, increase from 
$1\times 10^7$\,cm$^{-3}$\ for prestellar cores, to $7\times10^7$\,cm$^{-3}$\ for Class\,0 protostars, 
to $>8\times10^7$\,cm$^{-3}$\ for Class\,I YSOs. 
The extended envelopes of the Class\,I YSOs have a mean density of $7\times 10^4$\,cm$^{-3}$.

\item
At least two thirds (16 out of 24) of the star-forming globules 
studied here show evidence of forming multiple stars on scales between 
1,000 and 50,000\,AU, either as multiple star-forming cores, 
wide embedded binaries, or small star clusters. 
The fraction of closer binaries formed from unresolved mm cores 
might be higher, but remains unknown from this study. 

\item
We find that the large majority of these small prototstar and star groups in globules with 
multiple star formation are comprised of sources with different evolutionary stages.
This includes neighbouring mm sources with obviously 
different evolutionary stages, 
prestellar or protostellar cores with nearby IR sources, 
presumably more evolved protostars or Class\,I YSOs, 
as well as NIR star clusters next to 
large (sub)mm cores with the potential to form more stars.
In only three globules we find coeval pairs, 
ranging from multiple pre-stellar cores in CB\,246, 
embedded Class\,0 protostars in BHR\,71, to an embedded Class\,I YSOs pair in CB\,230.
This widespread non-coevality possibly suggests a picture of slow and sequential 
star formation in isolated globules.

\item
These findings also call for special attention when compiling 
SEDs and attempting to derive source properties from flux measurements with insufficient 
angular resolution. One may easily end up classifying the combined SED of a prestellar core 
and a nearby, more evolved YSO as a Class\,0 protostar.
\end{enumerate}

While this paper presents an extensive, though not complete inventory of star-forming cores 
in nearby Bok globules, it remains subject of other detailed studies to investigate 
the physical and chemical properties, multiplicity and evolutionaly stages of individual sources, 
and to evaluate if and how star formation in isolated globules differs from that in larger 
molecular cloud complexes.
We expect soon to obtain spatially resolved mid to far-infrared Herschel data 
for many of the globule cores presented here and plan to use these data to accurately measure the 
dust temparure profiles and derive more reliable density profiles. In particular for the 
prestellar cores, which characterize the initial conditions of the protostellar collapse, 
this may lead to significant corrections of previous estimates of mass and density 
distributions which were often hampered by the lack of temperature measurements.
Furthermore, we are currently working on detailed studies of two globule cores 
which characterize the stage of the onset of the protostellar collapse and which may shed 
light on the non-coeval evolution of different sub-cores within one globule (Schmalzl et al., in prep.).
Last, but not least, this paper shall provide some guidance for further follow-up studies
of individual globules, e.g., with Herschel, SCUBA2, and ALMA.

%%%%%%%%%%%%%%%%%%%%%%%%%%%%%%%%%%%%%%%%%%%%%%%%%%%%%%%%%%%%%

\acknowledgments{
The authors would like to thank the staff at all of the telescopes
where data for this paper were obtained. The
JCMT is operated by the JAC, Hawaii, on behalf of the UK STFC,
the Netherlands OSR, and the National Research Council (NRC) of Canada. 
This research used the facilities of the Canadian Astronomy Data Centre (CADC) operated 
by the NRC of Canada with the support of the Canadian Space Agency.
This paper is partially based on observations collected at the
European Southern Observatory, La Silla, Chile
This work is also based in part on observations made with the Spitzer Space Telescope, 
which is operated by the Jet Propulsion Laboratory, California Institute of 
Technology under a contract with NASA.
This publication makes use of data products from the Two Micron All Sky Survey, which is a joint 
project of the University of Massachusetts and the Infrared Processing and Analysis Center/California 
Institute of Technology, funded by the National Aeronautics and Space Administration and the 
National Science Foundation.
Partial support for T.L.B. was provided by NASA through contracts
1279198 and 1288806 issued by the Jet Propulsion Laboratory,  
California Institute of Technology, to the Smithsonian Astronomical Observatory.
We also thank J. Steinacker and A. Stutz for helpful discussions and critical reading of 
the manuscript. Last, but not least, we also acknowledge helpful comments by the anonymous 
referee, which helped to improve the clarity of the paper.}

Facilities:
\facility{IRAM:30m(MPIfR 1-channel, 7-channel, 19-channel bolometers)},
\facility{SEST(SIMBA)},
\facility{JCMT(SCUBA)},
\facility{CAO:3.5m(MAGIC, $\Omega$PRIME)},
\facility{Max Plank:2.2m(IRAC-2B)},
\facility{Spitzer(IRAC,MIPS)}

%%%%%% APPENDIX %%%%%%%%%%%%%%%%%%%%%%%%%%%%%%%%%%%%%%%%%%

\appendix
\section{Optical depth conversion for JCMT} \label{app-jcmt}

For the JCMT data, the optical depths quoted in Table\,\ref{tbl-obsmm} refers to 225\,GHz.
To convert to $\tau_{850}$ and $\tau_{450}$, the following conversions 
\citep{2002MNRAS.336....1A} 
can be used for dates prior to October 1999:
\begin{equation}
\tau_{850}=3.99(\tau_{225GHz}-0.004),\\
\end{equation}
\begin{equation}
\tau_{450}=23.5(\tau_{225GHz}-0.012).
\end{equation}
For dates post December 1999, the following conversions apply:
\begin{equation}
\tau_{850}=4.02(\tau_{225GHz}-0.001),\\
\end{equation}
\begin{equation}
\tau_{450}=26.2(\tau_{225GHz}-0.014).
\end{equation}
The change is due to the different filters which were installed in SCUBA in 1999.

%%%%%% Tables %%%%%%%%%%%%%%%%%%%%%%%%%%%%%%%%%%%%%%%%%%%%%%%%%%%%%%%%%%%%%%%%%%%%%%%%%%%%%%%%%%%%%%%%
\clearpage

%%%%% Table 1: General properties of Bok globules
\begin{deluxetable}{lll}
\tablecolumns{3} 
\tablewidth{0pc} 
\tablecaption{Typical properties of Bok globules and their dense cores\label{tbl-globprop}}
\tablehead{\colhead{} & \colhead{Globule} & \colhead{Dense Core}}
\startdata 
Mass             & 5 - 50\,M$_{\odot}$  & 0.5 - 5\,M$_{\odot}$ \\
Size             & 0.2 -- 1\,pc         & 0.02 - 0.05\,pc      \\
Mean density     & $10^3$\,cm$^{-3}$    & $10^7$\,cm$^{-3}$    \\
Gas temperature  & 15\,K                & 10\,K                        \\
Line widths      & 0.5 -- 2\,km\,s$^{-1}$ & 0.4 -- 0.7\,km\,s$^{-1}$     \\ 
\enddata 
\end{deluxetable} 
%%%%% End Table 1

%%%%% Table 2: Source list
\begin{deluxetable}{lllllllll}
\tabletypesize{\scriptsize}
\tablecolumns{9} 
\tablewidth{0pc}  
\tablecaption{Coordinates and distances of the observed globules\label{tbl-sourcelist}}
\tablehead{
 \colhead{Name} & \colhead{Other names\tablenotemark{a}} & \colhead{IRAS source} & 
 \colhead{R.A.\tablenotemark{b}} & \colhead{Dec.\tablenotemark{b}} & \colhead{{\it D}~ (Ref.)} & 
 \multicolumn{3}{c}{Observation codes\tablenotemark{c}}\\
 & & & \colhead{(1950)}& \colhead{(1950)} & \colhead{[pc]~~~~~~} & \colhead{NIR} & 
 \colhead{450\&850\,$\mu$m} & \colhead{1.3\,mm}}
\startdata 
CB\,4      & \nodata         	& \nodata  	& 00:36:15.0 & $+$52:35:00 & \phn600 ~(11)  & CA02    & \nodata 	& I96/97 \\
CB\,6      & LBN\,613         	& 00465$+$5028 	& 00:46:33.8 & $+$50:28:25 & \phn600 ~(1)   & CA93    & J97a    	& I95    \\
CB\,6\,N   & \nodata         	& \nodata   	& 00:46:50.0 & $+$50:39:45 & \phn600 ~(1)   & \nodata & \nodata 	& I96/97 \\
CB\,17\,SW & L\,1388         	& \nodata  	& 03:59:16.4 & $+$56:42:27 & \phn250 ~(1)   & \nodata & \nodata 	& I96    \\
CB\,17     & L\,1389          	& 04005$+$5647 	& 04:00:31.3 & $+$56:47:58 & \phn250 ~(1)   & CA95    & J98d	& I95    \\
CB\,26     & L\,1439          	& 04559$+$5200 	& 04:55:54.7 & $+$52:00:15 & \phn140 ~(3)   & CA93    & J00a  	& I96    \\
CB\,34     & \nodata         	& 05440$+$2059 	& 05:44:04.7 & $+$20:58:54 & 1500 ~(2)   & CA93    & J97d/98a    & I95    \\
CB\,39     & (HD250550)       	& 05591$+$1630 	& 05:59:06.0 & $+$16:30:58 & \phn700 ~(4)   & CA93    & \nodata  	& I95    \\
CB\,52     & \nodata         	& 06464$-$1650 	& 06:46:25.3 & $-$16:50:38 & 1500 ~(2)   & CA93    & \nodata 	& I95    \\ 
CB\,54     & LBN\,1042        	& 07020$-$1618 	& 07:02:07.2 & $-$16:18:58 & 1100 ~(5)   & CA93    & J00a   	& I95    \\
CB\,58     & \nodata         	& 07161$-$2336 	& 07:16:08.3 & $-$23:36:19 & 1500 ~(2)   & CA93    & J97d 	& S02    \\
BHR\,12    & DC\,253.3$-$1.6  	& 08076$-$3556 	& 08:07:41.4 & $-$35:56:12 & \phn400 ~(2,6) & LS95    & J00a  	& S02    \\
BHR\,36    & DC\,267.4$-$7.5  	& 08242$-$5050 	& 08:24:16.8 & $-$50:50:49 & \phn400 ~(7,6) & LS95    & J00a    & S02    \\
BHR\,41    & DC\,267.7$-$7.4  	& 08261$-$5100 	& 08:26:11.5 & $-$51:00:39 & \phn400 ~(7,6) & LS95    & \nodata 	& S02    \\
BHR\,55    & DC\,275.9$+$1.9  	& 09449$-$5052 	& 09:44:56.9 & $-$50:52:07 & \phn300 ~(7,6) & LS95    & \nodata	& S02    \\
BHR\,58    & DC\,289.3$-$2.8  	& 10471$-$6206 	& 10:47:07.2 & $-$62:06:28 & \phn250 ~(7,6) & LS95    & \nodata 	& S02    \\
BHR\,71    & DC\,297.7$-$2.8  	& 11590$-$6452 	& 11:59:03.3 & $-$64:52:17 & \phn200 ~(8,6) & LS95    & \nodata 	& S95    \\
BHR\,86    & DC\,303.8$-$14.2 	& 13036$-$7644 	& 13:03:38.9 & $-$76:44:11 & \phn180 ~(9,6) & LS95    & \nodata 	& S02    \\
BHR\,87    & DC\,307.3$+$2.9  	& 13224$-$5928 	& 13:22:26.0 & $-$59:28:11 & \phn400 ~(7)   & LS95    & \nodata 	& S02    \\
CB\,68     & L\,146           	& 16544$-$1604 	& 16:54:27.6 & $-$16:04:48 & \phn160 ~(2)   & CA94    & J98b,d    	& I94/95 \\
BHR\,137   & DC\,344.6$-$4.3  	& 17181$-$4405 	& 17:18:08.9 & $-$44:06:02 & \phn700 ~(7,6) & LS95    & \nodata 	& S02    \\
CB\,130    & L\,507           	& \nodata  	& 18:13:40.0 & $-$02:34:00 & \phn200 ~(2)   & CA02    & J01a    	& I97    \\
CB\,188    & \nodata        	& 19179$+$1129 	& 19:17:53.9 & $+$11:29:54 & \phn300 ~(2)   & CA02    & J97b/03c  	& I95    \\
CB\,199    & B\,335, L\,663   	& 19345$+$0727 	& 19:34:35.3 & $+$07:27:24 & \phn100 ~(10)  & \nodata & J07e/98b,d/01a& \nodata \\
CB\,205    & L\,810           	& 19433$+$2743 	& 19:43:22.1 & $+$27:43:39 & 2000 ~(2)   & CA93    & J97e/03c    & I95    \\
CB\,224    & L\,1100          	& 20355$+$6343 	& 20:35:30.1 & $+$63:42:58 & \phn400 ~(5)   & CA93    & J03a,b 	& I94    \\
CB\,230    & L\,1177          	& 21169$+$6804 	& 21:16:52.7 & $+$68:05:09 & \phn400 ~(5)   & CA93    & J97b/01a 	& I94    \\
CB\,232    & B\,158           	& 21352$+$4307 	& 21:35:14.2 & $+$43:07:04 & \phn600 ~(2)   & CA93    & J98b/00b    & I95/97 \\
CB\,240    & L\,1192          	& 22317$+$5816 	& 22:31:44.1 & $+$58:16:21 & \phn500 ~(2)   & CA93    & J97e/00b    & I95    \\
CB\,243    & L\,1246          	& 23228$+$6320 	& 23:22:54.2 & $+$63:19:54 & \phn700 ~(2)   & CA93    & J97b,c/98c  & I94/95 \\
CB\,244    & L\,1262          	& 23238$+$7401 	& 23:23:40.0 & $+$74:01:30 & \phn200 ~(12)  & CA93    & J97b,c/98b/01a,b & I95\\ 
CB\,246    & L\,1253          	& \nodata  	& 23:54:05.0 & $+$58:18:00 & \phn140 ~(2)   & CA93    & J00b,c/01a  & I96/97 \\ 
\enddata
\tablenotetext{a}{~L= \citet{1962ApJS....7....1L};~~ LBN= \citet{1965ApJS...12..163L};~~ B= \citet{1927QB819.B3.......};~~ 
                  BHR= \citet{1988ApJS...68..257C}}
\tablenotetext{b}{Reference position for all Figures; not always identical to IRAS PSC position}
\tablenotetext{c}{Near-IR observing details, see Table \ref{tbl-obsnir}. (Sub)mm observing details, see Table \ref{tbl-obsmm}.}
\tablerefs{
(1) this paper (Sect.\,\ref{sec-dis}),
(2) \citet{1997A&A...326..329L},
(3) \citet{2001ApJ...562L.173L},
(4) \citet{1997A&A...320..159T},
(5) \citet{2003ApJ...592..233W},
(6) \citet{1998A&A...338..223H},
(7) \citet{1995MNRAS.276.1067B},
(8) \citet{1997ApJ...476..781B},
(9) \citet{1997A&A...327.1194W},
%(10) \citet{1979PASJ...31..407T},
(10) \citet{2009A&A...498..4550},
(11) \citet{1983ApJ...271..143D},
(12) \citet{1995A&AS..113..325H}}
\end{deluxetable}
%%%%% End Table 2

%%%%% Table 3: Millimetre Observing Parameters
\begin{deluxetable}{llccclcl}
\tabletypesize{\scriptsize}
\tablecolumns{8} 
\tablewidth{0pc}  
\tablecaption{Submillimetre and millimetre continuum observing parameters\label{tbl-obsmm}}
\tablehead{
 \colhead{Obs.} & \colhead{Telescope} & \colhead{Date} & \colhead{Detector} & 
 \colhead{$\lambda_0$} & \colhead{HPBW} & \colhead{$\tau_{\rm Zenith}$} & 
 \colhead{Flux calibrators\tablenotemark{a}}\\
 \colhead{code} & & & & \colhead{[mm]} & \colhead{[arcsec]} & & }
\startdata 
J97a & JCMT 15m & 06/97 & SCUBA~      	& 0.45~/~0.85 	& $8.6~/~14.9$\tablenotemark{d}	& $0.062\pm0.007$\tablenotemark{b}   	& Ur    	\\
J97b & JCMT 15m & 08/97 & SCUBA~      	& 0.45~/~0.85 	& $8.5~/~14.8$  		& $0.09\pm0.03$\tablenotemark{b}   	& Ur, Ma    	\\
J97c & JCMT 15m & 09/97 & SCUBA~      	& 0.45~/~0.85 	& $9.5~/~15.1$	 		& $0.105\pm0.008$\tablenotemark{b}   	& Ur      	\\
J97d & JCMT 15m & 10/97 & SCUBA~      	& 0.45~/~0.85 	& $8.6~/~14.9$\tablenotemark{d} & $0.17\pm0.03$\tablenotemark{b}   	& Ur, 618     	\\
J97e & JCMT 15m & 12/97 & SCUBA~      	& 0.45~/~0.85 	& $8.6~/~14.9$\tablenotemark{d}	& $0.042\pm0.006$\tablenotemark{b}   	& HL, 10216	\\
J98a & JCMT 15m & 02/98 & SCUBA~      	& 0.45~/~0.85 	& $8.6~/~14.9$\tablenotemark{d}	& $0.052\pm0.005$\tablenotemark{b}   	& 10216, 16293	\\
J98b & JCMT 15m & 04/98 & SCUBA~      	& 0.45~/~0.85 	& $8.2~/~14.6$			& $0.06\pm0.04$\tablenotemark{b}   	& Ur, 10216, 2668, 618\\
J98c & JCMT 15m & 07/98 & SCUBA~      	& 0.45~/~0.85 	& $8.0~/~14.4$	 	 	& $0.032\pm0.004$\tablenotemark{b}   	& Ur    	\\
J98d & JCMT 15m & 08/98 & SCUBA~      	& 0.45~/~0.85 	& $8.0~/~14.4$	 	 	& $0.096\pm0.006$\tablenotemark{b}   	& Ur, 16293	\\
J00a & JCMT 15m & 03/00 & SCUBA~        & 0.45~/~0.85 	& $8.6~/~14.9$\tablenotemark{d} & $~0.28\pm0.05$\tablenotemark{b}   	& 10216, 618, 16293 \\
J00b & JCMT 15m & 06/00 & SCUBA~      	& 0.45~/~0.85 	& $8.3~/~14.8$	 	 	& $0.107\pm0.004$\tablenotemark{b}   	& Ur, 16293	\\
J00c & JCMT 15m & 09/00 & SCUBA~      	& 0.45~/~0.85 	& $8.6~/~14.9$\tablenotemark{d} & $0.088\pm0.008$\tablenotemark{b}   	& 618    	\\
J01a & JCMT 15m & 09/01 & SCUBA~      	& 0.45~/~0.85 	& $8.5~/~14.9$	 	 	& $0.079\pm0.009$\tablenotemark{b}   	& Ur, 618	\\
J01b & JCMT 15m & 10/01 & SCUBA~      	& 0.45~/~0.85 	& $8.6~/~14.9$\tablenotemark{d} & $0.056\pm0.005$\tablenotemark{b}   	& Ma, 2668    	\\
J03a & JCMT 15m & 02/03 & SCUBA\tablenotemark{c}&0.45~/~0.85 & $9.3~/~15.8$	 	& $0.042\pm0.001$\tablenotemark{b}   	& Ur    	\\
J03b & JCMT 15m & 05/03 & SCUBA\tablenotemark{c}&0.45~/~0.85 & $8.6~/~14.9$\tablenotemark{d} & $0.037\pm0.001$\tablenotemark{b} & Ur    	\\
J03c & JCMT 15m & 09/03 & SCUBA~      	& 0.45~/~0.85 	& $9.3~/~15.8$~ 	 	& $0.080\pm0.005$\tablenotemark{b}   	& Ur    	\\
I94 & IRAM 30m & 04/94  & MPIfR 7-chan.  & 1.3 		& ~~~10.5   			& $0.26\pm0.12$ 		& Ur, Ma    	\\
I95 & IRAM 30m & 02/95  & MPIfR 7-chan.  & 1.3 		& ~~~10.5 		  	& $0.23\pm0.11$ 		& Ur, Ma    	\\
I96 & IRAM 30m & 02/96  & MPIfR 19-chan.~& 1.3 		& ~~~10.5  		 	& $0.20\pm0.10$ 		& Ur          	\\
I97 & IRAM 30m & 03/97  & MPIfR 19-chan.~& 1.3 		& ~~~10.5   			& $0.22\pm0.12$ 		& Ur          	\\
S95 & SEST 15m & 11/95  & MPIfR 1-chan.  & 1.3 		& ~~~~23     			& $0.19\pm0.06$ 		& Ur          	\\
S02 & SEST 15m & 11/02  & SIMBA          & 1.3 		& ~~~~24     			& $0.20\pm0.05$ 		& Ur, Np 	\\ 
\enddata
\tablenotetext{a}{Planets and standard calibration sources. 
   Ur\,=\,Uranus, Ma\,=\,Mars, Np\,=\,Neptune, 618\,=\,CRL\,618, HL\,=\,HL\,Tau, 
   10216\,=\,IRC+10216, 16293\,=\,IRAS\,16293-2422, 2668\,=\,CRL\,2668}
\tablenotetext{b}{Zenith optical depth at 225\,GHz, monitored by the Caltech Submillimeter Observatory radiometer. 
   See Sect.\,\ref{app-jcmt} for 450- and 850-$\mu$m conversions}
\tablenotetext{c}{SCUBA operated in scan-map mode}
\tablenotetext{d}{Planet beam map not available. Quoted beam-sizes are an average}
\end{deluxetable}
%%%%% End Table 3

%%%%% Table 4: NIR Observing Parameters
\begin{deluxetable}{llcccll}
\tabletypesize{\scriptsize}
\tablecolumns{7} 
\tablewidth{0pc}  
\tablecaption{NIR observing parameters\label{tbl-obsnir}}
\tablehead{
 \colhead{Obs. code} & \colhead{Telescope} & \colhead{Date} & \colhead{Camera} & \colhead{Pixel scale} & 
 \colhead{Filters\tablenotemark{a}} & \colhead{Seeing}}
\startdata 
CA93  & Calar Alto 3.5m & 12/93 & MAGIC         & 0$.\!\!^{\prime\prime}$32 & J, H, KS & 0$.\!\!^{\prime\prime}$9 -- 2$.\!\!^{\prime\prime}$2\\
CA94  & Calar Alto 3.5m & 08/94 & MAGIC      	& 0$.\!\!^{\prime\prime}$32 & KS       & 1$.\!\!^{\prime\prime}$4 \\
CA95  & Calar Alto 3.5m & 01/95 & MAGIC   	& 0$.\!\!^{\prime\prime}$32 & KS       & 0$.\!\!^{\prime\prime}$5 -- 1$^{\prime\prime}$\\
CA02  & Calar Alto 3.5m & 10/02 & $\Omega$Prime & 0$.\!\!^{\prime\prime}$40 & J, H, KS & 1$.\!\!^{\prime\prime}$4 \\
LS95  & La Silla 2.2m   & 03/95 & IRAC-2B       & 0$.\!\!^{\prime\prime}$50 & J, H, KP & 1$.\!\!^{\prime\prime}$1 -- 1$.\!\!^{\prime\prime}$4 \\ \hline
\enddata
\tablenotetext{a}{$\lambda_0$\,/\,$\Delta\lambda$\ in $\mu$m:
                  J: 1.25\,/\,0.30; H: 1.65\,/\,0.30; KS: 2.16\,/\,0.32; KP: 2.10\,/\,0.34}
\end{deluxetable}
%%%%% End Table 4

%%%%% Table 5: (Sub)mm results
\begin{deluxetable}{lcccccccc}
\tabletypesize{\scriptsize}
\tablecolumns{9} 
\tablewidth{0pc}  
\tablecaption{Results of the millimetre and submillimetre continuum mapping\label{tbl-resmm}}
\tablehead{
 \colhead{Name}                                & 
 \colhead{Peak position\tablenotemark{a}}      & 
 \multicolumn{2}{c}{------ 450\,$\mu$m ------} & 
 \multicolumn{2}{c}{------ 850\,$\mu$m ------} & 
 \multicolumn{2}{c}{------ 1.3\,mm ------}     &
 \colhead{Fig.}                                \\
                                                    & 
 \colhead{(B1950)}                                  &
 \colhead{$I_{\nu}^{\rm peak}$\ ($\Omega_{\rm b}$)} & 
 \colhead{$S_{\nu}^{\rm tot}$\tablenotemark{b}}     & 
 \colhead{$I_{\nu}^{\rm peak}$\ ($\Omega_{\rm b}$)} & 
 \colhead{$S_{\nu}^{\rm tot}$\tablenotemark{b}}     & 
 \colhead{$I_{\nu}^{\rm peak}$\ ($\Omega_{\rm b}$)} & 
 \colhead{$S_{\nu}^{\rm tot}$\tablenotemark{b}}     & 
                                                   \\
 &
 \colhead{[\,h\,:\,m\,:\,s, $^{\circ}$\,:\,\arcmin\,:\,\arcsec\,]} & 
 \colhead{[Jy/beam]} & \colhead{[Jy]} & \colhead{[Jy/beam]} & \colhead{[Jy]} & \colhead{[Jy/beam]} & \colhead{[Jy]} & }
\startdata 
CB\,4    & \nodata	                & \nodata    & \nodata & \nodata 	& \nodata & $<$0.013\tablenotemark{c}\ (10.5)& \nodata & \nodata \\
CB\,6N   & \nodata		        & \nodata    & \nodata & \nodata 	& \nodata & $<$0.014\tablenotemark{c}\ (10.5)& \nodata & \nodata \\
CB\,17SW & \nodata		        & \nodata    & \nodata & \nodata 	& \nodata & $<$0.022\tablenotemark{c}\ (10.5)& \nodata & \nodata \\
CB\,39   & \nodata		        & \nodata    & \nodata & \nodata 	& \nodata & $<$0.027\tablenotemark{c}\ (10.5)& \nodata & \nodata \\
CB\,52   & \nodata		        & \nodata    & \nodata & \nodata 	& \nodata & $<$0.023\tablenotemark{c}\ (10.5)& \nodata & \nodata \\
BHR\,41  & \nodata		        & \nodata    & \nodata & \nodata 	& \nodata & $<$0.060\tablenotemark{c}\ (24)  & \nodata & \nodata \\[1mm]
CB\,6    & 00:46:33.8, $+$50:28:25 	& 1.0~~(10)  & 8.0     & 0.18~~(14.9)  	& 0.8	  & 0.058 (10.5)   	             & 0.18    & \ref{fig-cb6} \\[1mm]
CB\,17   & \nodata                      & \nodata                      & \nodata & \nodata    & 1.3 & \nodata              & 0.55    & \ref{fig-cb17} \\[1mm]
~~SMM1   & 04:00:33.6, $+$56:47:52      & $<$1.2\tablenotemark{c} (8.0)& \nodata & 0.17~~(15) & \nodata & 0.039 (10.5)         & \nodata & \ref{fig-cb17} \\
~~SMM2   & 04:00:31.5, $+$56:47:58      & $<$1.2\tablenotemark{c} (8.0)& \nodata & 0.13~~(15) & \nodata & 0.029 (10.5)         & \nodata & \ref{fig-cb17} \\[1mm]
CB\,26   & 04:55:54.6, $+$52:00:15 	& 3.2~~(8.6) & 6.7     & 0.48~~(14.9)  	& 0.6     & 0.16 (10.5) 	             & 0.24    & \ref{fig-cb26} \\[1mm]
CB\,34   & \nodata                	& \nodata    & 40      & \nodata     	& 5.5     &  \nodata   		             & 2.0     & \ref{fig-cb34} \\
~~SMM1   & 05:44:06.4, $+$20:59:38 	& 1.2~~(8.6) & 10      & 0.44~~(14.9)   & 1.5     & 0.090 (10.5) 	             & 0.70    & \ref{fig-cb34} \\
~~SMM2   & 05:44:02.5, $+$20:59:05 	& 1.0~~(8.6) & 5       & 0.29~~(14.9)   & 0.7 	  & 0.071 (10.5)  	             & 0.40    & \ref{fig-cb34} \\
~~SMM3   & 05:43:59.9, $+$20:59:31 	& 0.5~~(8.6) & 2       & 0.13~~(14.9)   & 0.2 	  & 0.045 (10.5) 	             & 0.30    & \ref{fig-cb34} \\[1mm]
CB\,54   & 07:02:06.0, $-$16:18:51 	& 9.3~~(8.6) & 74.0    & 1.5~~(14.9) 	& 7.3     & 0.34 (10.5)                      & 2.05    & \ref{fig-cb54} \\[1mm]
CB\,58   & 07:16:09.4, $-$23:35:52 	& \nodata    & \nodata &$<$0.27\tablenotemark{c}\ (14.9) & \nodata  & 0.12 (10.5)    & 0.48    & \ref{fig-cb58} \\[1mm]
BHR\,12  & \nodata                 	& \nodata    & 27.5    & \nodata     	& 6.9     & 0.67 ~~(24)   	             & 1.54    & \ref{fig-bhr12} \\
~~SMM1   & ~08:07:40.3, $-$35:56:05\tablenotemark{d}& 3.0~~(8.6)  & 5.9 & 1.2~~(14.9) & 1.7 & \nodata\tablenotemark{e} & 0.49\tablenotemark{e} & \ref{fig-bhr12}\\
~~SMM2   & ~08:07:40.0, $-$35:56:26\tablenotemark{d}& 3.7~~(8.6)  & 6.2 & 1.0~~(14.9) & 1.2 & \nodata\tablenotemark{e} & 0.36\tablenotemark{e} & \ref{fig-bhr12}\\[1mm]
BHR\,36  & 08:24:15.8, $-$50:50:44 	& \nodata    & \nodata 	& 1.2~~(16) & 5.0     & 0.28 ~~(24)  & 0.85  & \ref{fig-bhr36} \\[1mm]
BHR\,55  & 09:44:57.2, $-$50:52:11 	& \nodata    & \nodata 	& \nodata   & \nodata & 0.19 ~~(24)  & 0.42  & \ref{fig-bhr55} \\[1mm]
BHR\,58  & 10:47:01.0, $-$62:06:50 	& \nodata    & \nodata 	& \nodata   & \nodata & 0.13 ~~(24)  & 0.12  & \nodata \\[1mm]
BHR\,71  & 11:59:03.0, $-$64:51:58 	& \nodata    & \nodata	& \nodata   & \nodata & 1.83 ~~(23)  & 3.80  & \ref{fig-bhr71} \\[1mm]
BHR\,86  & 13:03:41.6, $-$76:44:23 	& \nodata    & \nodata 	& \nodata   & \nodata & 0.26 ~~(24)  & 1.33  & \ref{fig-bhr86} \\[1mm]
BHR\,87  & 13:22:24.5, $-$59:27:57 	& \nodata    & \nodata 	& \nodata   & \nodata & 0.11 ~~(24)  & 1.00  & \nodata \\[1mm]
CB\,68   & 16:54:27.2, $-$16:04:44 	& 2.5~~(8.6) & 21.0     & 0.53~~(14.6) & 4.0     & 0.15 (10.5) 	& 1.55    & \ref{fig-cb68} \\[1mm]
BHR\,137 & 17:18:08.5, $-$44:06:17 	& \nodata    & \nodata  & \nodata      & \nodata & 0.38 ~~(24)  & 1.25    & \ref{fig-bhr137} \\[1mm]
CB\,130  & \nodata                 	& \nodata    & 12       & \nodata      & 2.2     & \nodata    	& 1.20    & \ref{fig-cb130} \\
~~SMM1   & 18:13:39.6, $-$02:33:44 	& 1.1~~(8.6) & 3.0 	& 0.27~~(14.9) & 0.8 	 & 0.051 (10.5) & 0.36	  & \ref{fig-cb130} \\
~~SMM2   & 18:13:37.6, $-$02:33:46 	& 0.8~~(8.6) & 1.5 	& 0.14~~(14.9) & 0.3     & 0.033 (10.5) & 0.12	  & \ref{fig-cb130} \\[1mm]
CB\,188  & \nodata                      & \nodata    & 8.4	& \nodata      & 1.6     & \nodata	& 0.30    & \ref{fig-cb188} \\
~~SMM1   & 19:17:54.1, $+$11:30:02	& 1.0~~(9.3) & 2.0	& 0.21~~(15.8) & 0.3     & 0.062 (10.5)	& 0.10    & \ref{fig-cb188} \\[1mm]
CB\,199  & 19:34:35.3, $+$07:27:24\tablenotemark{d}& 4.2~~(8.6) & 15.1 & 1.1~~(14.9)  & 4.0     & \nodata   	& \nodata & \ref{fig-b335} \\[1mm]
CB\,205  & \nodata                	& \nodata    & 21.4     & \nodata      & 3.5     & \nodata    	& 0.85    & \ref{fig-cb205} \\
~~SMM1   & 19:43:22.1, $+$27:43:44	& 2.4~~(9.3) & \nodata 	& 0.48~~(15.8) & \nodata & 0.078 (10.5) & \nodata & \ref{fig-cb205} \\
~~SMM2   & 19:43:19.2, $+$27:43:19 	& 1.3~~(9.3) & \nodata 	& 0.30~~(15.8) & \nodata & 0.044 (10.5) & \nodata & \ref{fig-cb205} \\[1mm]
CB\,224  & \nodata               	& \nodata    & 6.0 	& \nodata      & 0.8	 & \nodata    	& 0.30    & \ref{fig-cb224} \\
~~SMM1   & 20:35:30.6, $+$63:42:47 	& 1.2~~(9.3) & 5.7 	& 0.27~~(15.8) & 0.7 	 & 0.067 (10.5) & 0.24    & \ref{fig-cb224} \\
~~SMM2   & 20:35:34.6, $+$63:43:10 	& 0.4~~(9.3) & 0.4 	& 0.14~~(15.8) & 0.1 	 & 0.040 (10.5) & 0.06    & \ref{fig-cb224} \\[1mm]
CB\,230  & 21:16:53.7, $+$68:04:55 	& 3.9~~(8.6) & 16.5     & 0.92~~(14.9) & 2.9     & 0.23 (10.5) 	& 0.88    & \ref{fig-cb230} \\[1mm]
CB\,232  &  \nodata	                & \nodata    & 27       & \nodata      & 2.4     & \nodata    	& 0.45    & \ref{fig-cb232} \\
~~SMM    & 21:35:13.3, $+$43:07:13 	& 1.9~~(12)  & 3.5      & 0.32~~(14.9) & 0.9     & 0.070 (10.5) & 0.20    & \ref{fig-cb232} \\
~~IRS1   & 21:35:14.5, $+$43:07:07 	& 1.6~~(12)  & 3.3      & \nodata      & 0.3     & \nodata    	& 0.08    & \ref{fig-cb232} \\[1mm]
CB\,240  & 22:31:45.9, $+$58:16:18 	&$<$0.5\tablenotemark{c}\ (8.6)	& \nodata & 0.14~~(14.9)  & 2.4 & 0.028 (10.5) & 0.39  & \ref{fig-cb240} \\[1mm]
CB\,243  & \nodata                	& \nodata    & \nodata 	& \nodata      & 2.0   & \nodata    		& 0.68  & \ref{fig-cb243} \\
~~SMM1   & 23:22:52.1, $+$63:20:13 	& 1.6~~(8.6) & 6.0      & 0.37~~(14.9) & 1.2   & 0.083 (11)    	& 0.38  & \ref{fig-cb243} \\
~~SMM2   & 23:23:03.4, $+$63:20:20 	&$<$0.5\tablenotemark{e}\ (8.6)	& \nodata & 0.15~~(14.9)  & 0.5   & 0.031 (11)  & 0.19  & \ref{fig-cb243} \\[1mm]
CB\,244  & \nodata                	& \nodata    & \nodata 	& \nodata      & 5.1	& \nodata    		& 2.0   & \ref{fig-cb244}\\
~~SMM1   & 23:23:48.5, $+$74:01:08 	& 1.8~~(8.6) & 13.0     & 0.43~~(14.9) & 1.6   & 0.11 (10.5) 		& 0.8   & \ref{fig-cb244}\\
~~SMM2   & 23:23:29.5, $+$74:01:54 	&$<$1.3\tablenotemark{c}\ (8.6)	& \nodata  & 0.32~~(14.9)  & 3.5 & 0.06 (10.5)  & 1.2   & \ref{fig-cb244}\\[1mm]
CB\,246  & \nodata			& \nodata    & 30       & \nodata      & 3.1   & \nodata      	& 2.56  & \ref{fig-cb246} \\ 
~~SMM1   & 23:54:11.0, $+$58:17:15	& 1.3~~(15)  & 22	& 0.12~~(16)   & 1.8	& 0.039 (15)		& 0.59  & \ref{fig-cb246} \\ 
~~SMM2   & 23:54:03.7, $+$58:18:36	& 1.6~~(15)  &  8	& 0.11~~(16)   & 1.3	& 0.041 (15)		& 0.58  & \ref{fig-cb246} \\ 
\enddata
\tablenotetext{a}{Derived from gaussian fit to the component in the 1.3\,mm map (if not stated otherwise).}
\tablenotetext{b}{Flux density measured in a polygon enclosing the 2\,$\sigma$ contour.}
\tablenotetext{c}{3\,$\sigma$\ detection limit; source not detected.}
\tablenotetext{d}{Source position derived from the 850\,$\mu$m SCUBA map.}
\tablenotetext{e}{The two sub-cores are not resolved in the 1.3\,mm map.}
\end{deluxetable}
%%%%% End Table 5

%%%%% Table 6: Morphology and multiplicity
\begin{deluxetable}{lcccclll}
\tabletypesize{\scriptsize}
\tablecolumns{8} 
\tablewidth{0pc}  
\tablecaption{Globule morphology and multiplicity\label{tbl-resmult}}
\tablehead{
 \colhead{Name}                                  & 
 \multicolumn{2}{c}{-- Size\tablenotemark{a} --} &
 \colhead{Aspect}                                &
 \colhead{$A_V$\tablenotemark{b}}                &
 \colhead{Globule}                               &
 \colhead{Multiplicity of embedded sources}      &
 \colhead{Projected separations}                 \\ 
 & \colhead{[\arcmin]} & \colhead{[pc]} & \colhead{Ratio} & \colhead{[mag]} & 
 \colhead{Morphology\tablenotemark{c}} & & \colhead{[AU]}}
\startdata
%CB\,4    & 1.6 & 0.28 & 1.2 & 0.5  & Roundish with small diffuse tail & Not detected  & -- \\
CB\,6    & 5.5 & 0.96 & 2.6  & 0.3  & I, T, N      & Unresolved                  & \nodata \\
CB\,17   & 2.2 & 0.16 & 1.0  & 0.4  & I, T         & Two mm cores and IRS        & 2,000 -- 5,000 \\
CB\,26   & 4.8 & 0.20 & 2.0  & 1.5  & CG, C, BR    & Unresolved\tablenotemark{d} & \nodata \\
CB\,34   & 3.1 & 1.35 & 2.0  & 1.0  & I, N         & Three mm cores and NIR star clusters    & 6,000 -- 80,000 \\
CB\,54   & 4.4 & 1.40 & 1.6  & 0.5  & I, N         & Unresolved mm core and NIR star cluster & 10,000 \\
CB\,58   & 5.1 & 2.22 & 2.3  & 1.0  & CG, BR       & Two mm cores and NIR star cluster       & 7,500 - 50,000 \\
BHR\,12  & 2.4 & 0.28 & 1.5  & 1.2  & CG, BR, N    & Two mm cores                & 8,000 \\
BHR\,36  & 3.5 & 0.41 & 1.3  & 1.5  & CG, BR, N    & Unresolved                  & \nodata \\
BHR\,55  & 3.5 & 0.30 & 1.3  & 1.2  & I, C         & Unresolved                  & \nodata \\
BHR\,71  & 4.9 & 0.25 & 2.7  & 3.4  & I            & Unresolved mm core, two IRS & 3,400 \\
BHR\,86  & 7.7 & 0.43 & 1.7  & 1.3  & CG, BR       & Unresolved                  & \nodata \\
CB\,68   & 3.9 & 0.18 & 1.3  & 1.2  & I, T         & Unresolved                  & \nodata \\
BHR\,137 & 3.9 & 0.79 & 1.7  & 2.0  & I, N, T      & Unresolved mm core and IRS  & 2,000 -- 15,000 \\
CB\,130  & 3.5 & 0.20 & 2.5  & 3.7  & I            & Two mm cores and IRS        & 3,000 -- 6,000 \\
CB\,188  & 1.6 & 0.14 & 2.0  & 3.4  & I            & Unresolved                  & \nodata \\
CB\,199  & 5.0 & 0.14 & 1.2  & 0.6  & I            & Unresolved                  & \nodata \\
CB\,205  & 8.2 & 4.76 & 1.5  & 1.5  & I, N         & Two mm cores and NIR star cluster & 4,000 -- 60,000 \\
CB\,224  & 3.4 & 0.39 & 1.0  & 1.3  & I            & Two mm cores                & 14,000 \\
CB\,230  & 2.7 & 0.31 & 1.5  & 1.5  & I, BR, N     & Unresolved mm core, two IRS & 4,000 \\
CB\,232  & 3.5 & 0.61 & 2.5  & 1.7  & CG           & MM core and IRS             & 7,000 \\
CB\,240  & 2.7 & 0.39 & 1.5  & 2.4  & I            & MM core and NIR star cluster & 1,200 -- 10,000 \\
CB\,243  & 4.7 & 0.96 & 4.6  & 2.0  & F, T         & Two mm cores and IRS        & 7,000 -- 50,000 \\
CB\,244  & 7.9 & 0.46 & 2.0  & 2.0  & I            & Two mm cores                & 18,000 \\
CB\,246  & 5.9 & 0.24 & 1.7  & 2.7  & I, T         & Two mm cores                & 13,000 \\
\enddata
\tablenotetext{a}{Characteristic or mean angular size ($\sqrt{a\,b}$), derived from the optical globule 
   sizes listed by
   \cite{1988ApJS...68..257C} for CB (northern) globules and in 
   \cite{1995MNRAS.276.1067B} for BHR (southern) globules.  
   Mean linear size derived by using the distances listed in Table\,\ref{tbl-sourcelist}.
\tablenotetext{b}{Peak visual extinction, measured with a resolution of 6\,arcmin \citep{2005PASJ...57S...1D}.}
   At the globule centres, the extinction is likely to be higher than this.} 
\tablenotetext{c}{Globule morphology, based on visual inspection of DSS2-red optical images:
   (I) isolated; (CG) cometary globule; (F) filamentary; (C) complex, multiple cores; 
   (BR) bright rim; (N) nebula associated; (T) tail.}
\tablenotetext{d}{CB\,26: contains possible a 20\,AU binary with a circumbinary disk \citep{lau2009}.}
\end{deluxetable}
%%%%% End Table 6

%%%%% Table 7: Source properties
\begin{deluxetable}{lcccccccccll}
\tabletypesize{\scriptsize}
\rotate
\tablecolumns{11} 
\tablewidth{0pc}  
\tablecaption{Physical parameters of selected sources\label{tbl-physprop}}
\tablehead{
 \colhead{Name}                                & 
 \colhead{$L_{\rm bol}$}                       &
 \colhead{$\frac{L_{\rm smm}}{L_{\rm bol}}$}   & 
 \colhead{$T_{\rm bol}$}                       & 
 \colhead{$\alpha_{\rm IR}$$\tablenotemark{a}$}& 
 \colhead{$T_{\rm d}$\tablenotemark{b}}        & 
 \colhead{$M_{\rm H}$\tablenotemark{c}}        &  
 \colhead{$n_{\rm H}$\tablenotemark{d}}        & 
 \colhead{$N_{\rm H}$\tablenotemark{e}}        & 
 \colhead{Size\tablenotemark{f}}               & 
 \colhead{Evol.}                               &
 \colhead{Remark}                              \\
                                               &
 \colhead{[L$_{\odot}$]}                       &
 \colhead{[\%]}                                &
 \colhead{[K]}                                 &
 \colhead{}                                    &
 \colhead{[K]}                                 &
 \colhead{[M$_{\odot}$]}                       &
 \colhead{[cm$^{-3}$]}                         &
 \colhead{[cm$^{-2}$]}                         &
 \colhead{[AU]}                                &
 \colhead{stage}                               &
                                               }
\startdata
% name            & Lbol & Lsubmm/Lbol & Tbol  & alpha_IR & mass & size    & remarks               & class
CB\,6            & 6.4     &  2.9           &  178 & $+$0.6  & 20 & 1.0\,/\,1.8 & $>$3E7\,/\,3eE4 & 1.1E23 & $<$3,000\,/\,36,000 & Class\,I & NIR nebula, MIR source, outflow \\[1mm]
CB\,17\,-\,SMM1+2& 0.07    & $>$30          & 15   & \nodata & 10 & 4.0 & 6E6 & 2.0E23 & 8,000 & pre-stellar & No IR counterpart, 8\,$\mu$m shadow \\
CB\,17\,-\,IRS   & 0.6     & $<$2.6         & $>$55& $+$0.8  & 20 & \nodata & \nodata & \nodata    & \nodata   & Class\,I & NIR nebula, MIR source, low-velocity outflow? \\[1mm]
CB\,26           & 0.25    &  2.5           &  137 & $-$0.1  & 20 & 0.1\,/\,0.1 & $>$2e8\,/\,1E5 & 2.9E23 & $<$700\,/\,7,500  & Class\,I & NIR nebula, edge-on disk, outflow \\[1mm]
BHR\,12\,-\,SMM1 & 11      &  1.6           & 126  & $+$1.1  & 20 & 0.7    & 6E7 & 2.6E23 &  2,000  & Class\,I & NIR source and jet, outflow \\
BHR\,12\,-\,SMM2 & 1.9     &  5.3           &  50  & $+$0.1  & 15 & 0.8    & 7E7 & 3.4E23 &  2,000  & Class\,0 & No NIR counterpart, MIR source, outflow \\[1mm]
BHR\,36          & 16      &  2.4           & 149  & $+$0.9  & 20 & 5.9    & 4E6 & 9.8E22 & 10,000  & Class\,I & NIR jet, MIR source, outflow \\[1mm]
BHR\,55          & 2.2     &  4.4           &  39  & $+$1.1  & 15 & 2.4    & 6E7 & 9.8E22 & $<$3,000 & Class\,0 & No NIR counterpart, MIR source, no outflow data \\[1mm]
BHR\,71\tablenotemark{g}& 10 &  3.4         &  53  & $+$1.7  & 15 & 9.7    & 1E8 & 1.0E24 & 3,600   & Class\,0 & No NIR counterpart, 2 MIR sources, 2 outflows \\[1mm]
BHR\,86          & 1.5     &  6.2           &  61  & $+$1.4  & 15 & 3.4    & 2E6 & 1.3E23 & 10,000  & Class\,0 & No NIR counterpart, MIR source, outflow \\[1mm]
CB\,68           & 1.3     &  3.9           &  50  & $+$2.4  & 15 & 0.6\,/\,1.9 & 5E7\,/\,1E6 & 4.1E23 & 2,000\,/\,10,000  & Class\,0 & No NIR counterpart, MIR source, outflow \\[1mm]
BHR\,137         & \nodata &  \nodata       &  \nodata & \nodata & 20 & 26     & 5E6 & 1.3E23 & 16,000  & \nodata  & No NIR or IRAS counterpart, no {\it Spitzer} data  \\[1mm]
CB\,130\,-\,SMM1 & 0.15    &  11            &  55  & $+$1.0  & 15 & 0.9\,/\,1.8 & 5E6\,/\,2E5 & 1.4E23 &  5,000\,/\,20,000  & Class\,0 & NIR nebula, MIR source, no outflow data \\
CB\,130\,-\,SMM2 & \nodata &  \nodata       &  \nodata & \nodata & 10\,/\,15 & 0.6\,/\,1.8 & 3E7\,/\,2E5 & 1.7E23 &  2,400\,/\,20,000  & prestellar & No IR counterpart, no outflow data \\
CB\,130\,-\,IRS  & $\ge$0.04     &  \nodata       &  \nodata & $-$0.7  & 20 & \nodata  & \nodata & \nodata & \nodata  & Class\,I & Red star with nebulosity, outside mm core \\[1mm]
CB\,188          & 1.5           &  1.3           & 307  & $+$0.2  & 20 & 0.3\,/\,1.0 & $>$4E7\,/\,8E4 & 1.1E23 & $<$1500\,/\,20,000 & Class\,I & Red star, outflow          \\[1mm]
CB\,199          & 0.5           &  4.8           & 37   & $+$1.6  & 15 & 0.6     & 1E8 & 3.7E23 &  1500   & Class\,0 & No NIR counterpart, MIR source, outflow \\[1mm]
CB\,224\,-\,SMM1 & 6.6           &  2.2           & 56   & $+$1.8  & 15 & 2.4     & 4E7 & 1.8E23 &  3600   & Class\,0 & Very faint NIR nebula, no {\it Spitzer} data, outflow \\
CB\,224\,-\,SMM2 & 2.4           &  0.8           & 337  & $+$0.2  & 20 & 0.4     & $>$4E7 & 7.7E22 & $<$2000 & Class\,I & Red star, disk, no outflow detected                 \\[1mm]
CB\,230\,-\,IRS1\tablenotemark{g}& 9  &  3.5      & 214  & $-$0.3  & 20 & 6.1     & 7E7 & 4.2E23 &  4000   & Class\,I & NIR nebula, MIR source, outflow \\
CB\,230\,-\,IRS2\tablenotemark{g}& $\ge$0.1& \nodata & \nodata & $-$0.7 & 20 & \nodata & \nodata & \nodata &  \nodata & Class\,I & NIR nebula, MIR source, outflow \\[1mm]
CB\,232\,-\,SMM  & \nodata       &  \nodata           & \nodata  & \nodata  & 10 & 9       & 2E7 & 3.5E23 &  7000   & prestellar & No IR counterpart, close proximity to IRS \\
CB\,232\,-\,IRS1 & 13            & 1.0            & 166  & $+$1.7  & 20 & 1.0     & \nodata & \nodata    &  \nodata    & Class\,I   & Red star, outflow, close proximity to SMM1\\[1mm]
CB\,243\,-\,SMM1 & $<$7.2\tablenotemark{h} & 4.1\tablenotemark{h}& 123\tablenotemark{h} & \nodata  & 10 & 22 & 2E7 & 4.0E23 & 10,000  & prestellar & No IR counterpart, close proximity to IRS \\
CB\,243\,-\,IRS  & $<$7.2\tablenotemark{h} & 4.1\tablenotemark{h}& 123\tablenotemark{h} & $+$0.8  & 20 & \nodata   & \nodata & \nodata    &  \nodata    & Class\,I   & Red star, outflow, close proximity to SMM1 \\
CB\,243\,-\,SMM2 & \nodata       &  \nodata           & \nodata  & \nodata  & 10 & 11     & 2E6 & 1.4E23 & 15,000  & prestellar & No IR source, 8\,$\mu$m shadow \\[1mm]
CB\,244\,-\,SMM1 & 1.3           &  4.0               & 69       & $+$2.4   & 15 & 3.1    & 7E7 & 3.0E23 &  2,700  & Class\,0   & NIR nebula, MIR source, outflow     \\ 
CB\,244\,-\,SMM2 & \nodata       &  \nodata           &  \nodata & \nodata  & 10 & 5.7    & 8E6 & 3.0E23 & 8,000   & prestellar & No IR counterpart, 8\,$\mu$m shadow  \\[1mm]
CB\,246\,-\,SMM1 & \nodata       &  \nodata           &  \nodata & \nodata  & 10 & 1.4    & 8E6 & 1.0E23 & 5,000   & prestellar & No IR counterpart, 8\,$\mu$m shadow  \\  
CB\,246\,-\,SMM2 & \nodata       &  \nodata           &  \nodata & \nodata  & 10 & 1.4    & 2E6 & 1.0E23 & 7,500   & prestellar & No IR counterpart, 8\,$\mu$m shadow  \\ 
\enddata
\tablenotetext{a}{NIR spectral index $\alpha_{\rm IR} = - \frac{\rm d\,log(\nu\,F_{\nu})}{\rm d\,log(\nu)}$\ 
                   \citep{1987ApJ...312..788A} and 
                   $\lambda_{\rm 1} = 2.2\,\mu$m, $\lambda_{\rm 2} = 24\,\mu$m, 
                   except for BHR\,12-SMM2, BHR\,55, BHR\,71, BHR\,86, CB\,68, CB\,199 ($\lambda_{\rm 1} = 3.6\,\mu$m), 
                   CB\,230-IRS2 ($\lambda_{\rm 2} = 5.8\,\mu$m), 
                   and CB\,224-SMM1 ($\lambda_{\rm 2} = 60\,\mu$m).}
\tablenotetext{b}{Assumed mass-averaged dust temperature $T_{\rm d}$\ (see Sect.\,\ref{sec-res-physprop}).}
\tablenotetext{c}{$M_{\rm H}$\ computed from integrated 1.3\,mm flux (Table\,\ref{tbl-resmm}), 
                  assuming $\kappa_{\rm dust}(1.3{\rm mm}) = 0.5$\,g\,cm$^{-3}$, H-to-dust mass ratio 110, and 
                  mass-averaged dust temperature from col. 6 (see Sect.\,\ref{sec-res-physprop}). 
                  If two values are given, the first one refers to the core, the second one to the envelope only.
                  For CB\,199, we used the 850\,$\mu$m flux with $\kappa_{\rm dust}(850\mu{\rm m}) = 1.1$\,g\,cm$^{-3}$.}
\tablenotetext{d}{Source-averaged number density, $n_{\rm H}$, computed from $M_{\rm H}$\ (col.\,7) and FWHM size (col.\,10) 
                  (see Sect.\,\ref{sec-res-physprop}). If two values are given, the first one referes to the core, 
                  the second one to the envelope only.}
\tablenotetext{e}{Beam-averaged peak column density, $N_{\rm H}$, computed from 1.3\,mm peak flux densities and beam sizes 
                  listed in Table\,\ref{tbl-resmm} (see Sect.\,\ref{sec-res-physprop}).}
\tablenotetext{f}{FWHM size (mean diameter of the 50\% contour), deconvolved with the beam and projected at the 
                  distance listed in Table\,\ref{tbl-obsmm}.
                  If two values are given, the first one referes to the core only, the second one to the envelope.}
\tablenotetext{g}{The dust cores in BHR\,71 and CB\,230 both have two embedded IR sources each, which are not resolved in the (sub)mm maps.}
\tablenotetext{h}{Values refer to the combined SED and are not representative of the individual sources.}
\end{deluxetable}
%%%%% End Table 7

%%%%%% FIGURES %%%%%%%%%%%%%%%%%%%%%%%%%%%%%%%%%%%%%%%%%%%%%%%%%%%%%%%%%%%%%%%%%%%%%%%%%%%%%%%%%%%%%%%%

\clearpage
\begin{figure*}[htbp]
\includegraphics[width=1.0\textwidth]{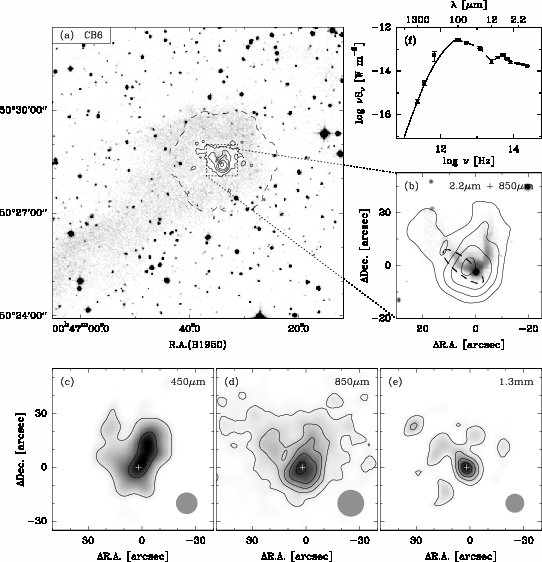}
\caption{\label{fig-cb6} \footnotesize
   CB\,6:
   a) Optical image (DSS2 red) with contours of the 850\,$\mu$m dust continuum
      emission overlaid.
   b) NIR K-band image with 850\,$\mu$m dust continuum contours.
      The IRAS PSC position is marked by a dashed ellipse.
   c) 450\,$\mu$m dust continuum emission; grey scale and contours at
      1.1 and 1.5\,Jy/beam.
   d) 850\,$\mu$m dust continuum emission; grey scale and contours at
      35 to 175 by 35\,mJy/beam.
   e) 1.3\,mm dust continuum emission; grey scale and contours at
      12 to 48 by 12\,mJy/beam. Beam sizes are indicated as
      grey ellipses. The cross marks the position of the 1.3\,mm peak.
   f) Spectral energy distribution showing the IRAM 1.3\,mm, SCUBA 850 and 450\,$\mu$m, 
      IRAS PSC, Spitzer MIPS and IRAC, and ground-based NIR data. 
   }
\end{figure*}

%\clearpage
\begin{figure*}[htbp]
\includegraphics[width=1.0\textwidth]{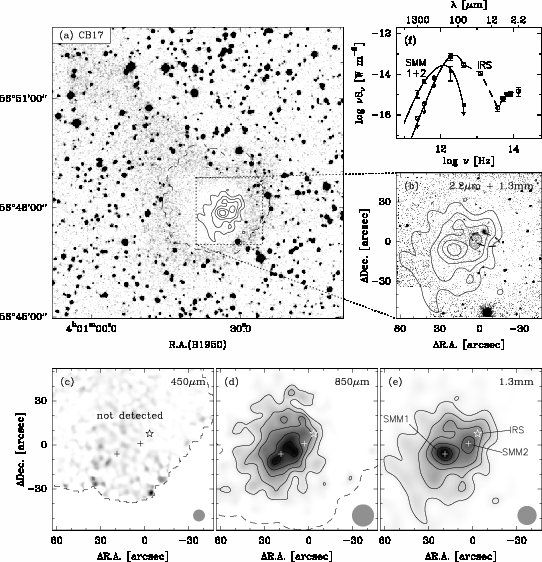}
\caption{\label{fig-cb17} \footnotesize
   CB\,17:
   a) Optical image (DSS2 red) with contours of the 1.3\,mm dust continuum
      emission overlaid.
   b) NIR K-band image with 1.3\,mm dust continuum contours.
      The IRAS PSC position is marked by a dashed ellipse.
   c) 450\,$\mu$m dust continuum map; the source is not detected at rms 
      400\,mJy/beam.
   d) 850\,$\mu$m dust continuum emission; grey scale and contours at
      50 to 170 by 30\,mJy/beam.
   e) 1.3\,mm dust continuum emission; grey scale and contours at
      12 to 44 by 8\,mJy/beam. Beam sizes are indicated as
      grey ellipses. Crosses mark the positions of SMM1 and SMM2 in the 1.3\,mm map.
      The asterisk marks the position of the Spitzer source IRS.
   f) Spectral energy distributions of SMM(1+2) and IRS, showing the IRAM 1.3\,mm, 
     SCUBA 850\,$\mu$m, IRAS PSC, Spitzer MIPS and IRAC, and ground-based NIR data. 
      Note that the 160\,$\mu$m data points result only from a coarse flux splitting 
      (85\% IRS and 15\% SMM) since the two sources are not resolved in the MIPS3 map.
   }
\end{figure*}

%\clearpage
\begin{figure*}[htbp]
\includegraphics[width=1.0\textwidth]{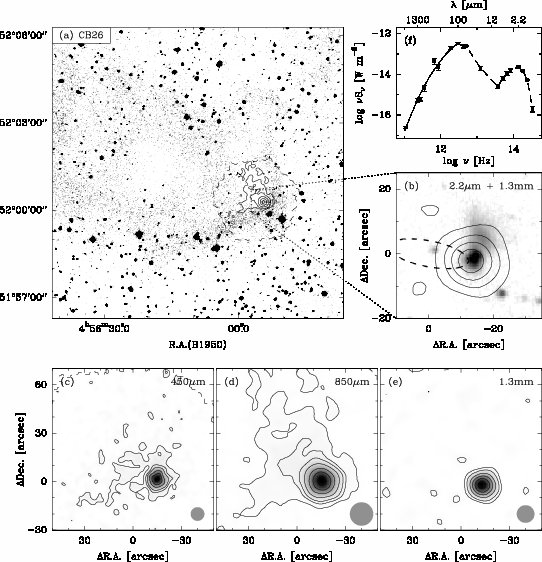}
\caption{\label{fig-cb26} \footnotesize
   CB\,26:
   a) Optical image (DSS2 red) with contours of the 850\,$\mu$m dust continuum
      emission overlaid.
   b) NIR K-band image with 1.3\,mm dust continuum contours.
      The IRAS PSC position is marked by a dashed ellipse.
   c) 450\,$\mu$m dust continuum emission; grey scale and contours at
      350, 700, 1050 to 3150 by 700\,mJy/beam.
   d) 850\,$\mu$m dust continuum emission; grey scale and contours at
      30, 60, 90, 150 to 400 by 100\,mJy/beam.
   e) 1.3\,mm dust continuum emission; grey scale and contours at
      15, 30 to 150 by 30\,mJy/beam. Beam sizes are indicated as
      grey ellipses.
   f) Spectral energy distributions, showing the OVRO 3\,mm, IRAM 1.3\,mm, 
      SMA 1.1\,mm, SCUBA 850 and 450\,$\mu$m, CSO 350\,$\mu$m, IRAS PSC, Spitzer MIPS and IRAC, 
      and ground-based NIR data. 
   }
\end{figure*}

%\clearpage
\begin{figure*}[htbp]
\includegraphics[width=1.0\textwidth]{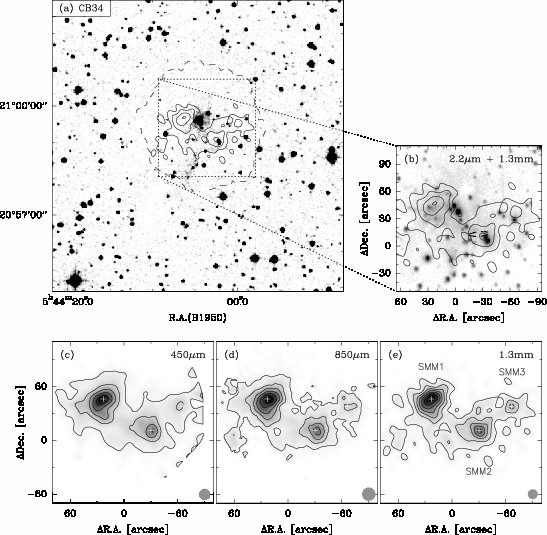}
\caption{\label{fig-cb34} \footnotesize
   CB\,34:
   a) Optical image (DSS2 red) with contours of the 1.3\,mm dust continuum
      emission overlaid.
   b) NIR K-band image with 1.3\,mm dust continuum contours.
      The IRAS PSC position is marked by a dashed ellipse.
   c) 450\,$\mu$m dust continuum emission; grey scale and contours at
      500 to 1700 by 300\,mJy/beam.
   d) 850\,$\mu$m dust continuum emission; grey scale and contours at
      75 to 450 by 75\,mJy/beam.
   e) 1.3\,mm dust continuum emission; grey scale and contours at
      18 to 108 by 15\,mJy/beam. Beam sizes are indicated as
      grey ellipses. Crosses mark the positions of the three 1.3\,mm peaks.
      No spectral energy distribution is shown since many different, partially 
      unresolved sources contribute to the measured flux densities.
   }
\end{figure*}

%\clearpage
\begin{figure*}[htbp]
\includegraphics[width=1.0\textwidth]{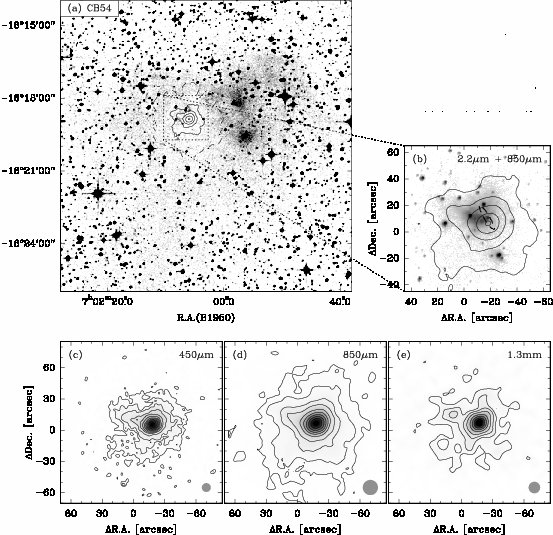}
\caption{\label{fig-cb54} \footnotesize
   CB\,54:
   a) Optical image (DSS2 red) with contours of the 850\,$\mu$m dust continuum
      emission overlaid.
   b) NIR K-band image with 850\,$\mu$m dust continuum contours.
      The IRAS PSC position is marked by a dashed ellipse.
   c) 450\,$\mu$m dust continuum emission; grey scale and contours at
      0.4, 0.8, 1.2 to 9 0.8\,Jy/beam.
   d) 850\,$\mu$m dust continuum emission; grey scale and contours at
      50, 100, 200, 300, to 1500 by 200\,mJy/beam.
   e) 1.3\,mm dust continuum emission; grey scale and contours at
      20, 40, 60, 90, 140, 190, 240, and 290\,mJy/beam. Beam sizes are indicated as
      grey ellipses.
      No spectral energy distribution is shown since many different, partially 
      unresolved sources contribute to the measured flux densities.
   }
\end{figure*}

%\clearpage
\begin{figure*}[htbp]
\includegraphics[width=1.0\textwidth]{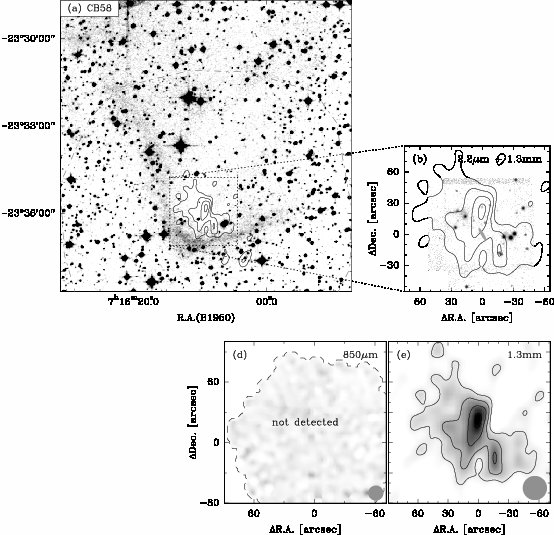}
\caption{\label{fig-cb58} \footnotesize
   CB\,58:
   a) Optical image (DSS2 red) with contours of the 1.3\,mm dust continuum
      emission overlaid.
   b) NIR K-band image with 1.3\,mm dust continuum contours.
      The IRAS PSC position is marked by a grey ellipse.
   d) 850\,$\mu$m dust continuum map; the source is not detected at rms 
      90\,mJy/beam.
   e) 1.3\,mm dust continuum emission; grey scale and contours at
      33 to 132 by 33\,mJy/beam. Beam sizes are indicated as
      grey ellipses.
      No spectral energy distribution is shown since many different, partially 
      unresolved sources contribute to the measured flux densities.
   }
\end{figure*}

%\clearpage
\begin{figure*}[htbp]
\includegraphics[width=1.0\textwidth]{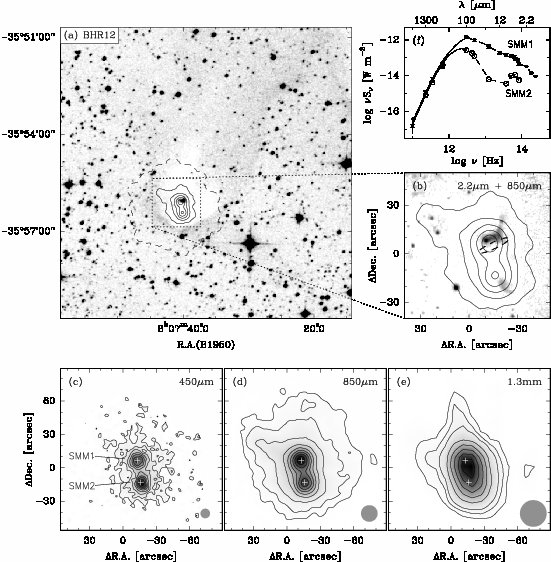}
\caption{\label{fig-bhr12} \footnotesize
   BHR\,12:
   a) Optical image (DSS2 red) with contours of the 850\,$\mu$m dust continuum
      emission overlaid.
   b) NIR K-band image with 850\,$\mu$m dust continuum contours.
      The IRAS PSC position is marked by a dashed ellipse.
   c) 450\,$\mu$m dust continuum emission; grey scale and contours at
      0.3, 0.6, 1 to 3.5 by 0.5\,Jy/beam.
   d) 850\,$\mu$m dust continuum emission; grey scale and contours at
      40, 100, 200 to 1200 by 150\,mJy/beam.
   e) 1.3\,mm dust continuum emission; grey scale and contours at
      25, 60, 100 to 600 by 100\,mJy/beam. Beam sizes are indicated as
      grey ellipses. Crosses mark the positions of the two 850\,$\mu$m peaks.
   f) Spectral energy distributions of SMM1 (filled squares) and SMM2 (empty circles)  
      showing the ATCA 3\,mm, SEST 1.3\,mm, SCUBA 850 and 450\,$\mu$m,  
      IRAS PSC, Spitzer MIPS and IRAC, and ground-based NIR data. 
      The 1.3\,mm fluxes (sources not resolved in the SEST map) are derived by 
      fitting two Gaussian sources to the map.
   }
\end{figure*}

%\clearpage
\begin{figure*}[htbp]
\includegraphics[width=1.0\textwidth]{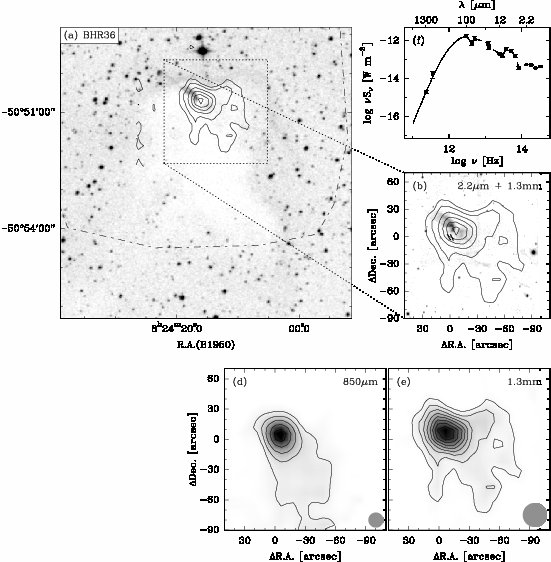}
\caption{\label{fig-bhr36} \footnotesize
   BHR\,36:
   a) Optical image (DSS2 red) with contours of the 1.3\,mmm dust continuum
      emission overlaid.
   b) NIR K-band image with 1.3\,mm dust continuum contours.
      The IRAS PSC position is marked by a dashed ellipse.
   d) 850\,$\mu$m dust continuum emission; grey scale and contours at
      0.15, 0.3, 0.5, 0.75, and 1\,Jy/beam.
   e) 1.3\,mm dust continuum emission; grey scale and contours at
      30 to 280 by 30\,mJy/beam. Beam sizes are indicated as
      grey ellipses.
   f) Spectral energy distribution showing the SEST 1.3\,mm, SCUBA 850\,$\mu$m
      IRAS PSC, Spitzer MIPS and IRAC, and ground-based NIR data. 
   }
\end{figure*}

%\clearpage
\begin{figure*}[htbp]
\includegraphics[width=1.0\textwidth]{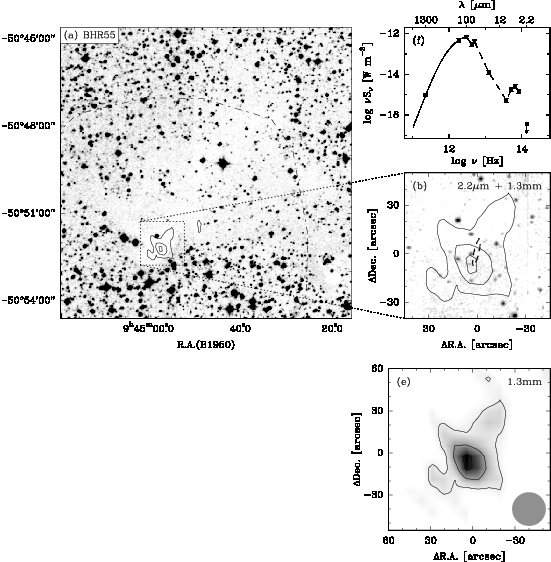}
\caption{\label{fig-bhr55} \footnotesize
   BHR\,55:
   a) Optical image (DSS2 red) with contours of the 1.3\,mmm dust continuum
      emission overlaid.
   b) NIR K-band image with 1.3\,mm dust continuum contours.
      The IRAS PSC position is marked by a dashed ellipse.
   e) 1.3\,mm dust continuum emission; grey scale and contours at
      55, 110, 165\,mJy/beam. The beam size is indicated as a
      grey ellipse.
   f) Spectral energy distribution showing the SEST 1.3\,mm, 
      IRAS PSC, Spitzer MIPS and IRAC, and ground-based NIR data. 
      Note that the association of the IRAC 
      fluxes with the FIR/mm source is somewhat uncertain.
   }
\end{figure*}

%\clearpage
\begin{figure*}[htbp]
\includegraphics[width=1.0\textwidth]{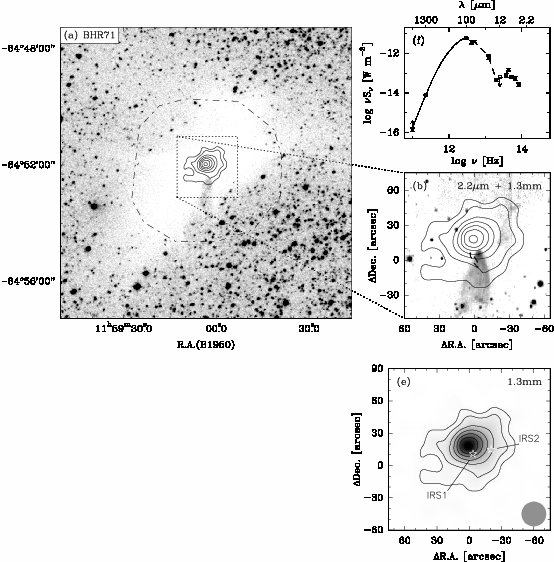}
\caption{\label{fig-bhr71} \footnotesize
   BHR\,71:
   a) Optical image (DSS2 red) with contours of the 1.3\,mmm dust continuum
      emission overlaid \citep[see][]{1997ApJ...476..781B}.
   b) NIR K-band image with 1.3\,mm dust continuum contours.
      The IRAS PSC position is marked by a dashed ellipse.
   e) 1.3\,mm dust continuum emission; grey scale and contours at
      100, 250 to 1750 by 300\,mJy/beam. The beam size is indicated as a
      grey ellipse. Embedded infrared sources IRS1 and IRS2 refer to 
      \cite{2001ApJ...554L..91B}.
   f) Spectral energy distribution showing the ATCA 3\,mm, SEST 1.3\,mm, 
      SMA 1.1\,mm, SCUBA 850 and 450\,$\mu$m, CSO 350\,$\mu$m,  
      IRAS 100 to 12\,$\mu$m, MIPS 70 and 24\,$\mu$m, IRAC 8 to 3.6\,$\mu$m, 
      as well as ISOCAM 15 and 7\,$\mu$m data. Note that this is a composite SED 
      containing the sum of flux contributions from both IRS1 and IRS2.
   }
\end{figure*}

\clearpage
\begin{figure*}[htbp]
\includegraphics[width=1.0\textwidth]{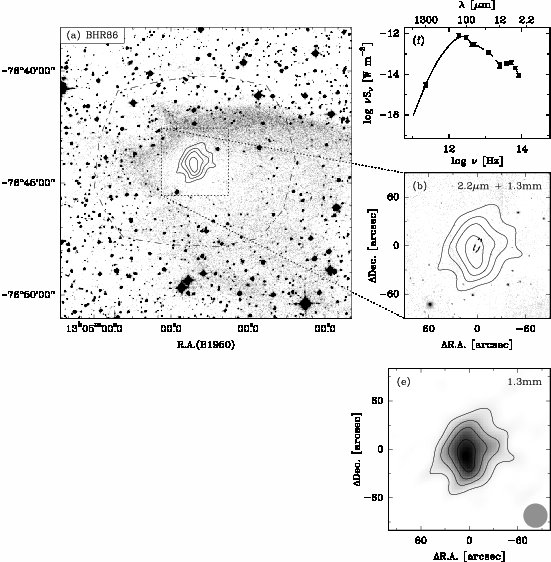}
\caption{\label{fig-bhr86} \footnotesize
   BHR\,86:
   a) Optical image (DSS2 red) with contours of the 1.3\,mmm dust continuum
      emission overlaid.
   b) NIR K-band image with 1.3\,mm dust continuum contours.
      The IRAS PSC position is marked by a dashed ellipse.
   e) 1.3\,mm dust continuum emission; grey scale and contours at
      75 to 315 by 80\,mJy/beam. The beam size is indicated as a
      grey ellipse.
   f) Spectral energy distribution showing the SEST 1.3\,mm, 
      IRAS FSC, and Spitzer MIPS and IRAC data.
   }
\end{figure*}

%\clearpage
\begin{figure*}[htbp]
\includegraphics[width=1.0\textwidth]{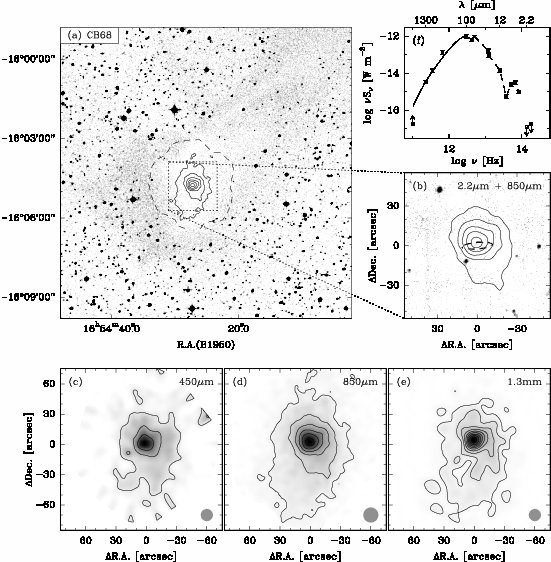}
\caption{\label{fig-cb68} \footnotesize
   CB\,68:
   a) Optical image (DSS2 red) with contours of the 850\,$\mu$m dust continuum
      emission overlaid.
   b) NIR K-band image with 850\,$\mu$m dust continuum contours.
      The IRAS PSC position is marked by a dashed ellipse.
   c) 450\,$\mu$m dust continuum emission; grey scale and contours at
      1.2, 2.2, and 3.2\,Jy/beam.
   d) 850\,$\mu$m dust continuum emission; grey scale and contours at
      90 to 540 by 90\,mJy/beam.
   e) 1.3\,mm dust continuum emission; grey scale and contours at
      18 to 162 by 18\,mJy/beam. Beam sizes are indicated as
      grey ellipses.
   f) Spectral energy distribution showing the OVRO 3\,mm, IRAM 1.3\,mm, 
      SCUBA 850 and 450\,$\mu$m, 
      IRAS PSC and FSC, Spitzer MIPS and IRAC, and ground-based NIR data.
   }
\end{figure*}

%\clearpage
\begin{figure*}[htbp]
\includegraphics[width=1.0\textwidth]{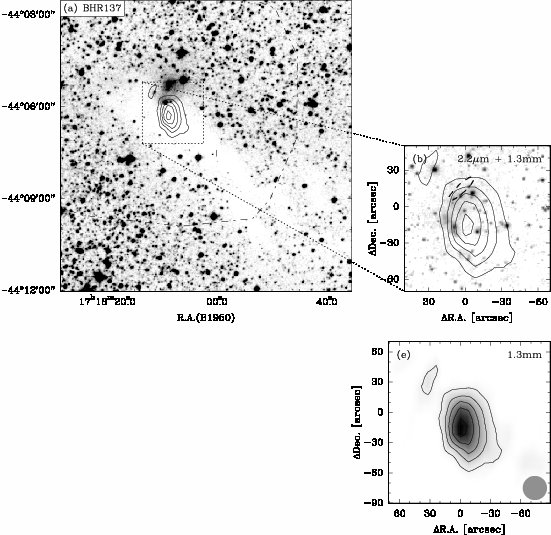}
\caption{\label{fig-bhr137} \footnotesize
   BHR\,137:
   a) Optical image (DSS2 red) with contours of the 1.3\,mmm dust continuum
      emission overlaid.
   b) NIR K-band image with 1.3\,mm dust continuum contours.
      The IRAS PSC position is marked by a dashed ellipse.
   e) 1.3\,mm dust continuum emission; grey scale and contours at
      70 to 350 by 70\,mJy/beam. The beam size is indicated as a
      grey ellipse.
      No spectral energy distribution is shown since it is not clear 
      which of the sources emitting at shorter wavelengths, if any, 
      is associated with the mm source.
   }
\end{figure*}

%\clearpage
\begin{figure*}[htbp]
\includegraphics[width=1.0\textwidth]{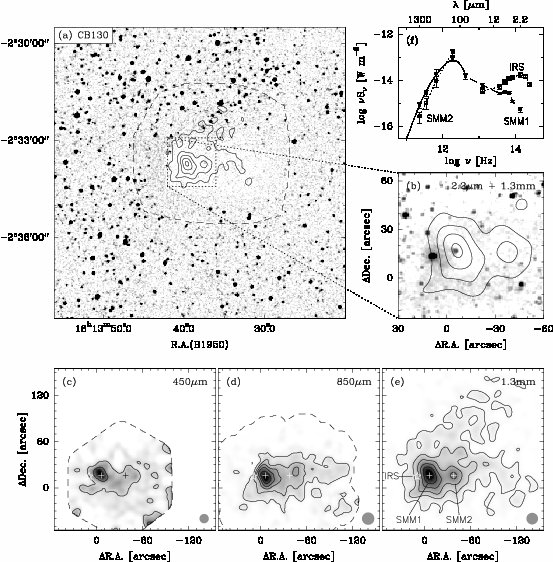}
\caption{\label{fig-cb130} \footnotesize
   CB\,130:
   a) Optical image (DSS2 red) with contours of the 1.3\,mm dust continuum
      emission overlaid.
   b) NIR K-band image with 1.3\,mm continuum contours.
   c) 450\,$\mu$m dust continuum emission; grey scale and contours at
      0.5 and 1\,Jy/beam.
   d) 850\,$\mu$m dust continuum emission; grey scale and contours at
      45 to 270 by 45\,mJy/beam.
   e) 1.3\,mm dust continuum emission; grey scale and contours at
      12 to 72 by 12\,mJy/beam. Beam sizes are indicated as
      grey ellipses.
   f) Spectral energy distribution of SMM1 (filled squares), 
      SMM2 (submm only, empty squares), and IRS (NIR to MIR only, open circles), 
      showing the IRAM 1.3\,mm, SCUBA 850 and 450\,$\mu$m, 
      Spitzer MIPS and IRAC, and ground-based NIR data. 
   }
\end{figure*}

%\clearpage
\begin{figure*}[htbp]
\includegraphics[width=1.0\textwidth]{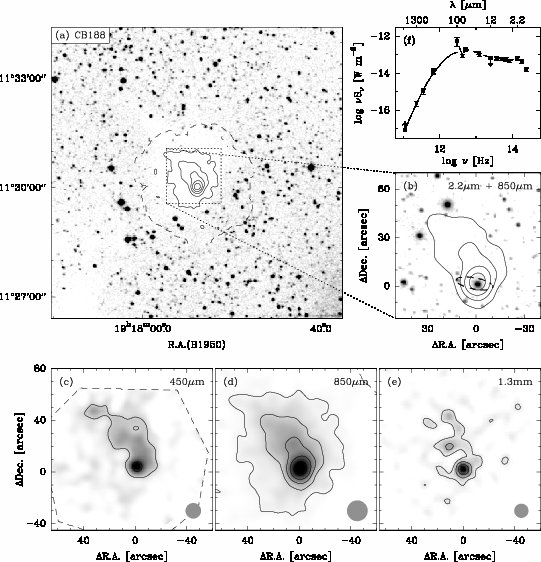}
\caption{\label{fig-cb188} \footnotesize
   CB\,188:
   a) Optical image (DSS2 red) with contours of the 850\,$\mu$m dust continuum
      emission overlaid.
   b) NIR K-band image with 850\,$\mu$m dust continuum contours.
      The IRAS PSC position is marked by a dashed ellipse.
   c) 450\,$\mu$m dust continuum emission; grey scale and contours at
      0.45 and 0.9\,Jy/beam.
   d) 850\,$\mu$m dust continuum emission; grey scale and contours at
      40 to 160 by 40\,mJy/beam.
   e) 1.3\,mm dust continuum emission; grey scale and contours at
      15 to 60 by 15\,mJy/beam. Beam sizes are indicated as
      grey ellipses.
   f) Spectral energy distribution showing the OVRO 3\,mm, IRAM 1.3\,mm, 
      SCUBA 850 and 450\,$\mu$m,
      IRAS PSC, Spitzer MIPS and IRAC, and ground-based NIR data. 
   }
\end{figure*}

%\clearpage
\begin{figure*}[htbp]
\includegraphics[width=1.0\textwidth]{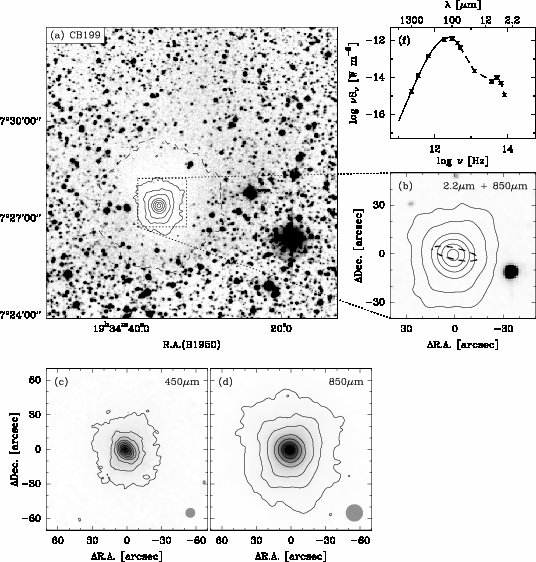}
\caption{\label{fig-b335} \footnotesize
   CB\,199 (B\,335):
   a) Optical image (DSS2 red) with contours of the 850\,$\mu$m dust continuum
      emission overlaid.
   b) NIR K-band image with 850\,$\mu$m dust continuum contours.
      The IRAS PSC position is marked by a dashed ellipse.
   c) 450\,$\mu$m dust continuum emission; grey scale and contours at
      0.4 to 4.4 by 0.5\,Jy/beam.
   d) 850\,$\mu$m dust continuum emission; grey scale and contours at
      60, 120, 200 to 1000 by 200\,mJy/beam. Beam sizes are indicated as
      grey ellipses.
   f) Spectral energy distribution showing the IRAM 1.3\,mm \citep{2001A&A...365..440M}, 
      SCUBA 850 and 450\,$\mu$m, 
      IRAS PSC, and Spitzer MIPS and IRAC data.
   }
\end{figure*}

%\clearpage
\begin{figure*}[htbp]
\includegraphics[width=1.0\textwidth]{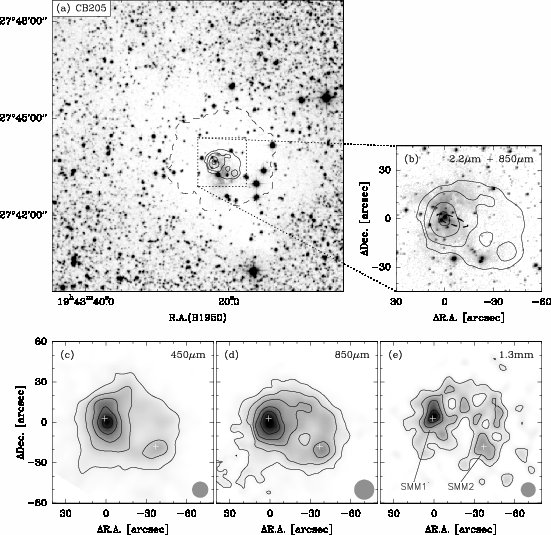}
\caption{\label{fig-cb205} \footnotesize
   CB\,205:
   a) Optical image (DSS2 red) with contours of the 850\,$\mu$m dust continuum
      emission overlaid.
   b) NIR K-band image with 850\,$\mu$m dust continuum contours.
      The IRAS PSC position is marked by a dashed ellipse.
   c) 450\,$\mu$m dust continuum emission; grey scale and contours at
      0.65 to 3.05 by 0.6\,Jy/beam.
   d) 850\,$\mu$m dust continuum emission; grey scale and contours at
      120 to 470 by 70\,mJy/beam.
   e) 1.3\,mm dust continuum emission; grey scale and contours at
      16 to 64 by 16\,mJy/beam. Beam sizes are indicated as
      grey ellipses. Crosses mark the positions of the two main 1.3\,mm peaks.
      No spectral energy distribution is shown since many different, partially 
      unresolved sources contribute to the measured flux densities.
   }
\end{figure*}

%\clearpage
\begin{figure*}[htbp]
\includegraphics[width=1.0\textwidth]{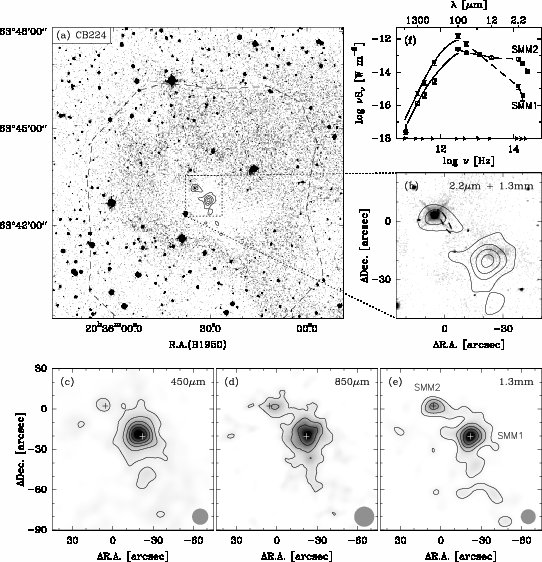}
\caption{\label{fig-cb224} \footnotesize
   CB\,224:
   a) Optical image (DSS2 red) with contours of the 1.3\,mm dust continuum
      emission overlaid.
   b) NIR K-band image with 1.3\,mm dust continuum contours.
      The IRAS PSC position is marked by a dashed ellipse.
   c) 450\,$\mu$m dust continuum emission; grey scale and contours at
      0.24 to 1.44 by 0.24\,Jy/beam.
   d) 850\,$\mu$m dust continuum emission; grey scale and contours at
      60 to 240 by 60\,mJy/beam.
   e) 1.3\,mm dust continuum emission; grey scale and contours at
      12 to 60 by 12\,mJy/beam. Beam sizes are indicated as
      grey ellipses. Crosses mark the positions of the two 1.3\,mm peaks.
   f) Spectral energy distribution of SMM1 (filled squares) 
      and SMM2 (open circles) showing the OVRO 3\,mm, IRAM 1.3\,mm, 
      SCUBA 850 and 450\,$\mu$m, 
      IRAS PSC and FSC, and ground-based NIR data.
   }
\end{figure*}

%\clearpage
\begin{figure*}[htbp]
\includegraphics[width=1.0\textwidth]{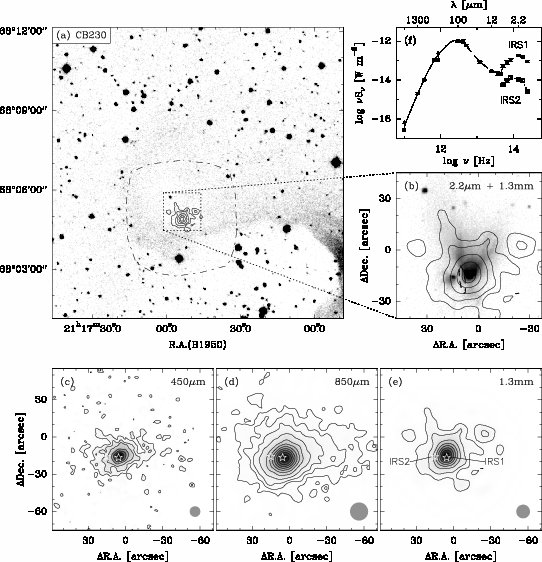}
\caption{\label{fig-cb230} \footnotesize
   CB\,230:
   a) Optical image (DSS2 red) with contours of the 1.3\,mm dust continuum
      emission overlaid.
   b) NIR K-band image with 1.3\,mm dust continuum contours.
      The IRAS PSC position is marked by a dashed ellipse.
   c) 450\,$\mu$m dust continuum emission; grey scale and contours at
      350, 700, 1050 to 3850 by 700\,mJy/beam.
   d) 850\,$\mu$m dust continuum emission; grey scale and contours at
      30, 60, 90, 150 to 900 by 100\,mJy/beam.
   e) 1.3\,mm dust continuum emission; grey scale and contours at
      15, 30 to 210 by 30\,mJy/beam. Beam sizes are indicated as
      grey ellipses.
   f) Spectral energy distributions of IRS1 (filled squares) and IRS2 
      (open circles, NIR to MIR only), showing the OVRO 3\,mm, IRAM 1.3\,mm, 
      SCUBA 850 and 450\,$\mu$m, UKT14 350\,$\mu$m,
      IRAS PSC and FSC, Spitzer MIPS and IRAC 8 to 3.6\,$\mu$m, 
      ISOCAM 6.7\,$\mu$m, and ground-based NIR data. 
      Note that contributions from both IRS1 and IRS2 are included in the submm SED 
      of SMM1 because the source is not resolved at wavelengths longward of 
      10\,$\mu$m. However, IRS1 dominates at all wavelengths. 
   }
\end{figure*}

%\clearpage
\begin{figure*}[htbp]
\includegraphics[width=1.0\textwidth]{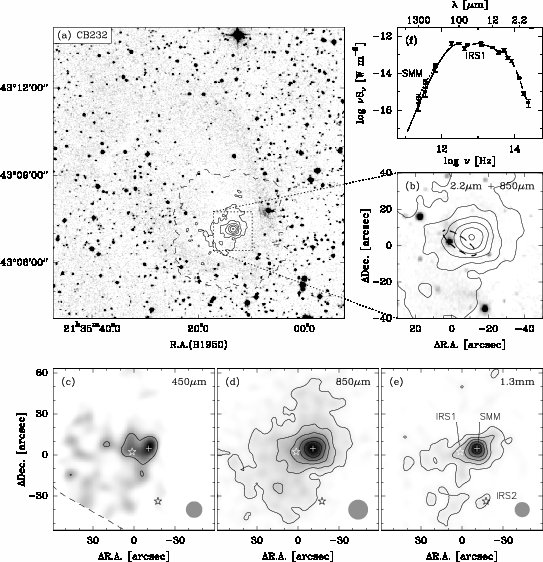}
\caption{\label{fig-cb232} \footnotesize
   CB\,232:
   a) Optical image (DSS2 red) with contours of the 850\,$\mu$m dust continuum
      emission overlaid.
   b) NIR K-band image with 850\,$\mu$m dust continuum contours.
      The IRAS PSC position is marked by a dashed ellipse.
   c) 450\,$\mu$m dust continuum emission; grey scale and contour at
      1.4\,mJy/beam.
   d) 850\,$\mu$m dust continuum emission; grey scale and contours at
      70 to 320 by 50\,mJy/beam.
   e) 1.3\,mm dust continuum emission; grey scale and contours at
      12 to 60 by 12\,mJy/beam. Beam sizes are indicated as
      grey ellipses. The cross marks the position of SMM1 in the 1.3\,mm map. 
      Asterisks mark the position of the two nearby Spitzer IRS.
   f) Spectral energy distribution of CB\,232\,IRS1 (filled squares) 
      and SMM (open circles, submm only), showing the IRAM 1.3\,mm, 
      SCUBA 850 and 450\,$\mu$m, IRAS PSC, Spitzer MIPS and IRAC, and ground-based NIR data. 
   }
\end{figure*}

\clearpage
\begin{figure*}[htbp]
\includegraphics[width=1.0\textwidth]{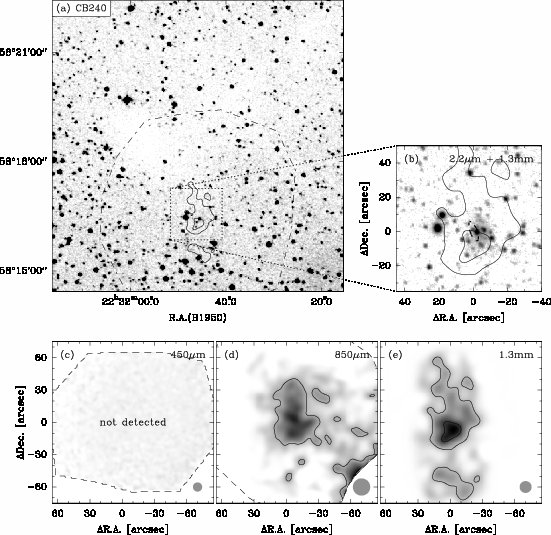}
\caption{\label{fig-cb240} \footnotesize
   CB\,240:
   a) Optical image (DSS2 red) with contours of the 1.3\,mm dust continuum
      emission overlaid.
   b) NIR K-band image with 1.3\,mm dust continuum contours.
      The IRAS PSC position is marked by a dashed ellipse.
   c) 450\,$\mu$m dust continuum map; the source is not detected at rms 
      170\,mJy/beam.
   d) 850\,$\mu$m dust continuum emission; grey scale and contour at
      100\,mJy/beam.
   e) 1.3\,mm dust continuum emission; grey scale and contours at
      16 and 32\,mJy/beam. Beam sizes are indicated as grey ellipses.
      No spectral energy distribution is shown since many different, partially 
      unresolved sources contribute to the measured flux densities.
   }
\end{figure*}

%\clearpage
\begin{figure*}[htbp]
\includegraphics[width=1.0\textwidth]{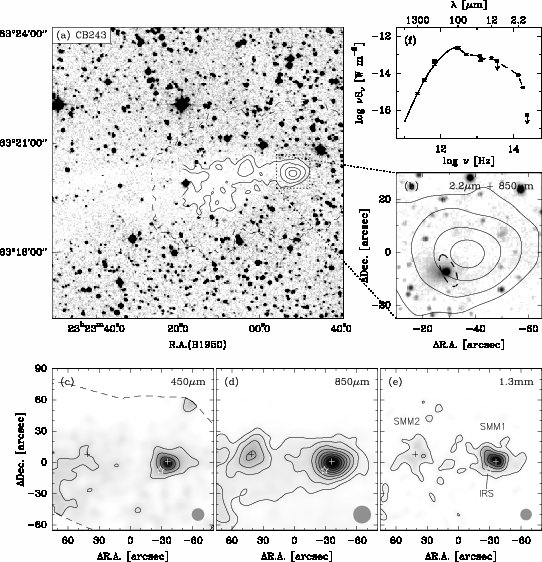}
\caption{\label{fig-cb243} \footnotesize
   CB\,243:
   a) Optical image (DSS2 red) with contours of the 1.3\,mm dust continuum
      emission overlaid.
   b) NIR K-band image with 1.3\,mm dust continuum contours.
      The IRAS PSC position is marked by a dashed ellipse.
   c) 450\,$\mu$m dust continuum emission; grey scale and contours at
      0.5 to 2 by 0.5\,Jy/beam.
   d) 850\,$\mu$m dust continuum emission; grey scale and contours at
      60 to 360 by 60\,mJy/beam.
   e) 1.3\,mm dust continuum emission; grey scale and contours at
      15 to 75 by 15\,mJy/beam. Beam sizes are indicated as
      grey ellipses. Crosses mark the positions of of SMM1 and SMM2.
      The asterisk marks the position of the reddened star and IRAS point source.
   f) Combined spectral energy distribution of SMM1 and IRS, showing the IRAM 1.3\,mm, 
      SCUBA 850 and 450\,$\mu$m, IRAS PSC, Spitzer MIPS, MSX, and ground-based NIR data. 
      Fluxes are dominated by SMM1 at $\lambda\ge 450\,\mu$m and 
      by IRS at $\lambda\le 100\,\mu$m, but the two SEDs could not be separated.
   }
\end{figure*}

%\clearpage
\begin{figure*}[htbp]
\includegraphics[width=1.0\textwidth]{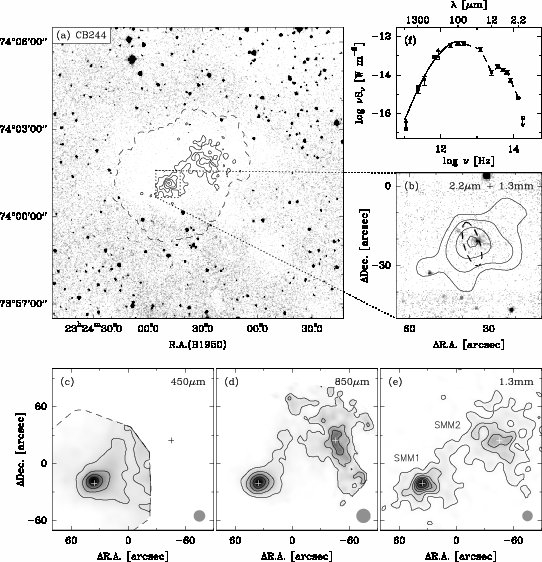}
\caption{\label{fig-cb244} \footnotesize
   CB\,244:
   a) Optical image (DSS2 red) with contours of the 1.3\,mm dust continuum
      emission overlaid.
   b) NIR K-band image with 1.3\,mm dust continuum contours.
      The IRAS PSC position is marked by a dashed ellipse.
   c) 450\,$\mu$m dust continuum emission; grey scale and contours at
      0.6 to 2.2 by 0.5\,Jy/beam.
   d) 850\,$\mu$m dust continuum emission; grey scale and contours at
      100 to 500 by 80\,mJy/beam.
   e) 1.3\,mm dust continuum emission; grey scale and contours at
      15 to 105 by 15\,mJy/beam. Beam sizes are indicated as
      grey ellipses. Crosses mark the positions of SMM1 and SMM2. 
   f) Spectral energy distribution of CB\,244\,MM\,1 showing the OVRO 3\,mm, IRAM 1.3\,mm, 
      SMA 1.1\,mm, SCUBA 850 and 450\,$\mu$m, UKT14 350\,$\mu$m,  
      IRAS PSC and FSC, Spitzer MIPS and IRAC, and ground-based NIR data. 
   }
\end{figure*}

%\clearpage
\begin{figure*}[htbp]
\includegraphics[width=1.0\textwidth]{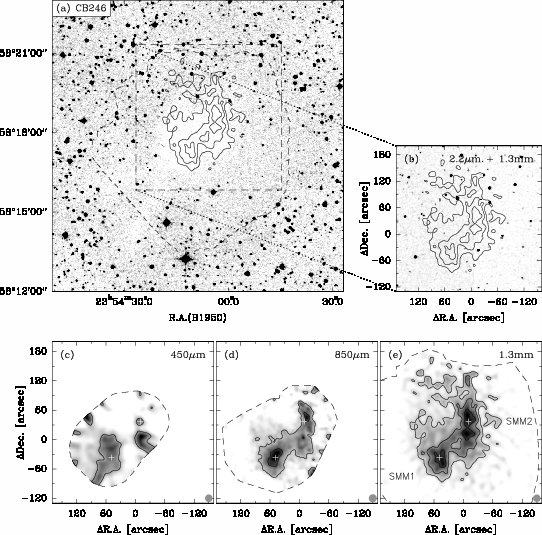}
\caption{\label{fig-cb246} \footnotesize
   CB\,246: 
   a) Optical image (DSS2 red) with contours of the 1.3\,mm dust continuum
      emission overlaid.
   b) NIR K-band image with 1.3\,mm dust continuum contours.
   c) 450\,$\mu$m dust continuum emission; grey scale and contour at
      1.0\,Jy/beam.
   d) 850\,$\mu$m dust continuum emission; grey scale and contours at
      60 and 100\,mJy/beam.
   e) 1.3\,mm dust continuum emission; grey scale and contours at
      15, 25, and 35\,mJy/beam. Beam sizes are indicated as
      grey ellipses. Crosses mark the approximate central positions of the two main 
      submm cores.
      No spectral energy distribution is shown because no fluxes are detected 
      at wavelengths shortward of 450\,$\mu$m.
   }
\end{figure*}

%\clearpage
\begin{figure*}[htbp]
\includegraphics[width=1.0\textwidth]{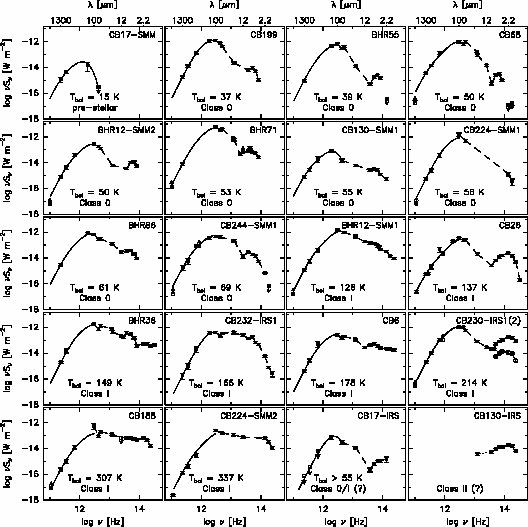}
\caption{\label{fig-sed-all} \footnotesize
Compilation of SEDs (same as in Figs.\,\ref{fig-cb6} through \ref{fig-cb244}), 
ordered by $T_{\rm bol}$\ and evolutionary stage (cf. Table\,\ref{tbl-physprop}).
Note that the classification of the last two sources (CB\,17-IRS and CB\,130-IRS) 
is uncertain because of missing long-wavelengths constraints to the SED.
}
\end{figure*}

%\clearpage
\begin{figure*}[htbp]
\includegraphics[width=0.8\textwidth]{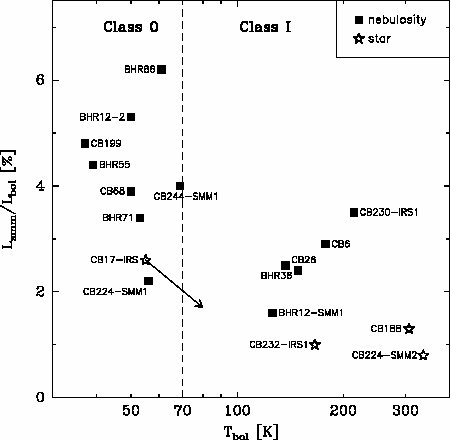}
\caption{\label{fig-tbol-lsubmm} \footnotesize
Ratio of submm to bolometric luminosity vs. bolometric temperature 
for the sources compiled in Table\,\ref{tbl-physprop}. 
The vertical dashed line marks the $T_{\rm bol} = 70$\,K boundary between Class\,0 
and Class\,I sources proposed by \citet{1998ApJ...492..703M}. 
Sources with NIR/MIR nebulosity only (i.e. no star-like point source)
are marked as filled squares.
Sources with NIR point source (star) are marked as open asterisks.
See Sect.\,\ref{sec-dis1-evol} for a discussion of evolutionary tracers.
}
\end{figure*}

\begin{figure*}[htbp]
\includegraphics[width=0.8\textwidth]{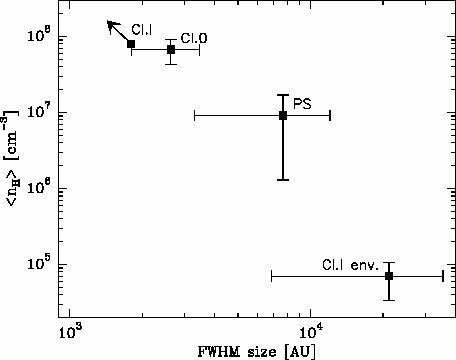}
\caption{\label{fig-glob_prop} \footnotesize
Mean FWHM core size and source-averaged density in globule cores as function of 
evolutionary stage (PS = prestellar core, Class\,0 protostars, Class\,1 YSO 
cores and extended envelopes).
Note that most Class\,I sources exhibit unresolved (sub)mm sources with presumably 
significant flux contributions from circumstellar disks.
}
\end{figure*}

%\clearpage
\begin{figure*}[htbp]
\includegraphics[width=1.0\textwidth]{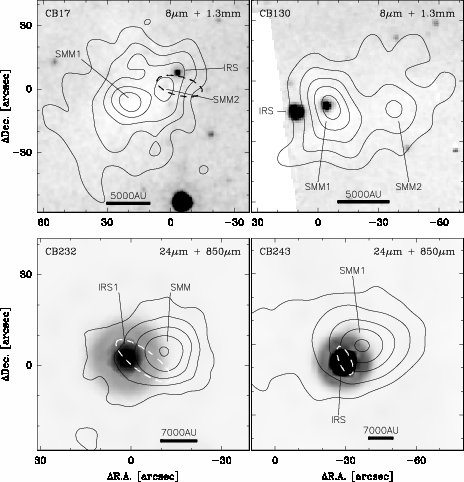}
\caption{\label{fig-spitzermaps} \footnotesize
Four examples of globule cores in which MIR and submm sources are 
not identical and which obviously contain two or more sources of different evolutionary stage 
with a few thousand AU:
{\it Top left:} CB\,17, {\it IRAC} 8\,$\mu$m image (greyscale) and 1.3\,mm dust 
  continuum contours (see Fig.\,\ref{fig-cb17}).
{\it Top right:} CB\,130, {\it IRAC} 8\,$\mu$m image (greyscale) and 1.3\,mm dust 
  continuum contours (see Fig.\,\ref{fig-cb130}).
{\it Bottom left:} CB\,232, {\it MIPS} 24\,$\mu$m image (greyscale) and 850\,$\mu$m dust 
  continuum contours (see Fig.\,\ref{fig-cb232}).
{\it Bottom right:} CB\,243, {\it MIPS} 24\,$\mu$m image (greyscale) and 850\,$\mu$m dust 
  continuum contours (see Fig.\,\ref{fig-cb243}).
IRAS PSC poitions are marked by dashed error ellipses, where applicable. 
Linear scale bars refere to the distances listed in Table\,\ref{tbl-sourcelist}.
See Sect.\,\ref{sec-dis1-mult} for discussion.
}
\end{figure*}

\clearpage

%%%%%%%%%%%%%%%%%%%%%%%%%%%%%%%%%%%%%%%%%%%%%%%%%%%%%%%%%%%%%

\bibliographystyle{apj}                      
\bibliography{globmaps}    %% includes the journal abbrevs

%%%%%%%%%%%%%%%%%%%%%%%%%%%%%%%%%%%%%%%%%%%%%%%%%%%%%%%%%%%%%

\end{document}